\documentclass[prd, aps, twocolumn, showpacs,superscriptaddress,nofootinbib]{revtex4-1}

\usepackage{amsbsy,amsmath,amssymb,amsthm,wasysym,amsfonts}
\usepackage{graphicx}
\usepackage{psfrag}
\usepackage{epstopdf}
\usepackage{color}
\usepackage{mathrsfs}
\usepackage{comment}
\usepackage[normalem]{ulem}
\DeclareSymbolFontAlphabet{\mathrsfs}{rsfs}
\DeclareMathAlphabet\mathbfcal{OMS}{cmsy}{b}{n}

\newcommand{\be}{\begin{equation}}  
\newcommand{\ee}{\end{equation}}
\newcommand{\bea}{\begin{eqnarray}}           
\newcommand{\eea}{\end{eqnarray}} 
\newcommand{\beqn}{\begin{eqnarray*}}
\newcommand{\eeqn}{\end{eqnarray*}}
\newcommand{\ba}{\begin{align}}
\newcommand{\ea}{\end{align}}

\newcommand{\bfx}{{\bf{x}}}
\newcommand{\bfv}{{\bf{v}}}

\def\ii{{\rm i}}

\def\ta{{\tilde{a}}}
\def\ta{{\tilde{a}}}

\def\p4{{\psi_4}} 

\usepackage{float}
\definecolor{cyan}{rgb}{0,0.9,0.9}
\definecolor{orange}{rgb}{0.9,0.5,0}
\definecolor{magenta}{rgb}{1,0,1}
\definecolor{purple}{rgb}{0.8,0.4,0.8}

\definecolor{dodgerblue}{rgb}{0.12, 0.56, 1.0}
\definecolor{alizarincrimson}{rgb}{0.82, 0.1, 0.26}

\begin{document}


\title{Impact of Numerical Relativity information on effective-one-body waveform models}
\author{Alessandro \surname{Nagar}}
\affiliation{Centro Fermi - Museo Storico della Fisica e Centro Studi e Ricerche ``Enrico Fermi'', Rome, Italy}
  \affiliation{INFN Sezione di Torino, Via P.~Giuria 1, 10125 Torino, Italy}
  \affiliation{Institut des Hautes Etudes Scientifiques, 91440
  Bures-sur-Yvette, France}
  \author{Gunnar \surname{Riemenschneider}}
  \affiliation{Dipartimento di Fisica, Universit\`a di Torino, via P. Giuria 1, I-10125 Torino, Italy}
  \affiliation{Physik-Department, Technische Universit\"{a}t M\"{u}nchen, Fames-Franck-Stra{\ss}e 1, 85748 Garching, Germany}
 \author{Geraint \surname{Pratten}}
 \affiliation{Universitat de les Illes Balears, IAC3---IEEC, 07122 Palma de Mallorca, Spain}
  
 \begin{abstract}
   We present a comprehensive comparison of the spin-aligned effective-one-body (EOB) waveform model
   of Nagar~et~al.~[Phys.~Rev.~D~93, 044046 (2016)], informed using 40~ numerical-relativity (NR) datasets,
   against a set of 149 $\ell=m=2$ NR waveforms freely available through the Simulation Extreme
   Spacetime (SXS) catalog. We find that, without further calibration, these EOBNR waveforms have
   unfaithfulness---at design Advanced-LIGO sensitivity and evaluated with total mass $M$
   varying as $10M_\odot\leq M \leq 200M_\odot$---always below $1\%$ against all NR waveforms
   except for three outliers, that still {\it never exceed} the $3\%$ level; with a
   minimal retuning of the (effective) next-to-next-to-next-to-leading-order spin-orbit
   coupling parameter for the non-equal-mass and non-equal-spin sector, that only
   needs {\it three more} NR waveforms, one is left with another two (though different) outliers,
   with maximal unfaithfulness of up to only $2\%$ for a total mass of $200M_\odot$.
   We show this is the effect of slight inaccuracies in the phenomenological description of the
   postmerger waveform of Del Pozzo and Nagar [arXiv:1606.03952] that was constructed by interpolating
   over only 40~NR simulations. We argue that this is easily fixed by using either an alternative
   ringdown description (e.g., the superposition of quasi-normal-modes) or an improved version of
   the phenomenological representation. By analyzing a NR waveform with mass ratio $8$ and dimensionless
   spins $+0.85$ obtained with the BAM code, we conclude that the model would benefit from NR simulations
   specifically targeted at improving the postmerger-ringdown phenomenological fits for mass ratios
   $\gtrsim 8$ and spins $\gtrsim 0.8$. We finally show that some of the longest SXS $q=7$
   waveforms suffer from systematic uncertainties in the postmerger-ringdown part that are interpreted
   as due to unphysical drifts of the center of mass: thus some care should be applied when
   these waveforms are used for informing analytical models.
\end{abstract}

\date{\today}


\maketitle

\section{Introduction}
Effective-one-body (EOB)~\cite{Buonanno:1998gg,Buonanno:2000ef,Damour:2001tu,Buonanno:2005xu}
waveforms informed by numerical relativity~(NR)
simulations~\cite{Damour:2002qh,Damour:2007xr,Damour:2007vq}
have played a central role in the detection, subsequent
parameter-estimation~\cite{TheLIGOScientific:2016wfe}
analyses and GR tests~\cite{TheLIGOScientific:2016src}
of the gravitational-wave (GW) observations GW150914~\cite{Abbott:2016blz} and
GW151226~\cite{Abbott:2016nmj,TheLIGOScientific:2016pea} announced in 2015.
EOB waveforms have also been employed to build frequency-domain, phenomenological models
for the inspiral, merger and ringdown stages of the BBH coalescence~\cite{Khan:2015jqa}.
Those models were also used to infer the properties~\cite{TheLIGOScientific:2016wfe}
and carry out tests~\cite{TheLIGOScientific:2016src} of GR with GW150914 and GW151226.
There are currently two avatars of the EOBNR models: the SEOBNRv* lineage,
that was used both in its SEOBNRv2 (spin-aligned, nonprecessing) and
SEOBNRv3 (generic, precessing, spins)~\cite{Pan:2013rra,Babak:2016tgq} realizations
to analyze LIGO/Virgo data~\cite{Abbott:2016izl}.
On the other hand, there is the {\tt SEOBNR\_ihes} avatar~\cite{Nagar:2015xqa},
that is currently restricted to aligned-spins 
(see however Ref.~\cite{Balmelli:2015zsa} for its extension to generically oriented spins) and has not yet been used in
the analysis of experimental data. Though both models embody the
same conceptual idea (i.e., the two body dynamics is represented as the
dynamics of an effective particle whose conservative part is driven
by some effective potential), there are several nontrivial differences.
Let us just briefly quote some: (i) the choice of the resummation of
the potential of the nonspinning sector; (ii) the treatment of
spin-spin interaction; (iii) the incorporation of the (spinning)
test-particle limit; (iv) the spin-gauge used for the spin-orbit
sector as well as the approach to resumming the gyro-gravitomagnetic
functions; (v) the structure of the next-to-quasi-circular (NQC) corrections;
(vi) the resummation of the residual waveform phases $\delta_{\ell m}$;
(vii) the setup of low-eccentricity initial data; (viii) the choice
of parameters that are informed through NR simulations.
The SEOBNRv2 model~\cite{Taracchini:2013rva} employed in the LIGO/Virgo
data analyses has recently been upgraded to its SEOBNRv4
realization~\cite{Bohe:2016gbl}. This new model features several analytical
upgrades with respect to SEOBNRv2, including: (i) the use of the full
4PN EOB interaction potential, as obtained in Ref.~\cite{Bini:2013zaa}
(and notably already incorporated in the {\tt SEOBNR\_ihes} model~\cite{Nagar:2015xqa})
and (ii) the next-to-next-to-leading order correction to the $(2,2)$ residual
amplitude\footnote{The performance of this term, obtained from the
  results of~\cite{Bohe:2013cla} was already discussed in Ref.~\cite{Nagar:2016ayt},
  where it was argued that additional resummation procedures might be needed to
  improve the behavior of the waveform amplitude towards merger. We recall that
  additional resummations (through Pad\'e approximants) were also found useful
  to be applied to the residual factorized correction to the waveform
  phase~\cite{Damour:2012ky}.}
correction and phase of the factorized waveform~\cite{Damour:2007xr,Damour:2008gu,Pan:2010hz,Faye:2014fra}
that have been recasted from recent PN results~\cite{Bohe:2013cla}.
In addition, the model is visibly improved with respect to SEOBNRv2
by calibrating it to a set of 141 NR waveforms, instead of the original 
38 NR waveforms that were available in the calibration of SEOBNRv2.
This procedure allows the new EOBNR waveforms to have faithfulness
(at Advanced-LIGO design sensitivity) above $99\%$ against all the
NR waveforms discussed in Ref.~\cite{Bohe:2016gbl}, including 16 additional
waveforms used for validation, when maximizing only on the initial phase and time.

Reference~\cite{Nagar:2015xqa}, hereafter Paper~I, discussed the performances of
the {\tt SEOBNR\_ihes} model. Although being analytically different from SEOBNRv2,
that model was similarly improved by using the same 38 NR waveforms used for
the calibration of SEOBNRv2 in Ref.~\cite{Taracchini:2013rva}.
Now that most of the 141 waveforms used for the NR calibration of SEOBNRv4
in~\cite{Bohe:2016gbl} are publicly available through the SXS catalog~\cite{SXS:catalog},
it is urgent to test the robustness of the EOBNR model of Paper~I on such an
extended sample of configurations. Doing this check is the main aim of this paper.
We shall show that the model of Paper~I, once interpolated to
arbitrary mass-ratios and spins, will provide a dominant $\ell=m=2$ waveform mode
whose unfaithfulness is always below $3\%$ against all available NR waveforms. 
More precisely, we will see that it is always below (or equal to) $1\%$ except for
two outliers that are slightly above this threshold. We then proceed to add four
more waveforms to the 39 used in Paper~I to slightly retune the model in order
to provide a slightly more reliable (effective) representation of the spin-orbit
interaction (as well as of the postmerger-ringdown in a few cases)
that yields values of the unfaithfulness always below $1\%$.

The paper is organized as follows. In Sec.~\ref{sec:NRdata} we briefly summarize
the the NR data used in this paper. The cross-check of the SEOBNR model of Paper~I
against the spin-aligned waveforms publicly available in the SXS catalog is
done in Sec.~\ref{sec:checkEOBnr}. The improvement of the model, that is needed
for just a few SXS NR waveforms, is discussed in Sec.~\ref{sec:new_informing},
where it is found that the largest source of inaccuracy comes from the phenomenological
description of the postmerger-ringdown phase based on the interpolating fit
of Ref.~\cite{Nagar:2016iwa}. Section~\ref{sec:q7} discusses in detail
the EOB/NR comparisons for $q=7$, notably pointing out systematic effects in the
SXS waveforms due to the unphysical drift of the center of mass of the system.
Finally, Sec.~\ref{sec:q8-bam}, explores the uncertainties of the model at
the boundary of the ``information'' domain, by performing explicit comparisons
with an equal-spin waveform with mass ratio $q\equiv m_1/m_2=8$ and dimensionless
spins $\chi_1=\chi_2=+0.85$ obtained using the BAM code. We argue that the
largest source of uncertainty comes from the imperfect modeling of the
postmerger-ringdown phase. This is the part of the model that needs the most
radical improvements. After the conclusions in Sec.~\ref{sec:conclusions},
the paper is complemented by an Appendix that summarizes some technical
material that is needed in the main text.

\section{Numerical relativity data}
\label{sec:NRdata}
The numerical waveforms are taken from two independent sets: the publicly available
SXS catalog~\cite{SXS:catalog} produced with the Spectral Einstein Code (SpEC) 
\cite{Scheel:2008rj,Szilagyi:2009qz,Hemberger:2012jz,Buchman:2012dw,
  Hemberger:2013hsa,Scheel:2014ina,Blackman:2015pia,Lovelace:2011nu,
  Lovelace:2010ne,Lovelace:2014twa,Chu:2015kft,Szilagyi:2015rwa,
  Lovelace:2016uwp,Abbott:2016nmj,Abbott:2016apu} 
as well as a single waveform produced by
the \rm{BAM} code~\cite{Bruegmann:2006at,Husa:2007hp,Gonzalez:2006md}. 

We use 149 waveforms from the public SXS catalog~\cite{Mroue:2013xna} to validate the model.
Conservative error bars are estimated by computing the phase difference at the NR merger
(defined as the peak of the $\ell=m=2$ waveform amplitude) of the highest-resolution (H) waveform 
between the highest and second highest (SH) resolutions
\be
\delta\phi^{\rm{NR}}_{\rm{mrg}} \equiv \left(\phi^{\rm{H}}(t)- \phi^{\rm{SH}}(t)\right)|_{t=t^{\rm H}_{\rm mrg}} .
\ee
This quantity is monotonically increasing up to merger and is zero at the start of the two simulations.
For those simulations for which we only have a single resolution available, no error bar is given.
For the SXS simulations, we use the asymptotic waveforms that have been extrapolated to future
null infinity. As in Paper~I, we use the $N=3$ extrapolation order for the highest available
resolutions in each simulation. The complete information for the NR runs used in this paper
(and in particular the uncertainties at merger) can be found in
Tables~\ref{tab:nospin}-\ref{tab:q3}.

In addition to the SXS simulations, we also use a BAM $(8,+0.85,+0.85)$ waveform corresponding
to approximately $15.7$ gravitational wave cycles before merger~\cite{Husa:2015iqa}.

\section{Checking the SEOBNR\_ihes model of Paper~I}
\label{sec:checkEOBnr}

\subsection{Summary of the available analytical flexibility}
The {\tt SEOBNR\_ihes} model NR informed in Paper~I has the following analytical flexibility:
(i) an effective 5PN coefficient $a_6^c$ entering the nonspinning sector of the model;
(ii) an effective next-to-next-to-next-to-leading-order (NNNLO) spin-orbit
parameter $c_3$, that is tuned in order to make the spin-orbit interaction stronger
or milder with respect to the simple NNLO (resummed) analytical prediction encoded in
the gyro-gravitomagnetic functions $G_S$ and $G_{S_*}$~\cite{Nagar:2011fx,Barausse:2011ys,Damour:2014sva},
see Eqs.~(41) and (42) of~\cite{Damour:2014sva}; (iii) a parameter $\Delta t_{\rm NQC}$
aimed at connecting the EOB and NR time axis, that is given by Eq.~(21) of Paper~I;
(iv) four next-to-quasi-circular (NQC)
parameters ($a_1,a_2,b_1,b_2)$ entering the $\ell=m=2$
waveform $h_{22}\equiv h_{22}^{\rm circ}\hat{h}_{22}^{\rm NQC}(a_1,a_2,b_1,b_2)$
(see Eq.~(95) of~\cite{Damour:2014sva}) that are determined
by imposing osculation between the EOB and NR amplitude and
frequency (and their time derivatives) close to
merger\footnote{More precisely, the NR waveform point used to do
so is located at $t_{\rm NQC}^{\rm NR}/M=(t_{\rm mrg}^{\rm NR}+2)/M$, where
$t_{\rm mrg}^{\rm NR}$ is the NR merger time defined as usual as the
peak of the $\ell=m=2$ waveform amplitude and $M\equiv m_1+m_2$
the total mass of the system.};
(v) the phenomenological description of the postmerger-ringdown
waveform part introduced in Ref.~\cite{Damour:2014yha}.
In Paper~I, in order to accomplish a very accurate description of both the NQC
corrections and the ringdown, the {\it exact} values of the
NR NQC-extraction point was used to determine the parameters $(a_i,b_i)$,
as well as the primary ringdown fit of~\cite{Damour:2014yha},
instead of some effective interpolation based on a sparse number
of simulations. This choice (that is behind the central result of Paper~I,
the computation of the unfaithfulness in Fig.~9 and Fig.~21)
was done in order to reduce the systematics in the determination
of $c_3$ (or $a_6^c$) related to the small sample of NR waveform
data available, with rather little information available outside
the equal-mass, equal-spin regime. By contrast, here we start
by using the same values of $(a_6^c,c_3)$ determined in Paper~I,
but we follow a different route to obtain the NQC parameters and
to model the postmerger-ringdown waveform.
More precisely: (i) the postmerger-ringdown waveform description adopted here
is given by the global interpolating fits of Ref.~\cite{Nagar:2016iwa}
for the spin-aligned case and a special fit (presented here for the first time)
for the nonspinning case (see Sec.~\ref{sec:nospin} below);
(ii) we similarly use a different treatment for the NQC corrections:
for the nonspinning case, the NQC parameters $(a_i,b_i)$ are obtained
as fits of the ``exact'' coefficients determined after iterations (because
then $\hat{h}^{\rm NQC}_{22}$ is included in the radiation
sector~\cite{Damour:2014sva}) by imposing osculation with
the actual NR waveform data point; on the contrary, for the spinning case,
such fits are not sufficiently accurate so we instead present two different
fits to the NR NQC-extraction point with the final NQC parameters
being obtained after several iterations.

\subsection{The nonspinning sector}
\label{sec:nospin}
We use the functional form of $a_6^c(\nu)$ obtained in Paper~I,
that is $a_6^c(\nu)=3097.3\nu^2 - 1330.6\nu+81.38$ where $\nu\equiv m_1 m_2/M^2$
is the symmetric mass ratio. This fit was determined using the first
nine datasets in Table~\ref{tab:nospin}, with mass ratios
$1\leq q \leq 9.99$, plus a $q=1$ 15-orbit long dataset mentioned in Table~I of Paper~I.
Here and below (as it was the case of Paper~I) we use waveforms that have been extrapolated
to future null infinity by using a polynomial of order $N=3$, using the highest resolution available
in the SXS catalog. In addition, for each of these datasets one can determine
(after a few iterations) the corresponding NQC parameters
$(a_1,a_2,b_1,b_2)$ that parametrize
the $\ell=m=2$ NQC factor as
\be
\hat{h}_{22}^{\rm NQC}=(1+a_1 n_1 + a_2 n_2)e^{{\rm i}(b_1 n_1'+b_2n_2') },
\ee
where $(n_1,n_2,n_1',n_2')$ are functions that explicitly depend on
the radial momentum and that are listed in Eq.~(96) of Ref.~\cite{Damour:2014sva}.
The obtained values  $(a_i,b_i)$ can then be accurately fit with
a rational function in terms of the variable $X_\nu\equiv 1-4\nu$. Using this procedure, we obtain
\begin{align}
a_1 &= -0.08052\,\dfrac{ 1 - 2.0033 X_\nu^2}{ 1 + 3.0859 X_\nu^2},\\
a_2 &=  1.52995\,\dfrac{ 1 + 1.1643 X_\nu^2}{ 1 + 1.9203 X_\nu^2},\\
b_1 &= 0.146768( 0.0742 X_\nu + 1.01691256),\\
b_2 &= 0.896911(-0.6107 X_\nu + 0.94295129).
\end{align}
For the phenomenological description of the postmerger-ringdown part,
we (i) first produce the postmerger primary fits as in~\cite{Damour:2014yha}
and then (ii) fit the coefficients as functions of $\nu$.
Referring the reader to Eqs.~(4)-(10) of Ref.~\cite{Damour:2014yha}
for the coefficients entering the fit, we obtain the following
$\nu$ dependences:
\begin{align}
  \alpha_{21}(\nu)    & = -0.3396\nu^2 + 0.0165\nu + 0.1817,\\
  \alpha_1(\nu)      & = -0.1809\nu^2 + 0.0216\nu + 0.0872,\\
  c_3^A(\nu)         & =  0.9211\nu  - 0.58477,\\
  c_3^\phi(\nu)       & = -2.225\nu  + 4.612,\\
  c_4^\phi(\nu)       & = -7.2958\nu  + 3.9702,\\
  \Delta\omega(\nu)  & =  1.13387\nu^2 - 0.006827\nu + 0.114279,\\
  A_{22}^{\rm mrg}(\nu) & =  1.232\nu^2 + 0.3113\nu + 1.4208.                                       
\end{align}
Note that in order to reduce systematic uncertainties, the primary fit
behind these numbers have been obtained by using the values of the final black
hole mass and angular momentum $(M_f,j_f)$ as given in the {\tt metadata.txt}
file for each SXS simulation. Evidently, when interpolating all over the
parameter space we use $(M_f,j_f)$ as provided by the NR fits of
Healy et al.~\cite{Healy:2014yta}. By contrast, the QNMs frequencies
are not computed by using the interpolating fits of Ref.~\cite{Berti:2005ys}
but are directly interpolated (with a cubic spline interpolant)
from the the tables provided on E.~Berti's website~\cite{berti:web}.

Since in Paper~I we were not using interpolating fits for either the
NQC parameters or the postmerger-ringdown ones, it is
interesting to evaluate again the quality of the
nonspinning model in terms of unfaithfulness. The EOB/NR
unfaithfulness (as function of the total mass $M$ of the binary)
is defined as
\be
\label{eq:barF}
\bar{F}(M) \equiv 1 -\max_{t_0,\phi0}\dfrac{\langle h_{22}^{\rm EOB},h_{22}^{\rm NR}\rangle}{||h_{22}^{\rm EOB}||||h_{22}^{\rm NR}||},
\ee
where $(t_0,\phi_0)$ are the initial time and phase, $||h||\equiv \sqrt{\langle h,h\rangle}$,
and the inner product between two waveforms is defined as 
$\langle h_1,h_2\rangle\equiv 4\Re \int_{f_{\rm min}^{\rm NR}(M)}^\infty \tilde{h}_1(f)\tilde{h}_2^*(f)/S_n(f)\, df$,
where $\tilde{h}(f)$ denotes the Fourier transform of $h(t)$, $S_n(f)$ is the zero-detuned,
high-power noise spectral density of advanced LIGO~\cite{shoemaker} and
$f_{\rm min}^{\rm NR}(M)=\hat{f}^{\rm NR}_{\rm min}/M$ is the {\it starting frequency of the NR waveform}
(after the junk radiation initial transient). Both EOB and NR waveforms are tapered in the
time-domain so as to reduce high-frequency oscillations in the corresponding Fourier transforms.
\begin{figure}[t]
\center
\includegraphics[width=0.45\textwidth]{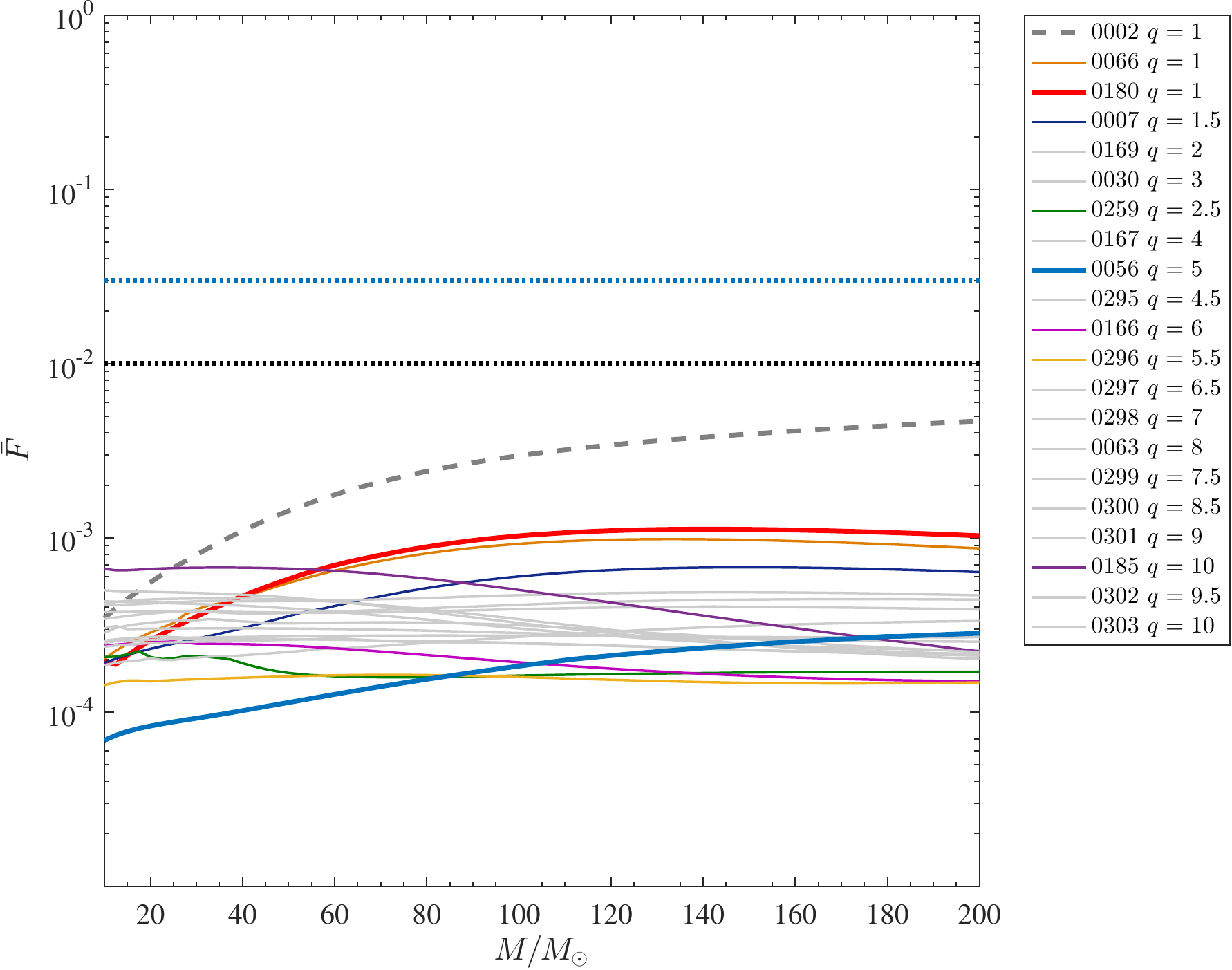}
\caption{\label{fig:barF_spin0} EOB/NR unfaithfulness, Eq.~\eqref{eq:barF},
  for all the nonspinning configurations of Table~\ref{tab:nospin}.
  The outlier behavior of the 0002 configuration is due to large
  inaccuracies in the postmerger region of the NR waveform, see Fig.~\ref{fig:q1_amp_omg}.}
\end{figure}
Figure~\ref{fig:barF_spin0} shows how $\bar{F}$, computed  for $10M_\odot\leq M \leq 200M_\odot$,
always remains well below the reference level of $1\%$, that translates into a
negligible loss of events. For all waveforms, we find that 
$10^{-4}\lesssim \bar{F}\lesssim 10^{-3}$, except for a 32-orbit
equal-mass simulation SXS:BBH:0002. The reason for this is that the NR merger and ringdown is rather inaccurate. 
This can be seen by comparing to $\bar{F}$ for the other two,
equal-mass, 28-orbit long, datasets which saturate at the $0.1\%$ level.
The $\bar{F}$ plot itself is not very informative for understanding what
is going on. Instead, we find a usual time-domain analysis, comparing the
EOB and NR amplitude and frequencies, more illustrative, as shown in Fig.~\ref{fig:q1_amp_omg}.
In this figure, as well as below, $\Psi_{22}$ indicates the Regge-Wheeler-Zerilli
normalized waveform~\cite{Nagar:2005ea}, defined as $\Psi_{22}\equiv {\cal R} h_{22}/\sqrt{24}$
where $h_{22}$ is the $\ell=m=2$ strain waveform in
$h_+ - {\rm i} h_\times=\dfrac{1}{\cal R}\sum_{\ell m}h_{\ell m}{}_{-2}Y_{\ell m}(\theta,\phi)$. 
\begin{figure}[t]
\center
\includegraphics[width=0.45\textwidth]{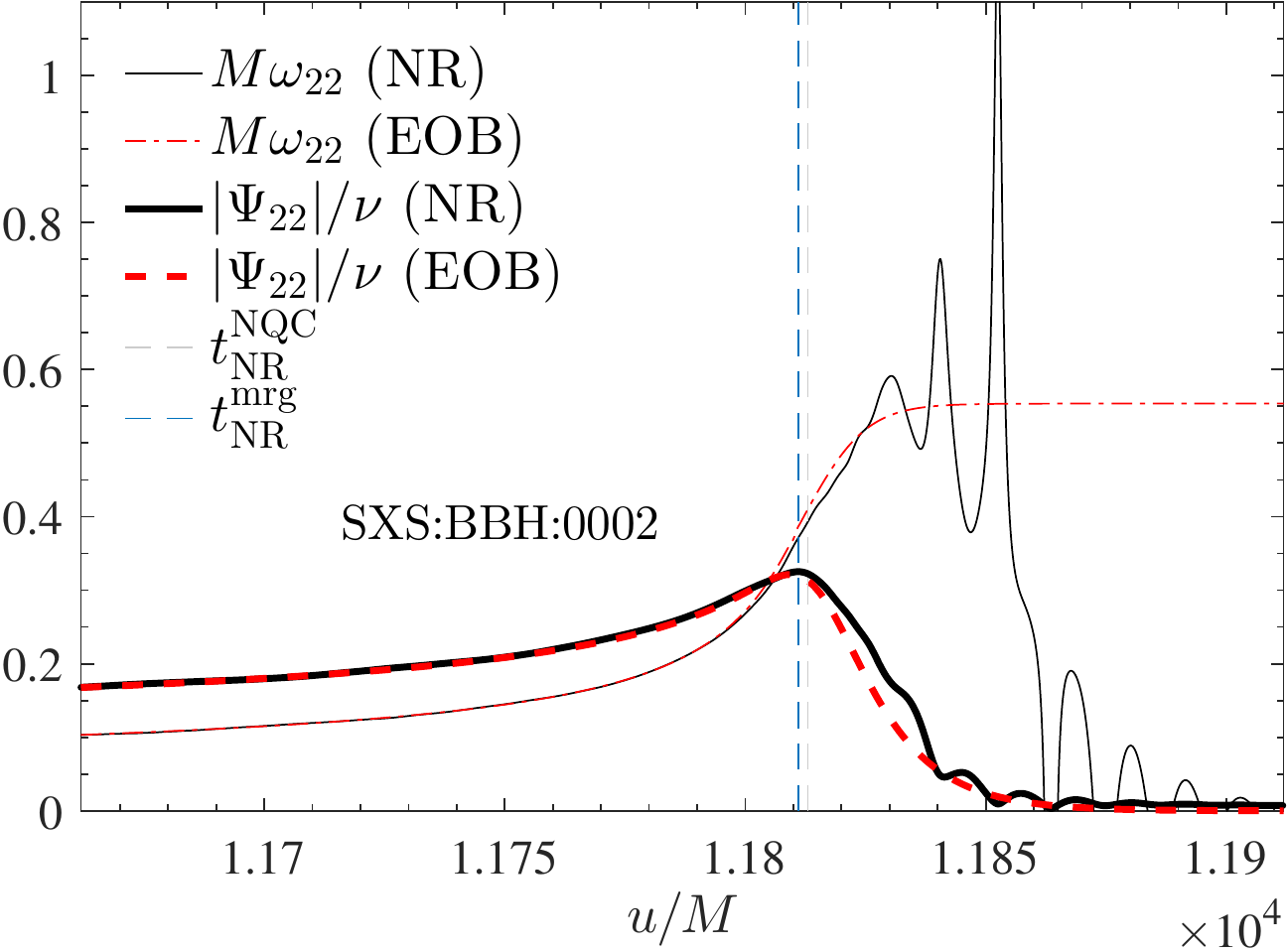}\\
\vspace{5mm}
\includegraphics[width=0.45\textwidth]{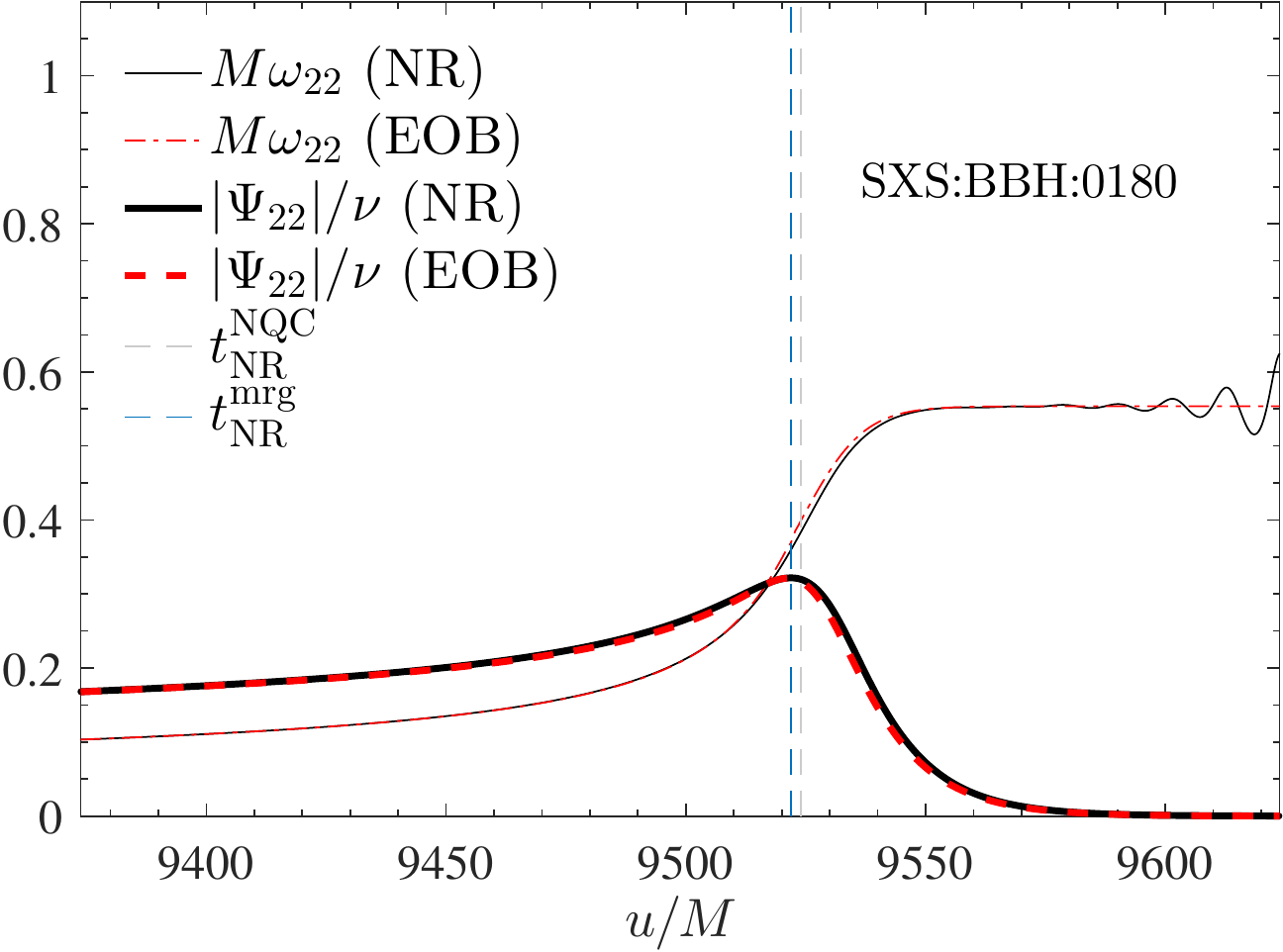}\\
\caption{\label{fig:q1_amp_omg}Equal-mass case: EOB/NR comparisons of amplitudes
  and frequencies for two equal-mass datasets: SXS:BBH:0002 (32~orbits)
  and SXS:BBH:180 (28~orbits). The numerical problem after the merger
  are evident in SXS:BBH:0002. Even if SXS:BBH:0180 was not used in
  calibration the model in Paper~I, it shows excellent consistency
  with the EOB waveform. The NR merger time $t_{\rm mrg}^{\rm NR}$ as
  well as the time location of the NQC-extraction point $t_{\rm NQC}^{\rm NR}/M\equiv (t_{\rm mrg}^{\rm NR}+2)/M$
  are explicitly marked on the plot.}
\end{figure}
The waves are aligned following our usual alignment
procedure (described in details in Sec.~VA of Ref.~\cite{Baiotti:2011am})
on the following, early inspiral, frequency intervals:
$[\omega_L,\omega_R]=[0.0248,0.0305]$
for SXS:BBH:0002 and $[\omega_L,\omega_R] = [0.0256,0.0317]$
for SXS:BBH:0180. The plot is done versus $u/M$, denoting the NR retarded time at future null infinity.
The numerical problems in postmerger phase of
SXS:BBH:0002 are strikingly evident in this plot. Still, the
EOB/NR agreement is excellent up to the NR merger point, which
is explicitly indicated on the figure\footnote{The NR point used
  for NQC determination, at $t_{\rm NQC}^{\rm NR}=t_{\rm mrg}^{\rm NR}+2$,
  is also explicitly marked for convenience.}.
As mentioned in Paper~I, this remains remarkable because
the value of $a_5^c$ was determined exploiting only a 14-orbit-long
waveform produced in one of the early SXS runs~\cite{Buchman:2012dw},
which is not part of the SXS catalogue but was used in the 
calibration of previous EOB models~\cite{Pan:2011gk,Damour:2012ky}. 
All considered, Fig.~\ref{fig:barF_spin0} illustrates that our
nonspinning EOB model is rather robust, simple and efficient
(thanks to the global fits mentioned above, there is no need
of iterating to generate waveforms for production purposes)
and the EOB/NR values of $\bar{F}$ are consistent with
the expected numerical uncertainty once expressed in terms
of this quantity~\cite{Chu:2015kft}. 

\subsection{The spinning sector}
\label{sec:spin0}
Let us now proceed to the spinning sector, that was informed, in Paper~I,
using a sample of 30 spin-aligned NR simulations. These spinning configurations
are listed, for convenience, in Table~\ref{tab:spinold}. Starting from this data,
Paper~I determined the following analytical representation for the NNNLO effective
spin-orbit parameter $c_3$ as function of the individual spins
\begin{align}
\label{eq:c3fun}
& c_3(\ta_1,\ta_2,\nu)  = p_0 \dfrac{1 + n_1(\ta_1+\ta_2) + n_2(\ta_1+\ta_2)^2}{1 + d_1(\ta_1+\ta_2)}\nonumber\\
               & + (p_1\nu  + p_2\nu^2 + p_2\nu^3) (\ta_1 + \ta_2)\sqrt{1-4\nu} \nonumber\\
              & + p_4 (\ta_1-\ta_2)\nu^2,
\end{align}
where $\tilde{a}_{1,2}\equiv X_{1,2} \,\chi_{1,2}$. Here we have also defined $X_1=\dfrac{1}{2}(1+\sqrt{1-4\nu})$ and $X_2=1-X1$. The dimensionless spins are denoted by
$\chi_i \equiv S_i/m_i^2$ with $i=1,2$. The coefficients for $c_3$ are given by
\footnote{Note that in Ref.~\cite{Nagar:2015xqa} there was a typo
  in the definition of $c_3$ as well as in the value of $p_0$, that should read $44.822889$
  instead of  $44.786477$.}
\begin{align}
\label{c3_p0}
p_0 & =  +44.822889,\\
\label{c3_n1}
n_1 & =  -1.879350,\\
\label{c3_n2}
n_2 & =  + 0.894242,\\
\label{c3_d1}
d_1 & =  -0.797702,\\
\label{c3_p1}
p_1 & =  +1222.36, \\
\label{c3_p2}
p_2 & =  -12764.4, \\
\label{c3_p3}
p_3 & =  +36689.6,\\
\label{c3_p4}
p_4 & = -358.086.
\end{align}
The function~\eqref{eq:c3fun} is composed of two parts: a part that accounts for equal-mass,
spin aligned binaries, given by a rational function of only $\hat{a}_0\equiv \tilde{a}_1+\tilde{a}_2$,
and a further two terms that parametrize the deviations away from the equal-mass, equal-spin regime.
The procedure to obtain the fitting parameters $\{p_0,n_1,n_2,d_1,p_1,p_2,p_3,p_4\}$ is as follows.
First, we focused on the equal-mass, equal-spin case and for each corresponding dataset of Table~\ref{tab:spinold}
$c_3$ wass determined by requiring that the phase difference between the EOB and NR waveforms, aligned
in the early inspiral, and computed at NR merger, is comparable or smaller than the corresponding
phase uncertainty at the NR merger point; interestingly, it was easy to find ``good'' values of
$c_3$ that approximately lie on a straight line (except for very large spins), which
eventually yielded to the rational function representation in Eq.~\eqref{eq:c3fun}.
A similar procedure was then followed for the unequal-mass, unequal-spin regime, similarly finding good
values of $c_3$ that were then globally fitted so as to obtain the other two terms of Eq.~\eqref{eq:c3fun}.
To give a few more explicit details about this procedure, Table~\ref{tab:c3uneq} lists our
preferred, ``first-guess'', values of $c_3$ for the 13 asymmetric datasets used as well as
the final values provided by the interpolating fit~\eqref{eq:c3fun}. We want to stress that obtaining
the first-guess values is very easy due to the controllable physical effect of $c_3$,
that enters as a parameter in the (resummed) gyro-gravitomagnetic functions of Ref.~\cite{Damour:2014sva},
see in particular Eqs.~(41)-(42), (45) and (52). Tuning this parameter essentially corresponds to
making the spin-orbit interaction stronger or weaker. In practical terms, this amounts to having
a larger or smaller number of GW cycles between the inspiral and merger.
Once the NR and EOB waveforms are aligned in the early inspiral, it is easy then to find ``good''
values of $c_3$ simply by hand. This is how the values in the third columns of Table~\ref{tab:c3uneq}
were actually obtained.
\begin{table}[t]
\caption{\label{tab:c3uneq}First-guess values of $c_{3}$ compared with the values 
  obtained from the interpolating fit.~\eqref{eq:c3fun} for the small
  sample of NR datasets used to inform the model of Paper~I away from
  the equal-mass, equal-spin limit.}
\begin{center}
\begin{ruledtabular}
  \begin{tabular}{lcccc}
    $(q,\chi_1,\chi_2)$ & $\hat{a}_0$ & $c_3^{\rm  first-guess}$ & $c_3^{\rm fit}$ & \\
    \hline
    \hline
  $(1,-0.5,0)$     & $-0.2500$  & 61.5 & 62.61 \\
  $(1.5,-0.5,0)$   & $-0.300$   & 62   & 61.74\\
  $(3,-0.5,0)$     & $-0.375$   & 63   & 63.69\\
  $(3,-0.5,-0.5)$  & $-0.500$   & 68   & 66.89\\
  $(5,-0.5,0)$     & $-0.4167$  & 62   & 62.02\\
  $(8,-0.5,0)$     & $-0.4444$  & 57   & 57.22\\
  $(1,+0.5,0)$     & $+0.2500$  & 26.5 & 27.22\\
  $(1.5,0.5,0)$    & $+0.3000$  & 26   & 28.17\\
  $(2,+0.6,0)$     & $+0.4000$  & 26.5 & 24.54\\
  $(3,+0.5,0)$     & $+0.3750$  & 26   & 26.38\\  
  $(3,+0.5,+0.5)$  & $+0.5000$  & 24   & 23.59\\
  $(5,+0.5,0)$     & $+0.4167$  & 28   & 28.17\\
  $(8,+0.5,0)$     & $+0.4444$  & 33   & 33.06\\
\end{tabular}
\end{ruledtabular}
\end{center}
\label{tab:fitting}
\end{table}
In addition to this, the model has to be completed with NQC corrections and an expression
of the NQC parameters over the full parameter space is required. Whilst obtaining a direct 
fit of $(a_i,b_i)$, as in the nonspinning case above, would ideally be the best solution
(because once the model is NQC-informed by a restricted sample of NR data sets there is
no need to iterate outside this calibration domain), in practice it proved difficult
to find simple and easy-to-fit behavior, notably due to the large magnitude 
of the NQC amplitude parameters for equal-mass,
large spin binaries, $\chi_1=\chi_2\gtrsim 0.9$. The reason behind this is that the
correction that has to be applied to the EOB waveforms might be rather larger towards
merger for these configurations. Since the value of the radial momentum is small,
one needs the NQC parameters to increase (they can be up to 4 or 5 times larger than the corresponding
values for milder, or anti-aligned, spins) in order to be able to correct the waveform.
As pointed out in Ref.~\cite{Nagar:2016ayt}, an improved resummation of the factorized waveform
amplitude can reduce the need of large NQC corrections and, eventually, help in obtaining
simpler but accurate global fits of the NQC parameter. Since we postpone this study to future
work, we have to find an alternative method to interpolate the NQC parameters outside the
domain of calibration. We do so by looking for a global representation of the NR
NQC-extraction point.
\begin{figure}[t]
\center
\includegraphics[width=0.5\textwidth]{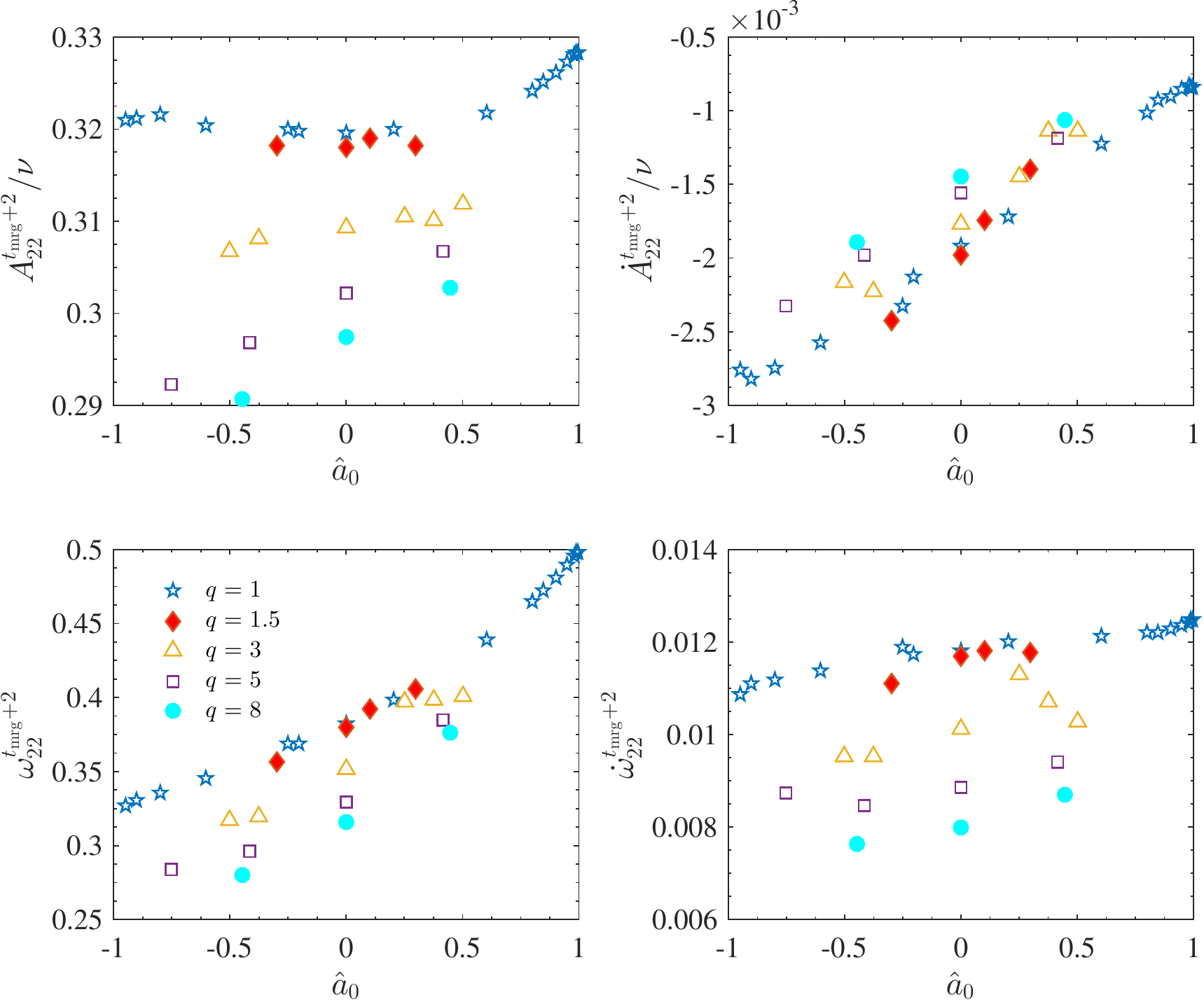}
\caption{\label{fig:NQC_0}Dependence of the NR NQC-extraction point
  on $\hat{a}_0\equiv X_1\chi_1+X_2\chi_2$. Left panels: amplitude (top)
  and frequency (bottom) at $t_{\rm NR}^{\rm NQC}/M=(t_{\rm mrg}^{\rm NR}+2)/M$.
  Right panels: corresponding time derivatives at the same time.
  The four points with $q=1.5$ behave inconsistently with respect to the $q=1$
  and $q=3$ data, and are thus discarded to compute the global interpolating
  fits, see Eq.~\eqref{eq:omg22NQC0} and
  Tables~\ref{NQC_hybrid_q1}-\ref{NQC_hybrid_ql1} in Appendix~\ref{app:nqc}.}
\end{figure}
Figure~\ref{fig:NQC_0} illustrates the behavior of the NR quantities
$(A_{22},\dot{A}_{22},\omega_{22},\dot{\omega}_{22})^{\rm NR}_{t^{\rm NR}_{\rm NQC}}$ at
the NQC-extraction time, $t_{\rm NQC}^{\rm NR}/M=(t_{\rm mrg}^{\rm NR}+2)/M$,
where $A_{22}\equiv |\Psi_{22}|/\nu$ and $\omega_{22}\equiv M\dot{\phi}_{22}$,
with $\phi_{22}$ the GW phase and the overdot indicating time derivatives.
The figure includes data taken from all configurations used in Paper~I
plus those extracted from dataset SXS:BBH:0208, with $(5,-0.90,0)$.
Inspecting the figure, one immediately
notices that the $q=1.5$ points behave somehow differently from the others.
This is rather evident in the behavior of $\dot{A}_{22}$ as well as $\omega_{22}$,
that are crossing the other $q=1$ and $q=3$ points instead of following the same trend
and being in the middle. To avoid inclusion of possible systematics,
we then drop all $q=1.5$ datasets. Then we proceed by doing a two-step fit:
(i) first, for each value of $\nu$ we fit the points versus $\hat{a}_0$
(with polynomials or rational functions, see below) then (ii) we fit the
coefficients of this linearly versus $\nu$.
In practice, since we have many points for $q=1$, we use two separate fits:
(i) a special fit, very accurate, obtained for $q=1$ only, that employs
fourth-order polynomials in $\hat{a}_0$
for $(A_{22},\dot{A}_{22},\omega_{22},\dot{\omega}_{22})$; (ii) a second fit
that takes into account all data, but that uses a two-parameters rational
function in $\hat{a}_0$ for $\omega_{22}$ and quadratic functions in $\hat{a}_0$
for the other quantities. More quantitative details about these fits 
are given in Appendix~\ref{app:nqc}.
Using this ``interpolated'' representation of the NR NQC functioning point,
$(A_{22},\dot{A}_{22},\omega_{22},\dot{\omega}_{22})^{\rm fit}$, together with the functions
for $c_3$ and $a_5^c$ discussed above, we are now in the position of checking the
faithfulness of the model all over the spin-aligned configurations available in
the SXS catalog. We note that these are actually a few {\it less} than those used
in Ref.~\cite{Bohe:2016gbl}, since specific dataset produced for that study are not
public yet. We generate EOB waveforms and we compute $\bar{F}$. The EOB 
waveforms are typically very long, since the relative initial separation is $r_{0}/M=24$
in terms of the EOB radial coordinates. Then, they are cut at some geometric
frequency corresponding to a time on the NR time-axis of a few hundreds
$M$'s after the junk radiation. Both NR and EOB waveforms are then tapered
before taking the Fourier transform. The total mass of the binary is varied
between $10M_\odot\leq M\leq 200M_\odot$; results are shown in Fig.~\ref{fig:barF_0}.
As in the nonspinning case discussed above, the horizontal, dotted, blue line
marks the $3\%$ limit, while the horizontal, black, dotted line marks the
$1\%$ limit. The figure illustrates that: (i) despite relying on the fact 
that $(a_6^c(\nu),c_3)$ as well as the NQC and ringdown parameters were informed using 
the rather small sample of 39 NR waveforms (and notably {\it only} the 13 asymmetric
NR datasets of Table~\ref{tab:c3uneq}, the unfaithfulness is always below $3\%$ 
over the {\it whole} NR catalog of publicly available 149 spin-aligned waveforms;
(ii) more precisely, $\max{(\bar{F})}$ turns out to be  {\it always below $1\%$}
(and when it is so the corresponding line is depicted grey in Fig.~\ref{fig:barF_0})
except for two dataset that just hit this level $(2,+0.85,+0.85)$
(with $\max{(\bar{F})}=1.04\%$)  and $(3,+0.73,-0.85)$ (with $\max{\bar{F}}=1.05\%$),
and another two that are larger $(1,+0.90,0)$ (with $\max{(\bar{F})}=1.39\%$) 
and more importantly $(1,+0.90,+0.50)$, that reaches $\max{(\bar{F})}=2.49\%$.
None of these datasets was available at the time of Paper~I; (iii) the results
of Paper~I, and in particular Fig.~9 there, are essentially confirmed, though
just slightly worsened because of the use of interpolating fits for both the
NQC point and the postmerger-ringdown part; (iii) the line depicted in colors
in the figure are datasets that were found ``problematic'' (i.e. with $\max{(\bar{F})}>3\%$)
for SEOBNRv2, as illustrated in Fig.~2 of Ref.~\cite{Bohe:2016gbl},
but they are below or comparable to the $1\%$ level here. It is worth commenting
in particular the case $(1,-0.90,+0.90)$: since the spins are equal and antialigned and $q=1$,
one can easily see inspecting analytically the EOB dynamics that the spin-orbit
interaction gets essentially canceled and it is only the effect of the spin-spin
interaction (that is subdominant) that cumulates with to the orbital dynamics.
Since we see (check Table~\ref{tab:nospin}) that there is a good match between
EOB/NR for $(1,0,0)$ one a priori expects that $(1,-0.90,+0.90)$ will work
similarly well. Not surprisingly, then, this is what we find in Fig.~\ref{fig:barF_0}
with $\bar{F}$ of the order of $0.1\%$ over the full mass range. This
looks rather different from the SEOBNRv2 case, where $\bar{F}$ gets
above the $3\%$ limit, see Fig.~2 of Ref.~\cite{Bohe:2016gbl}.
\begin{figure}[t]
\center
\includegraphics[width=0.42\textwidth]{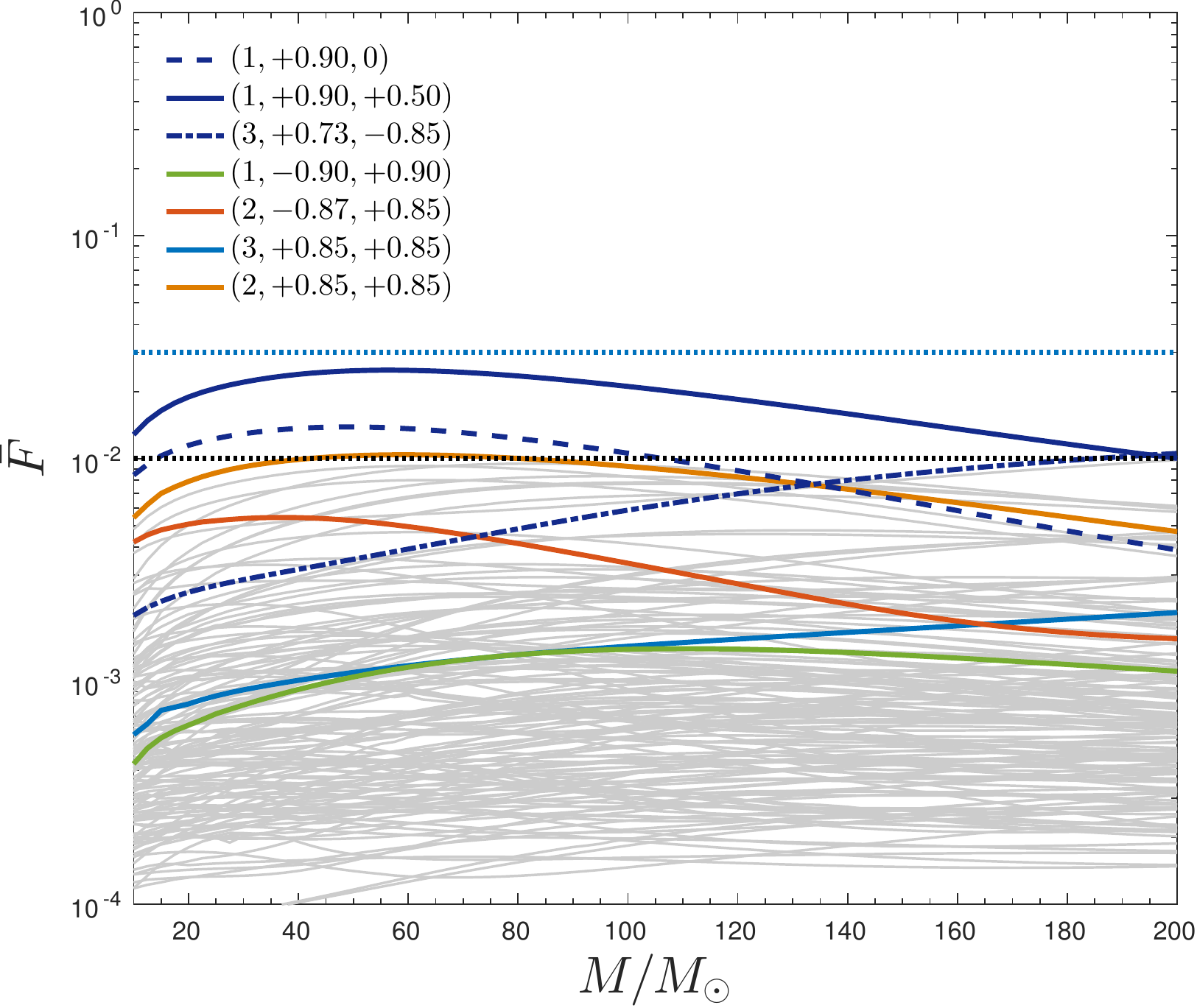}
\caption{\label{fig:barF_0} EOB/NR unfaithfulness computed all over the spin-aligned SXS NR catalog.
  The dark-blue curves are those with $1\%\lessapprox \max{(\bar{F})}< 3\%$. The grey ones correspond
  to $\max{(\bar{F})}<1\%$. The colored curves are those configurations where $\max{(\bar{F})}$
  computed between the corresponding NR data and the SEOBNRv2 EOB model is larger
  than $3\%$ as found in Ref.~\cite{Bohe:2016gbl}. Remarkably, except the case
  $(2,+0.85,+0.85)$ these configurations are all well below the $1\%$ level.}
\end{figure}
\begin{figure}[t]
\center
\includegraphics[width=0.42\textwidth]{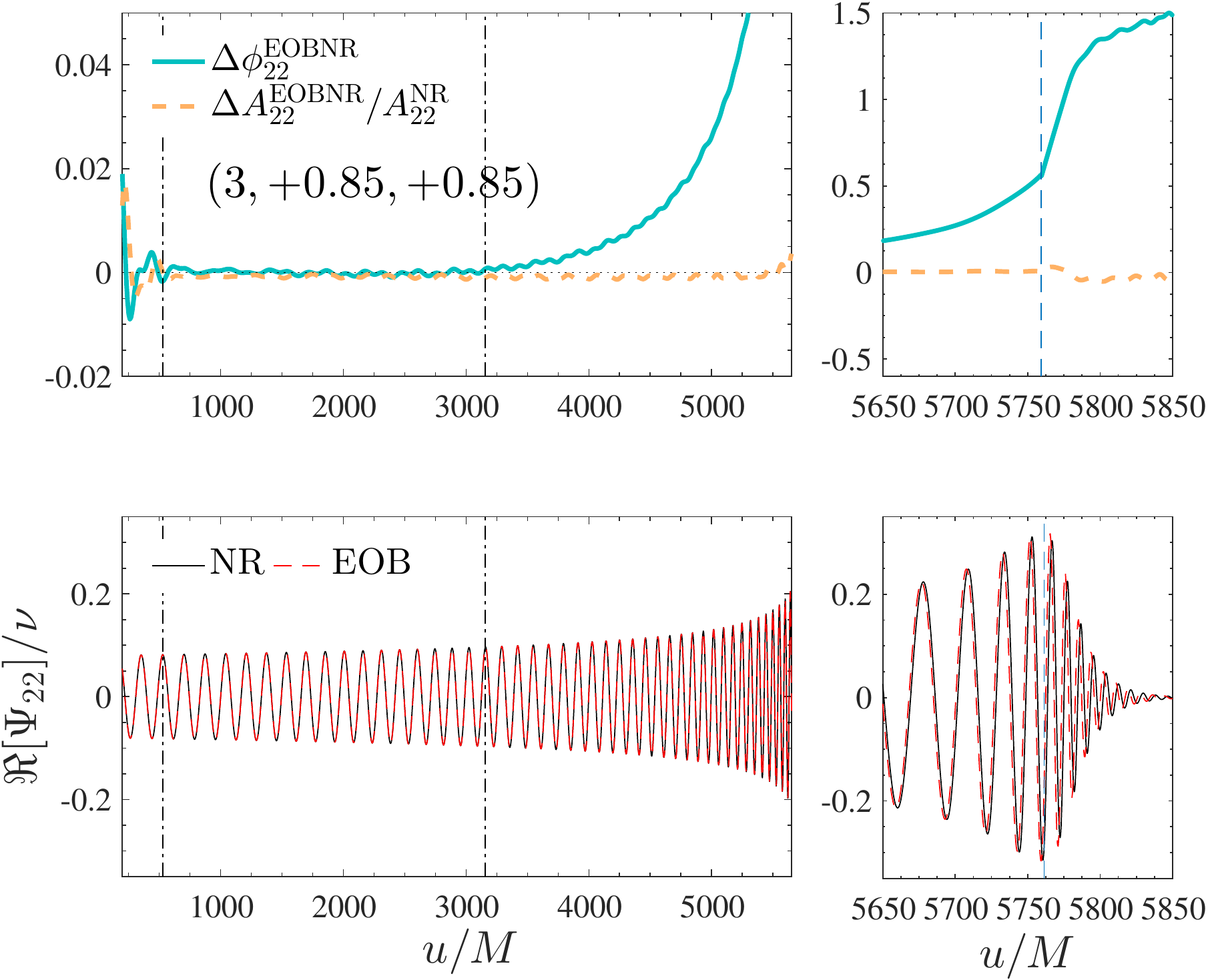}
\caption{\label{fig:q3p085} Time-domain phasing comparison for $(3,+0.85,+0.85)$,
  dataset SXS:BBH:0293. This corresponds to the blue curve in Fig.~\ref{fig:barF_0}.
  This phasing result is comparable to the one obtained using SEOBNRv4
  (see Fig.~4 of Ref.~\cite{Bohe:2016gbl}) although no new calibration
  of the model of Paper~I was needed here.}
\end{figure}
Before discussing the ``outliers'' with $1\%\lessapprox \max{(\bar{F})} <3\%$, we produce a
time-domain phasing comparison, in Fig.~\ref{fig:q3p085}, between the EOB and NR waveforms
for $(3,+0.85,+0.85)$, dataset SXS:BBH:0293.
This plot is the analogous of the time-domain check of SEOBNRv4 model, see Fig.~4 of
Ref.~\cite{Bohe:2016gbl}, although no additional calibration of the model of
Paper~I was needed to obtain such result. The top panels of the figure show the
fractional amplitude difference (orange) and the phase difference (blue), obtained
by aligning in time and phase the two waveforms with our standard technique that
relies on minimizing the phase difference between the frequencies 
corresponding to the time-interval marked by the two vertical lines.
We note two things: (i) from Table~\ref{tab:q3} one sees that the nominal NR
uncertainty at merger for this dataset is $\delta\phi^{\rm NR}_{\rm mrg}=1.15$~rad,
so the EOB/NR phase difference at merger $\Delta\phi^{\rm EOBNR}_{\rm mrg}=0.56$~rad
is fully compatible with this conservative numerical uncertainty;
(ii) the derivative of the phase difference in the top-right panel of the figure
is discontinuous at merger, where the EOB inspiral-plunge-merger description
is matched to the phenomenological description of the postmerger-ringdown
of Ref.~\cite{Nagar:2016iwa}. We recall that this phenomenological
ringdown representation is built interpolating among the same sample of NR data
(plus the case $(5,-0.9,0)$), mostly around the equal-mass, equal-spin configurations,
used in Paper~I. In this case, the inaccuracy visible in the time-domain analysis is
practically irrelevant for current unfaithfulnesses standard (i.e. $\max{(\bar{F})}<1\%$),
but the same effect can be more important in other configurations, 
eventually yielding larger values of $\bar{F}$, as we shall see below.
\begin{figure*}[t]
\center
\includegraphics[width=0.4\textwidth]{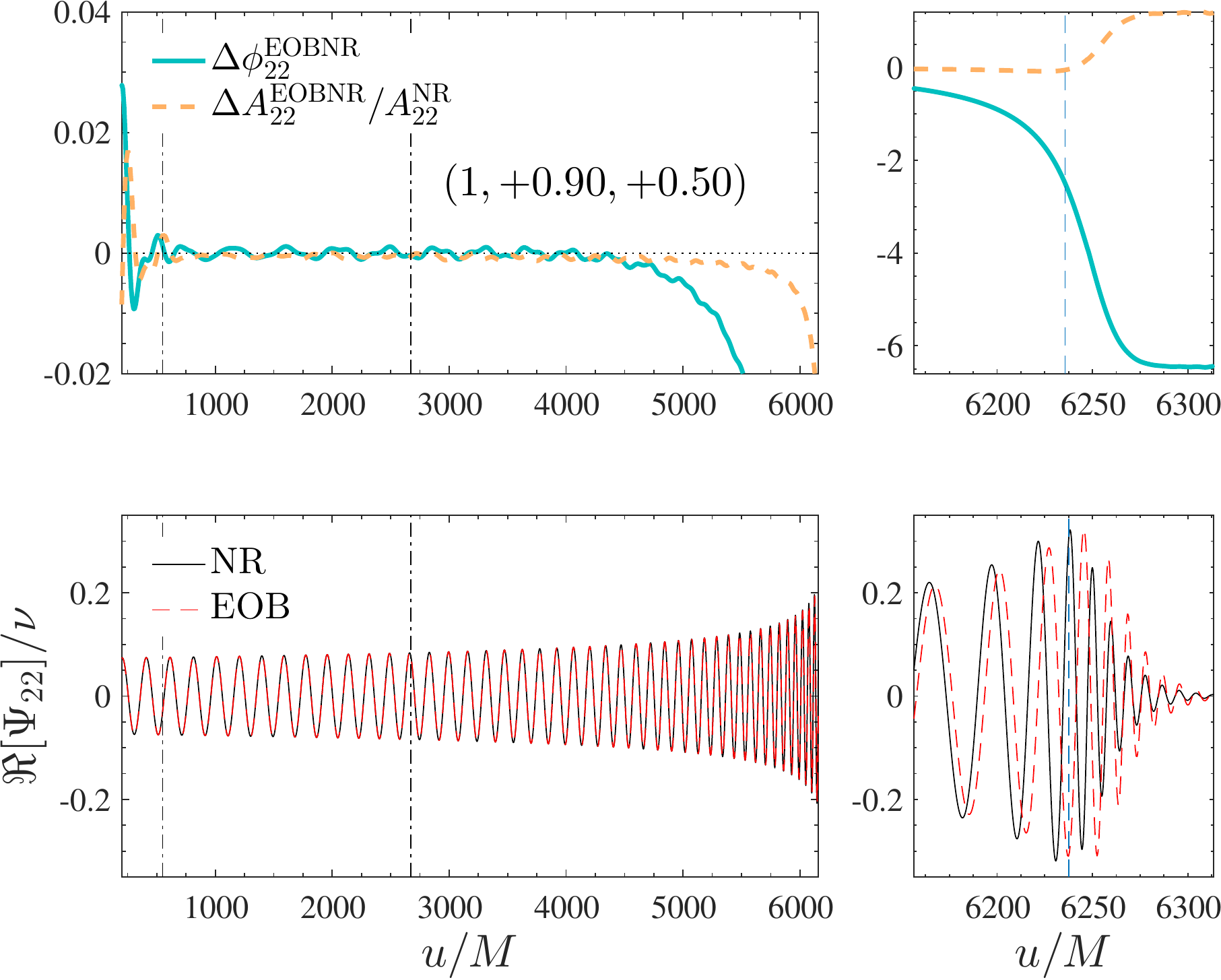}
\hspace{4.5mm}
\includegraphics[width=0.4\textwidth]{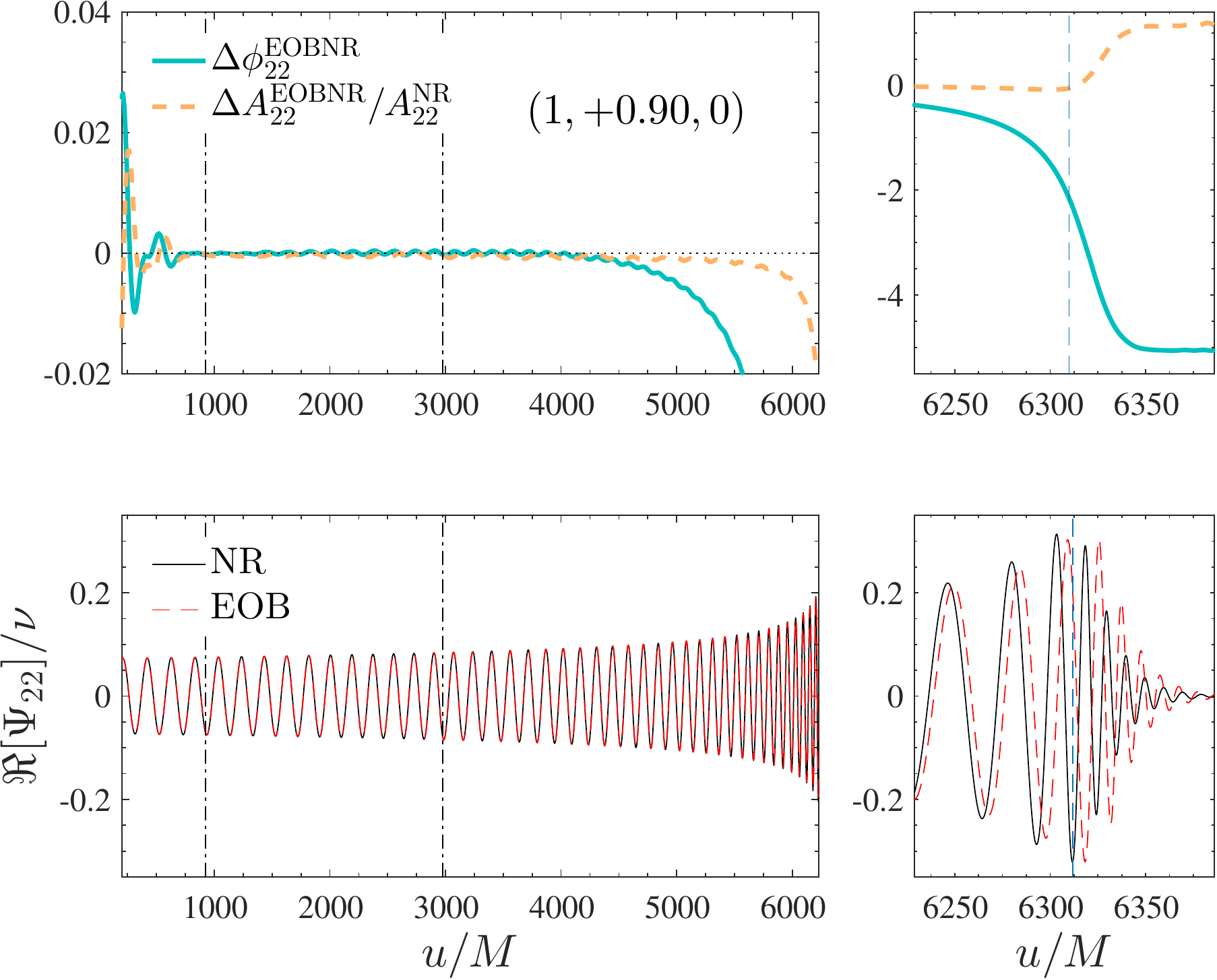}\\
\vspace{4mm}
\includegraphics[width=0.4\textwidth]{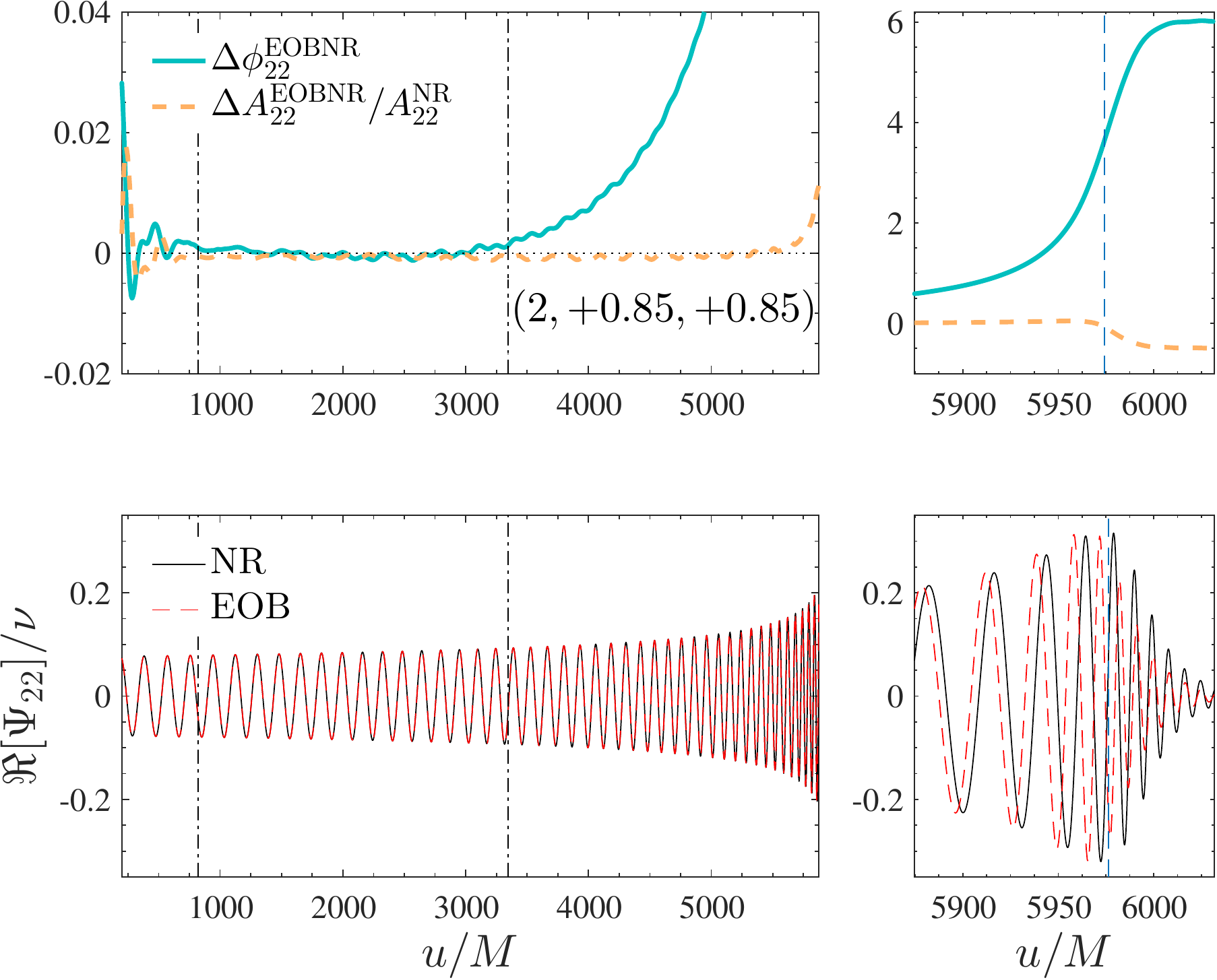}
\hspace{4.5mm}
\includegraphics[width=0.4\textwidth]{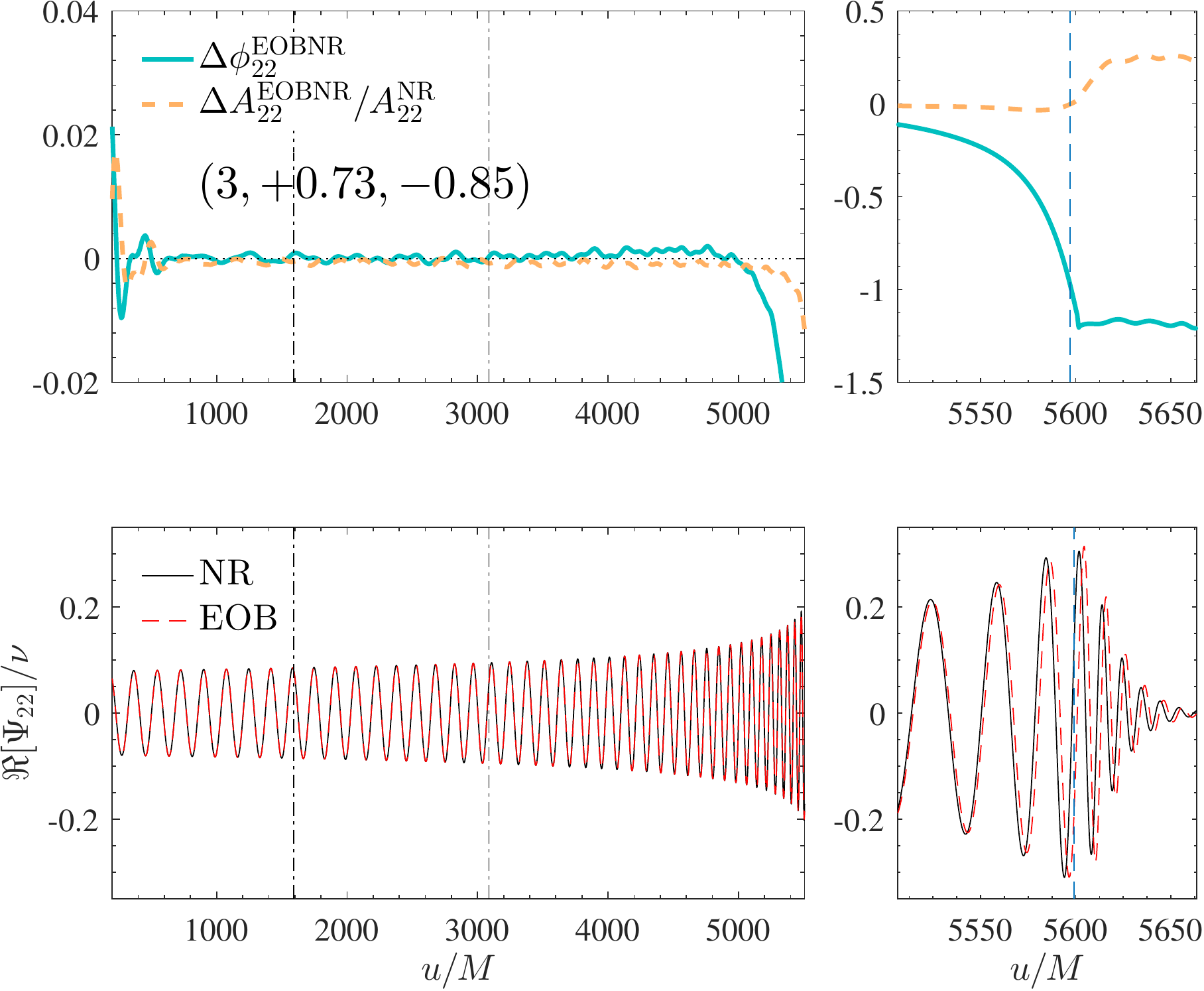} \\
\caption{\label{fig:outliers} Time-domain phasing analysis of four outliers (with $1\%\lesssim \max(\bar{F})< 3\%$)
  in Fig.~\ref{fig:barF_0}. The waveform are aligned using standard procedure in the time-window
  indicated on the plots. The phase difference between EOB and NR waveforms, $\Delta\phi^{\rm EOBNR}$,
  at merger is typically larger than the corresponding NR error.
  This calls for the improved determination of $c_3$ discussed in Sec.~\ref{sec:new_informing}.}
\end{figure*}

Let us now produce a similar analysis for the four outliers,
with $1\%\lesssim\max{(\bar{F})}<3\%$ in Fig.~\ref{fig:barF_0}, so to physically
understand what is (slightly) inaccurate and eventually improve it.
The time-domain waveforms comparisons are shown in Fig.~\ref{fig:outliers}.
As above, the vertical lines mark the alignment region.
One sees that each EOB waveform develops a secular phase difference with respect to the
NR one, that develops at latest $2000/3000M$ before merger. The effect
is visually analogous to what appears in Fig.~\ref{fig:q3p085}, though
the main difference is now that the EOB/NR phase differences at NR merger,
$\Delta\phi^{\rm EOBNR}_{\rm mrg}\equiv\phi^{\rm EOB}-\phi^{\rm NR}|_{t^{\rm NR}_{\rm mrg}}\approx(-2.5,-2.2,3.7,-1.4)$~rad,
are always {\it larger} (except for the the $(3,+0.73,-0.85)$ case)
than the numerical nominal NR uncertainties $\delta\phi^{\rm NR}_{\rm mrg}=(+1.7,+2.9,+0.3,-0.66)$
This suggests that some strong-field information  is (effectively) not
incorporated properly in the model. Interestingly, the ``worse'' result
in terms of phasing, for $(2,+0.85,+0.85)$, where the EOB waveform 
accumulates as much as 3.7~rad of dephasing at merger with respect 
to the NR one, is actually the one (among these four) with the smallest 
value of $\max{(\bar{F})}$, see Fig.~\ref{fig:barF_0}. 
This example illustrates that great care should be put in stating
the quality of an analytical waveform model using only $\bar{F}$
as a diagnostics. In fact, although $\bar{F}$ is a useful and
essential observable in addressing how good a waveform model is
for data-analysis purposes, it might hide important information
about the actual compatibility of an analytical (and usually
approximate) model versus a NR simulation that is supposed to
represent the ``exact'' result (modulo numerical uncertainties).
In other words, if it is true that having a small $\bar{F}$ (e.g $\lesssim 1\%$)
is a {\it necessary} criterion to define whether a waveform model
is viable for data-analysis purposes (in particular for detection,
since $\bar{F}<3\%$ translates into a loss of detections smaller
than $10\%$~\cite{Damour:1997ub}), it may not be a {\it sufficiently stringent}
one to conclude that the analytical model is {\it a fully reliable representation
  of the physics} encoded into a NR waveform. 

The opposite situation (i.e., with the phase difference is smaller/compatible with
the NR error) is instead found for $(3,+0.73,-0.85)$ configuration, but 
in this case $\max{(\bar{F})}$ is just at the border of the usually
accepted $1\%$ unfaithfulness region (the gorwth of $\bar{F}$ is actually due
to inaccuracies in the postmerger-ringdown EOB waveform, as we will discuss below
and just marginally from the inspiral-plunge).  
On the contrary, for the other two datasets, the accumulated phase difference is
rather large and lies, again, outside the nominal error bar (which is positive),
but now this is also clearly mirrored in the values of $\bar{F}$ above the $1\%$ threshold.
Inspecting the plots, one sees that for all datasets except $(2,+0.85,+0.85)$
the EOB waveform is {\it longer} than the NR one, with the effect more pronounced
in the $q=1$ cases. Within our EOB framework, we physically interpret this fact,
in simple terms, by saying that the spin-orbit part of the EOB Hamiltonian is too
strong, i.e. the repulsive effect of the positively aligned spin is too large. 
By contrast, for $(2,+0.85,+0.85)$ the spin-orbit coupling is too weak
and the system plunges faster than it should according to the NR prediction.
A simple, new, determination of $c_3$ is what it is needed to reduce the EOB/NR
phase differences and eventually obtain smaller phase differences as well
as smaller values of $\bar{F}$

\section{Improving the model with minimal amount of additional NR information}
\label{sec:new_informing}
Let us now investigate how the already good results discussed in the previous paragraph
can be improved by including a (minimal) additional amount of NR information to
the existing EOBNR model. In practice, we inform a new fit for the
$(p_1',p_2',p_3',p_4')$ entering Eq.~\eqref{eq:c3fun} by using the $q=1$
and $q=2$ outliers mentioned above to determine improved value of $c_3$
such to reduce the EOB/NR phase difference up to merger and make it
closer to the nominal NR error bar. In addition, we shall also use more NR
waveform data to compute an improved interpolating fit of the NR NQC
functioning point. By contrast, as mentioned above, here we {\it do not}
use more NR incorporate more datasets to construct an improved phenomenological
description of the postmerger-ringdown waveform (see Ref.~\cite{Bohe:2016gbl} in this respect).

\subsection{NR-driven modifications of $c_3(\nu,\ta_1,\ta_2)$ }
\label{sec:EOBc3B}
\begin{figure}[t]
\center
\includegraphics[width=0.42\textwidth]{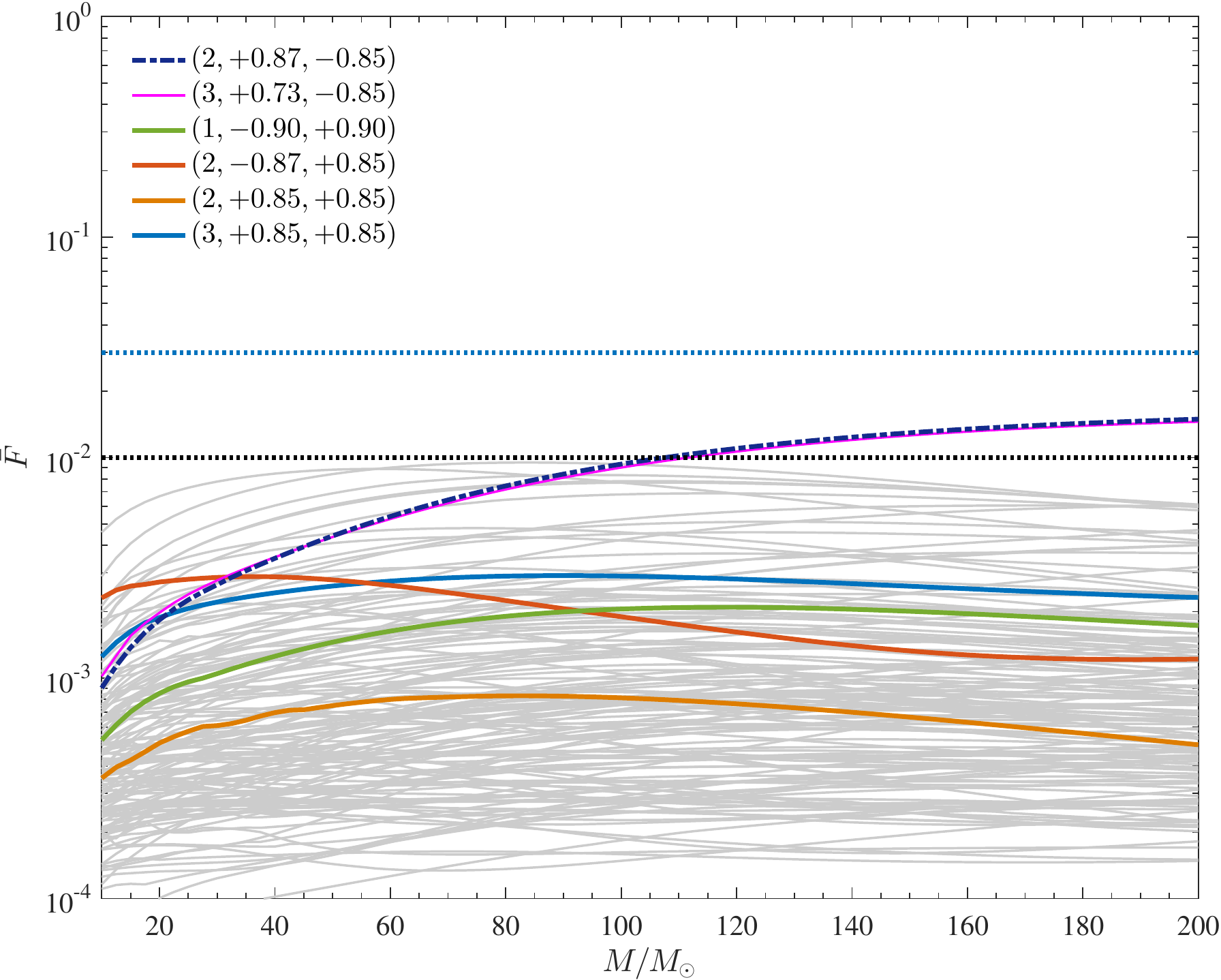}
\caption{\label{fig:barF_new} EOB/NR unfaithfulness obtaned with new determination of
  of $c_3$ yielded by Eqs.~\eqref{eq:p1_prime}-\eqref{eq:p4_prime}.
  Note the improvement for $(2,-0.87,+0.85)$ (cf. Fig.~\ref{fig:barF_0}) that is
  prompted by having determined a better value of $c_3$ for $(2,+0.85,+0.85)$.
  The two outliers (that in Fig.~\ref{fig:barF_0} where just below the $1\%$ level at $M=200M_\odot$)
  reach now approximately the $1.5\%$ level because of the relatively poor accuracy
  of the postmerger interpolating fit.}
\end{figure}

To determine an improved function for $c_3$ we proceed as in Paper~I and as
briefly reminded above. The EOB and NR waveforms are aligned in the early inspiral and
$c_3$ is chosen so to reduce the phase difference at merger in a way that $\Delta\phi_{\rm mrg}^{\rm EOBNR}$
is compatible (or below) the nominal NR error bar. The fact that, in Fig.~\ref{fig:outliers},
the EOB waveform is longer (shorter) than the NR one means that, for these configurations,
one has to increase (decrease) the value of the parameter $c_3$ with respect to the
interpolation provided by Eq.~\eqref{eq:c3fun} with the coefficients~\eqref{c3_p0}-\eqref{c3_p4}. 
Following this rationale, one concludes that some new, first-guess, 
$c_3$ values may be $c_3=25$ for $(1,+0.90.0)$ 
(instead of $c_3^{\rm old}=13.38$ coming from the previous fit),
$c_3=10.2$ for $(1,+0.90,+0.50)$ (instead of $c_3^{\rm old}=7.97$)
and $c_3=10.4$ for $(2,+0.85,+0.85)$ (instead of $14.21$).
We then put these values together with those listed in Table~\ref{tab:c3uneq}
and a new global fit using the same functional form of Eq.~\eqref{eq:c3fun}
is performed. The equal-mass, equal-spin part of
the fit for $c_3$ remains untouched, and we only
modify the coefficients $(p_1,p_2,p_3,p_4)$,
that now read
\begin{align}
\label{eq:p1_prime}
p^{\rm new}_1  &=  +978.873,\\
\label{eq:p4_prime}
p^{\rm new}_2  &= -9456.08,\\
\label{eq:p4_prime}
p^{\rm new}_3  &= +24304.1,\\
\label{eq:p4_prime}
p^{\rm new}_4  &= -149.342.
\end{align}
Using such improved analytical representation for
$c^{\rm new}_3(\nu,\tilde{a}_1,\tilde{a}_2)$ (and without
changing anything else with respect to the previous model)
we perform a novel computation of $\bar{F}$. The inclusion
of the three new ``information'' points makes $\bar{F}$
always below the $1\%$ level, though now with the exception 
of two dataset that reach the $\sim 1.5\%$ level for $M=200M_\odot$
due to inaccuracies in the modeling of the postmerger phase,
as we will illustrate in detail below.
Note in addition that for $(1,+0.90,+0.50)$ $\max{(\bar{(F)})}$
is just below $1\%$ ($0.89\%$): this was done somehow on purpose
to avoid overtuning, by simply requiring that the EOB/NR phase
difference at merger, $\Delta\phi^{\rm EOBNR}_{\rm mrg}$, is halved
with respect to the previous case (see in this respect line 71
and 72 of Table~\ref{tab:q1}). Seen our, very conservative,
estimate of the error bar on the NR phasing, we think this
is a reasonable choice to improve the EOB/NR agreement within
the NR phase uncertainty.
\begin{figure}[t]
\center
\includegraphics[width=0.42\textwidth]{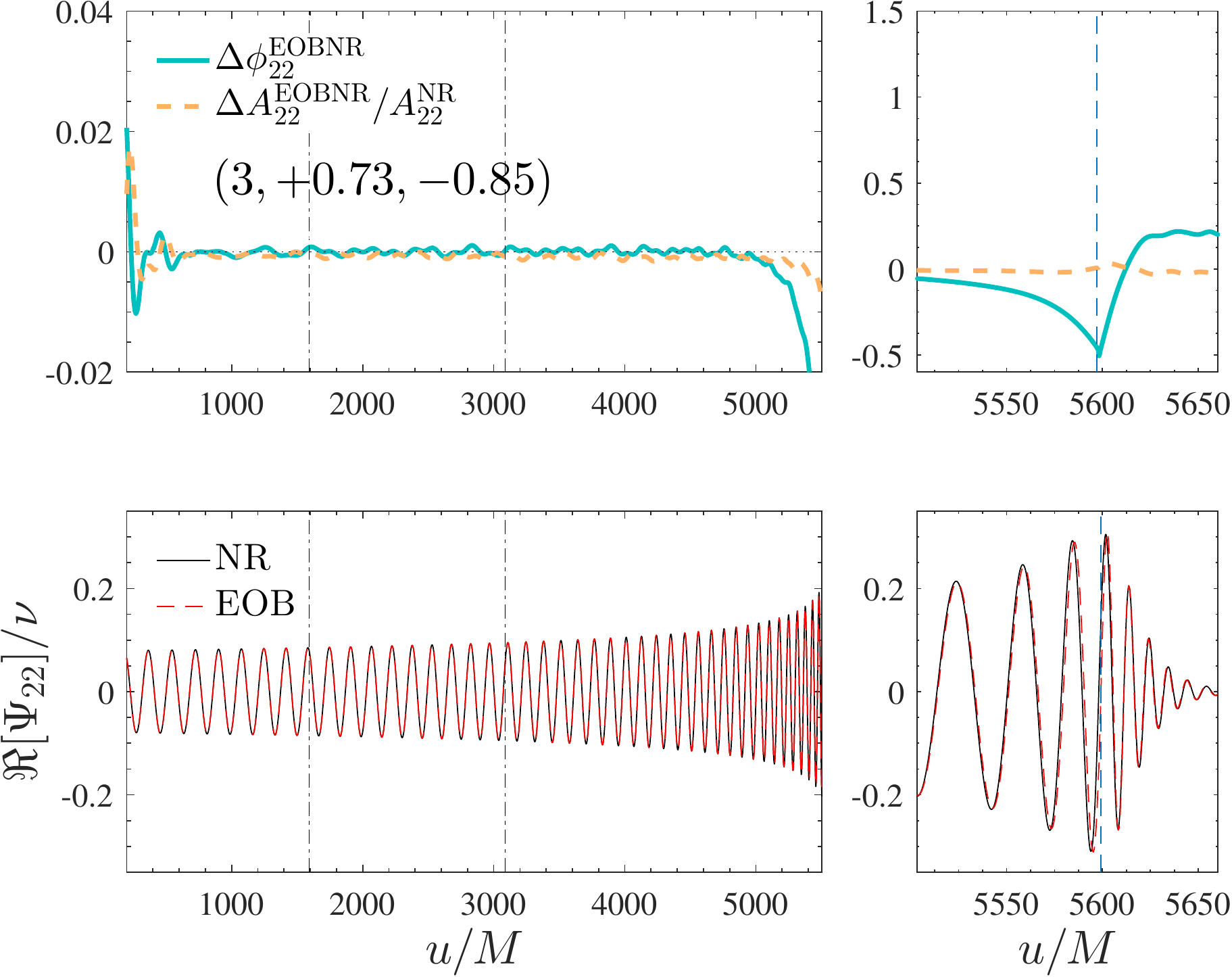}\\
\includegraphics[width=0.42\textwidth]{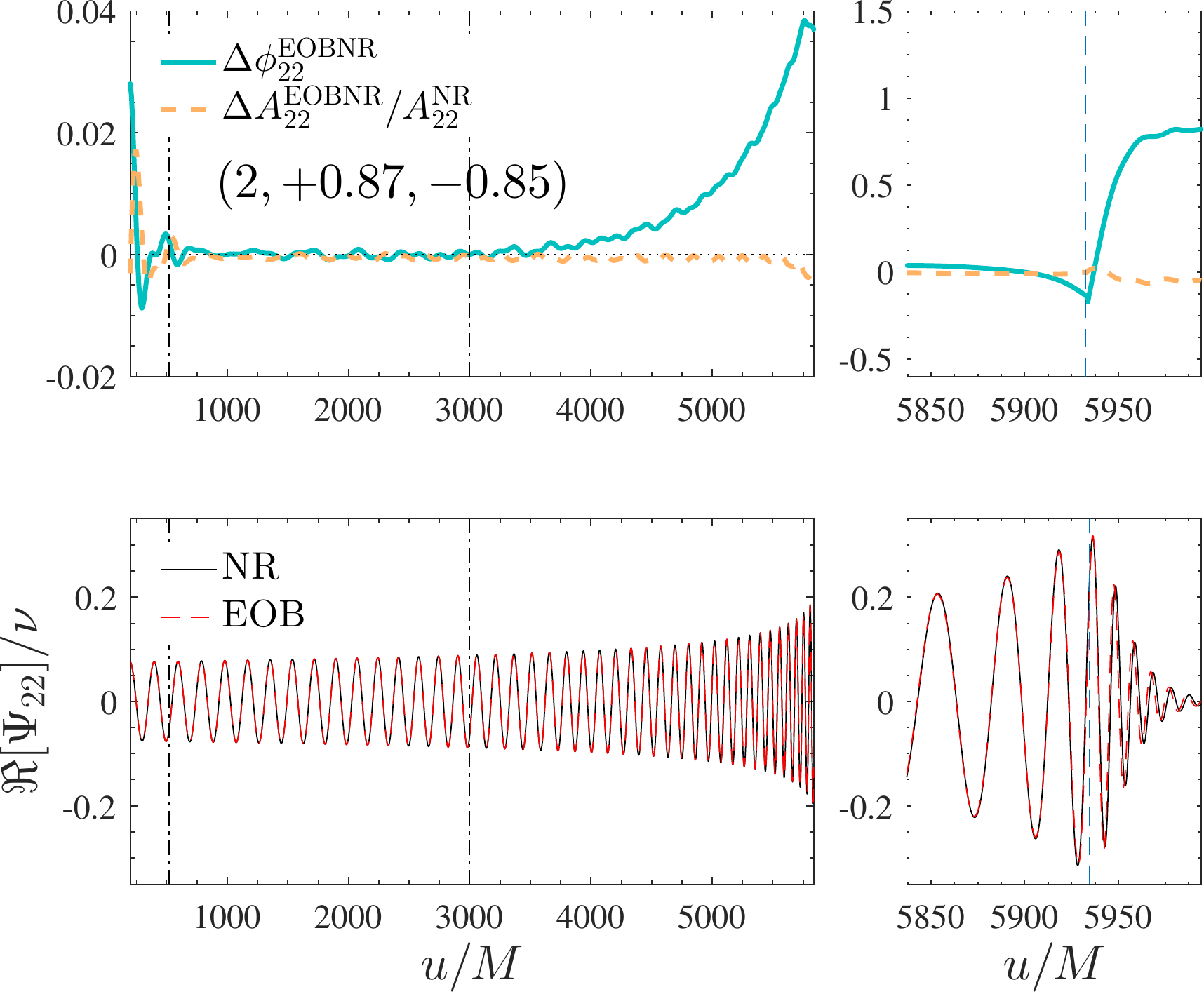}
\caption{\label{fig:q3p073_m085_new}Effect of the new determination of $c_3$ given by
  Eqs.~\eqref{eq:p1_prime}-\eqref{eq:p4_prime} on $(3,+0.75,-0.85)$ dataset, SXS:BBH:0292,
  and on $(2,+0.87,-0.85)$, SXS:BBH:0258.}
\end{figure}
\begin{figure}[t]
  \center
  \includegraphics[width=0.42\textwidth]{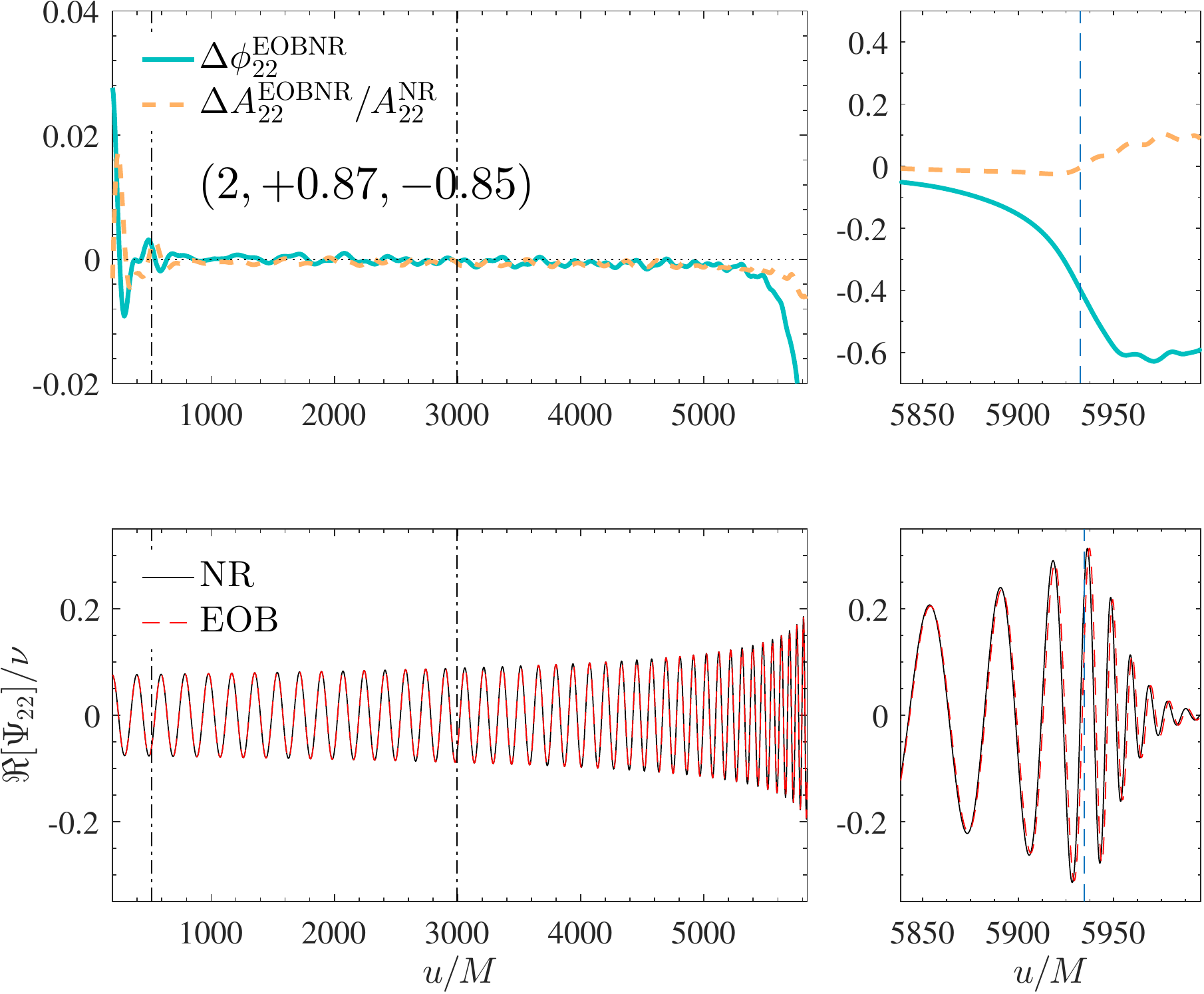}\\
  \includegraphics[width=0.42\textwidth]{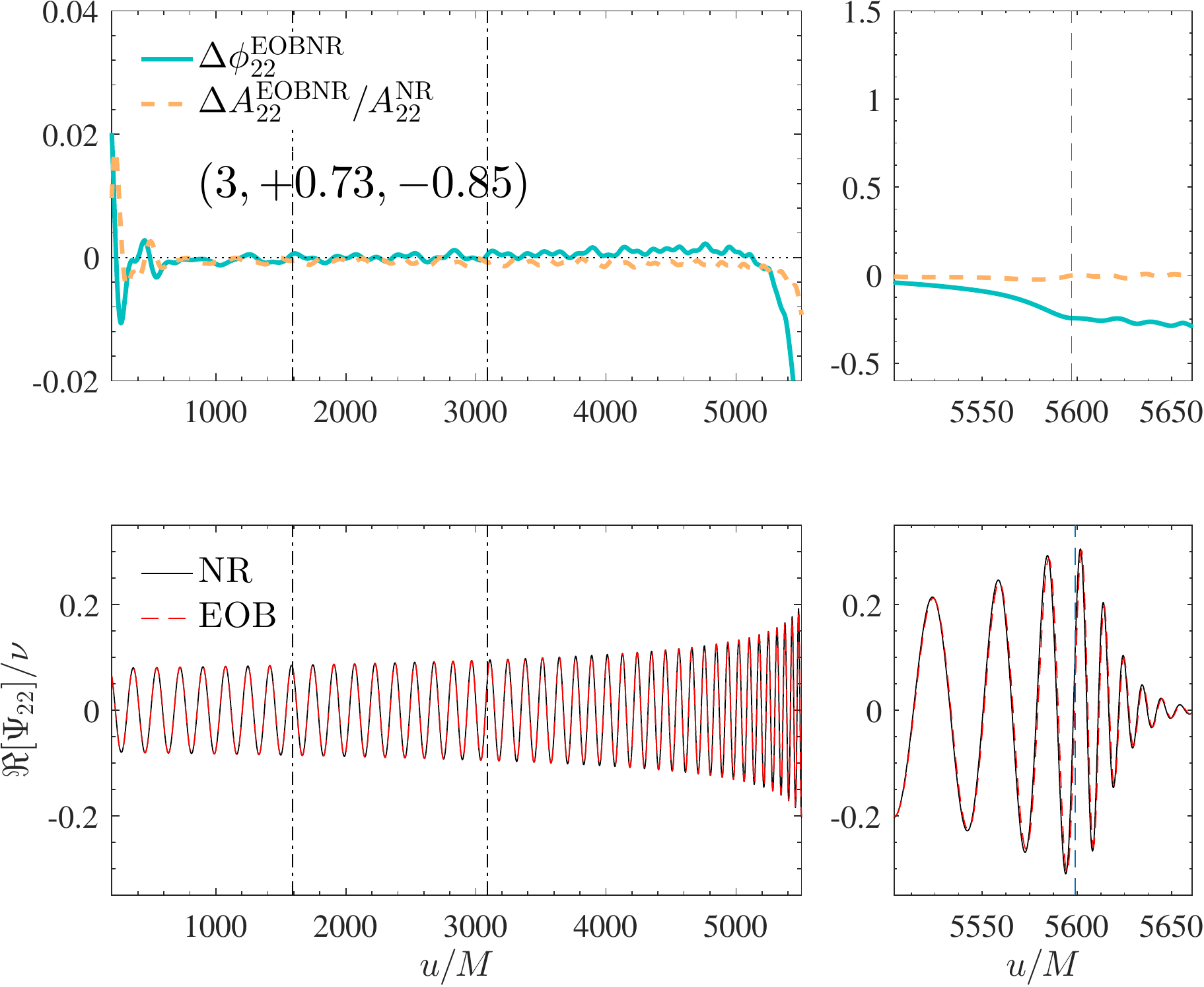}
  \caption{\label{fig:q3q2_exact}Same as Fig.~\ref{fig:q3p073_m085_new} but using (i) the ``exact'',
    NR values of the NQC functioning point (i.e., not the interpolating fit) and (ii) the description
    of the ringdown yielded by the primary fit. The value of $c_3$ is unchanged with respect to
    Fig.~\ref{fig:q3p073_m085_new}. This result is the best one achievable for this value of $c_3$
    and yields the solid, thick lines for $\bar{F}$ in Fig.~\ref{fig:outliers_bF}.}
\end{figure}
\begin{figure}[t]
\center
\includegraphics[width=0.42\textwidth]{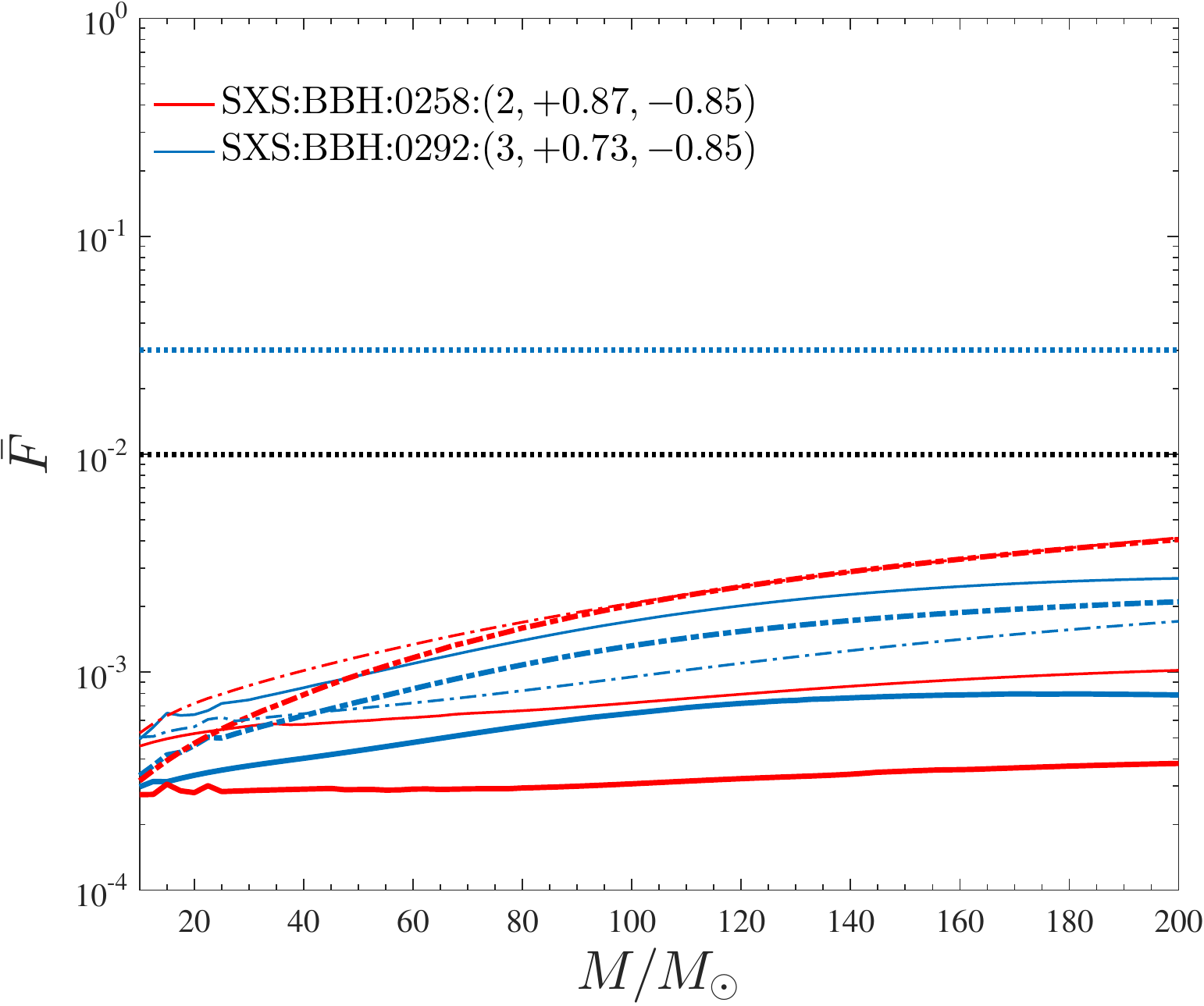}
\caption{\label{fig:outliers_bF}Improved unfaithfulness for the two
  outliers of Fig.~\ref{fig:barF_new} when the description of the
  NQC and/or of the postmerger waveform is either improved or changed:
  (i) thick solid: using the {\it exact} NR NQC point and primary NR ringdown fit;
  (ii) thin, solid: fitted NQC point but primary NR ringdown fit;
  (iii) thick, dash-dotted: exact NQC point but approximate ringdown description as the superposition of $8$ QNMs;
  (iv) thin, dashed-dotted: fitted NQC point and approximate ringdown description as the superposition of $8$ QNMs.
  This last case can be used as effective replacement of the inaccurate ringdown waveform part
  highlighted in Figs.~\ref{fig:barF_new} and \ref{fig:q3p073_m085_new}.}
\end{figure}
Let us turn now to discuss in detail the two outliers,
$(3,+0.73,-0.85)$ and $(2,+0.87,-0.85)$. The first configuration
was already shown to be ``quasi-problematic'' in Fig.~\ref{fig:barF_0}
and in the bottom-right panel of Fig.~\ref{fig:outliers}.
The second one was not labeled as problematic, because $\bar{F}<1\%$,
but in fact in Fig.~\ref{fig:barF_0} it is that grey line that
closely follows the behavior of the $(3,+0.73,-0.85)$ i.e. it is
monotonically growing with $M$ to reach a value just barely smaller
than $1\%$ at $M=200M_\odot$. The inspection of $\bar{F}$ alone
seems then to indicate that no improvement is brought here by
the new determination of $c_3$.
This is actually not true and it is hidden by the unfaithfulness
plot, that, being given by an integral, hides important local
properties of the waveform. To appreciate this, we show the
time-domain phasing comparisons in Fig.~\ref{fig:q3p073_m085_new}:
one sees here that, for $(3,+073,-0.85)$ the previous large
EOB/NR phase difference at merger of $\approx -0.66$~rad is
absorbed by the new value of $c_3$ ($c_3=27.9$) that is smaller
than the previous value ($c_3=22.9$) so to effectively reduce
the spin-orbit coupling and make the system plunge faster.
Similar consideration apply to the other dataset, where the
previous dephasing at merger $\Delta\phi^{\rm EOBNR}_{\rm mrg}=-1.8$~rad
is now reduced to $\approx -0.13$~rad, wich is compatible 
with the nominal error bar $\delta \phi^{\rm NR}_{\rm mrg}\simeq 0.7$~rad.

By contrast, in both cases most of the EOB/NR phase difference is
accumulated {\it after} the merger and mirrors the inaccuracy,
in these particular cases, of the interpolating postmerger waveform 
model that we are using here~\cite{Nagar:2016iwa}. 
This is not that surprising, since the model of Ref.~\cite{Nagar:2016iwa} 
was obtained through interpolation of a rather sparse sample of 
primary ringdown fits that do not include, for example, any $q=2$ waveform data.
To prove this is indeed the case, and to remove the inaccuracies coming from
the various interpolating fits, we show in Fig.~\ref{fig:q3q2_exact}
the level of phasing agreement that can be reached (without changing $c_{3}$), 
for both outliers, using both (i) the {\it exact} values for the NR NQC point to determine
$(a_i,b_i)$ and (ii) the postmerger representation obtained by 
the {\it primary fit} to the NR waveform. The corresponding values
of $\bar{F}$ are shown, as solid lines, in Fig.~\ref{fig:outliers_bF},
and remain below $10^{-3}$. This indicates that a fully acceptable model 
could be obtained by just improving the merger and ringdown description,
without changing the value of $c_3$. For example, this could be achieved 
by using more NR waveform data to produce a more accurate ringdown 
interpolating fit, as recently done in Ref.~\cite{Bohe:2016gbl}.
Here, we actually try  to explore a different route~\footnote{As mentioned in
  Ref.~\cite{Nagar:2016iwa}, the primary fitting template becomes
  progressively less accurate when the mass ratio is increased and
  it should be eventually modified. For this reason,  a detailed study 
  is postponed to future work.}, recalling that the
original EOB ``recipe'' is to construct the ringdown as a
superposition of QNMs~\cite{Buonanno:2000ef,Damour:2007xr}
This basic approach, though in general successful, might be subtle
(and inaccurate) for high positive spins, so that
pseudo-quasi-normal modes frequency were introduced and determined via comparison
with NR waveform data~\cite{Taracchini:2013rva,Pan:2013rra,Taracchini:2012ig,Pan:2009wj}.
However, in situations where the final black hole spin is not too extreme 
($a_{f}\lesssim +0.8$),
as in this case, the standard QNMs superposition is usually reliable,
and offers the possibility of computing accurately the postmerger waveform
even without any NR input except the values of the final black hole mass
and angular momentum. We then apply the ringdown matching procedure,
following Refs.~\cite{Damour:2007xr,Damour:2009kr}, by using
$N=8$ QNMs matched to the EOB waveform on a small
interval $\Delta t^{\rm QNM}=0.5M$.
This approach yields a sufficiently accurate description of the
ringdown waveform, as highlighted by the dot-dashed lines
in Fig.~\ref{fig:outliers_bF}. The figure also illustrates the
difference between  using the ``exact'' NR NQC point or 
the interpolating fit, something that also slightly worsens the 
global performance. This exercise, done here only on the 
two outliers, is interesting in that illustrates the power of the 
simplest approach to ringdown modelization and might eventually 
turn out useful in regions of the parameter space where the 
interpolating global fit is clearly inaccurate, provided the
final black hole spin is not too high ($a_{f\lesssim} +0.8$).
The inaccuracy of the phenomenological ringdown description 
can be easily spotted by inspecting the GW frequency, that 
develops a large discontinuity at the matching point. In our specific 
case, such a step in the frequency is behind the cusp appearing 
in the phase differences Fig.~\ref{fig:q3p073_m085_new}.
At a practical level, when this inaccuracy is found, the waveform
generation algorithm can simply switch from the phenomenological
fit to the superposition of QNMs. Although this looks like an easy way
out, one has to be aware that a comprehensive estimate of the accuracy
of this analytic ringdown representation as function of the binary parameter, 
and in particular its dependence on the number of QNMs, is currently 
lacking and will need a dedicated study.
In particular, it will be interesting to analyze the performance of
the recipe we present here ($\Delta t^{\rm QNM}=0.5M$, $N=8$)
all over the SXS catalogue.

We finally stress that the results displayed in Figs.~\ref{fig:barF_new}~and~\ref{fig:outliers_bF}
obtained with a new determination of $c_3$ with respect of Paper~I are
certainly {\it not  the best possible} ones that can be achieved within 
the current theoretical framework. For pedagogical reasons, here we choose 
to implement only  {\it minimal changes} in $c_3$ in order to easily 
accomodate {\it only} outliers above $1\%$ and contemporarily to test 
the robustness of the model of Paper~I. We did so by only asking that 
the phase difference, for those particular datasets, be smaller and/or 
comparable to the error bar. Actually, other datasets similarly show
phase differences at merger larger than $\delta\phi_{\rm mrg}^{\rm NR}$,
 for example $(3,+0.85,+0.85)$, though this is still yielding $\max(\bar{F})<1\%$. 
In principle, one could have retuned $c_3$ even more with respect to Paper~I 
so to further reduce the EOB/NR phase difference (compatibly with the
error bar), so to eventually obtain even smaller values of $\bar{F}$. 
This will be done in a future study that will take advantage
of an improved analytical description of the resummed waveform
amplitudes~\cite{Nagar:2016ayt}. The minimal changes we implemented 
here are mainly aimed at highlighting the simplicity of the tuning
procedure and the robustness of the original as well as improved model
all over the SXS waveform catalog.

\subsection{Effect of improved fits for the NR NQC-extraction point}
\label{sec:NQC_improved}

We saw in the previous section that it is possible to
improve the faithfulness of the EOB[NR] model of Paper~I against
the full SXS catalogue by just including three more NR configurations
that eventually provide a new fit for $c_3$. In addition, the analysis
above has also clearly highlighted that, though the phenomenological
description of the ringdown of Ref.~\cite{Nagar:2016iwa} may become
inaccurate already in important regions of the parameter space
with $2\leq q \leq 3$, old ideas and methods for modeling the
postmerger-ringdown part prove effective to improve the global accuracy
of the model without the actual need of additional NR information.
\begin{figure}[t]
\center
\includegraphics[width=0.45\textwidth]{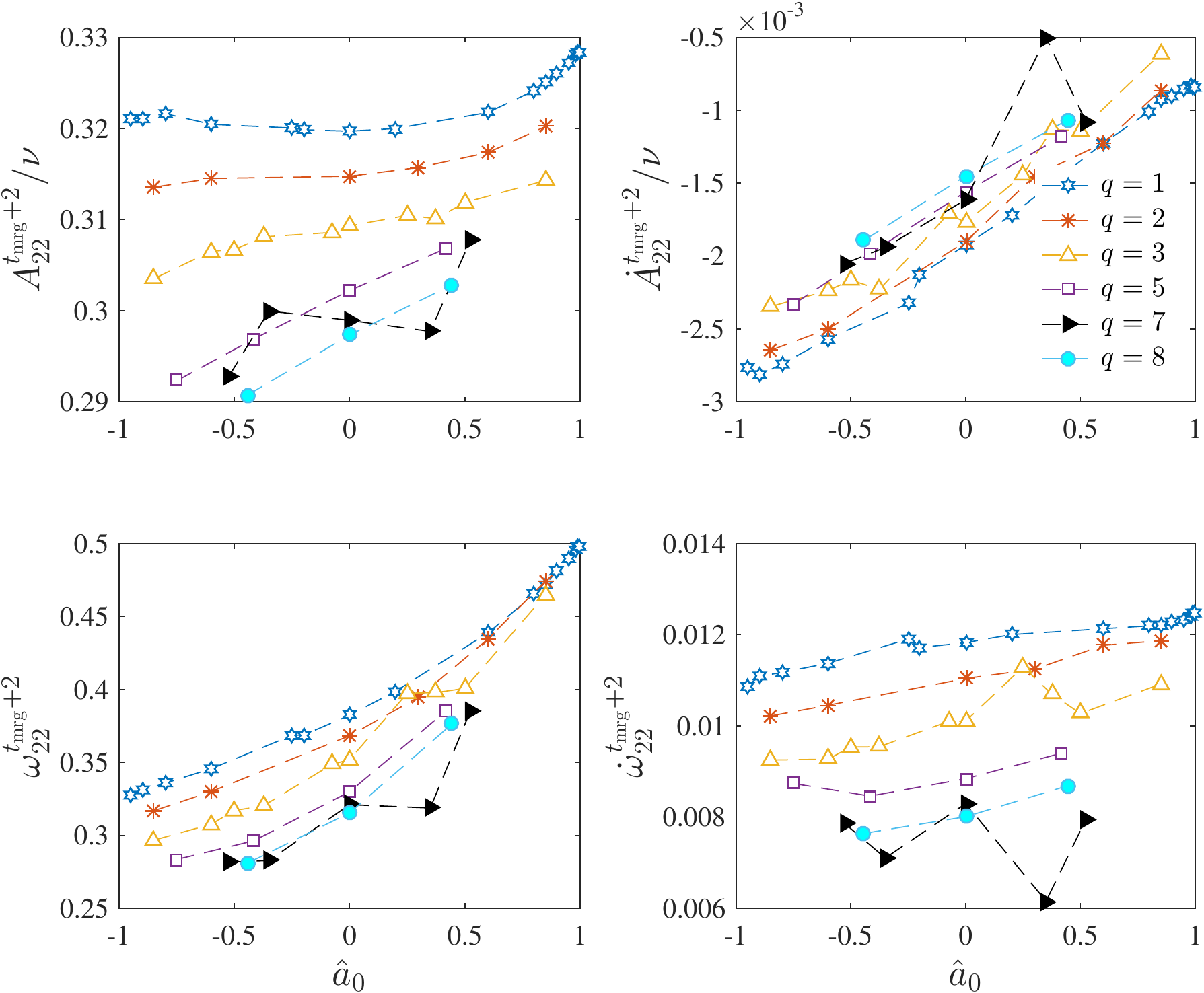}
\caption{\label{fig:NQC_point_all} Behavior of the NR points eventually
  used to determine improved values of the NQC parameters. These value are taken
  at $t^{\rm NR}_{\rm NQC}=t^{\rm NR}_{\rm mrg}+2$ where $t^{\rm NR}_{\rm mrg}$ corresponds
  to the peak of the modulus of the $\ell=m=2$ waveform. Some of the $q=7$ datapoints
  are clearly inconsistent with the others and are thus not used in the global fits.
  This weird behavior might be due to some unphysical drift of the center of mass.
  See Sec.~\ref{sec:q7} for discussion.}
\end{figure}
By contrast, the purpose of this Section is to explore how the
previous result change by using more NR waveform data to improve
one specific part of the EOBNR model, i.e. the determination
of the NQC parameters. As an exploratory analysis, we investigate
how $\bar{F}$ changes by just using more NR dataset to determine
the fit of the NR NQC extraction point. In doing so, both $c_3$
and the postmerger description remain unchanged with respect to previous section.
We compute here a new fit of $(A,\dot{A},\omega,\dot{\omega})^{\rm NR}_{t^{\rm NR}_{\rm NQC}}$
using the following data: (i) the same $q=1$ data used before;
(ii) all the $q=2$ and $q=3$ {\it equal spin} datasets, either aligned
or anti-aligned with the angular momentum. We do not consider unequal-spin
datasets, except for the ones with $(\pm 0.5,0)$ that were previously
included; (iii) we take into account the 4 datasets with $q=5$, with
spins $(-0.9,0)$, $(0.0)$ and $(\pm 0.5,0)$; (iv) we drop all
NR waveforms with $q>5$. In particular, we discard all the seven
available datapoints with $q=7$, since their behavior versus $\hat{a}_0$
appears qualitatively different compared to both $q=5$ and $q=8$ data,
and especially inconsistent with them for the longest $(7,\pm 0.4,0)$
waveforms. We will argue in Sec.~\ref{sec:q7} below, that it is
(partly) related to some (unphysical) drift of the center of mass
and, as such, we discard these points to avoid introducing systematics
in the fits. This is clearly illustrated in Fig.~\ref{fig:NQC_point_all},
that displays $(A_{22},\dot{A}_{22},\omega_{22},\dot{\omega}_{22})^{\rm NR}_{t^{\rm NR}_{\rm NQC}}$ versus $\hat{a}_0$
for all the configurations we take into account. Interestingly,
the plot also illustrates that our previous ansatz for $A_{22}$ and $\dot{A}_{22}$
to be simple quadratic functions of $\hat{a}_0$ for $q>1$ is incorrect,
since their actual behavior is qualitatively closer to the $q=1$ case.
For $q=5$ we cannot make any statement, but since we now have 4 points
for this $q$, we can use a cubic-in-$\hat{a}_0$ ansatz for the amplitude
and its derivative instead of the incorrect quadratic one we were obliged
to choose before because of lack of datapoints. The explicit values of the
fit coefficient (that are used for $q>1$ only, but are determined also
including the $q=1$ data) are given in Table~\ref{NQC_global}.
Evidently, in order not to lose accuracy, for $q=1$ we still use the
previusly determined best fit obtained with  4-th-order-in-$\hat{a}_0$
polynomials as given explicitly in Table~\ref{NQC_hybrid_q1}.
With these improved fits, we regenerate all EOB waveforms for $q>1$ with
improved NQC points and recompute the EOB/NR unfaithfulness.
The updated result is shown in Fig.~\ref{fig:barF_new_global},
where we are also using the QNMs-superposition description for
the postmerger-ringdown for $(2,+0.87,-0.85)$ and $(3,+0.73,-0.85)$.
\begin{figure}[t]
\center
\includegraphics[width=0.42\textwidth]{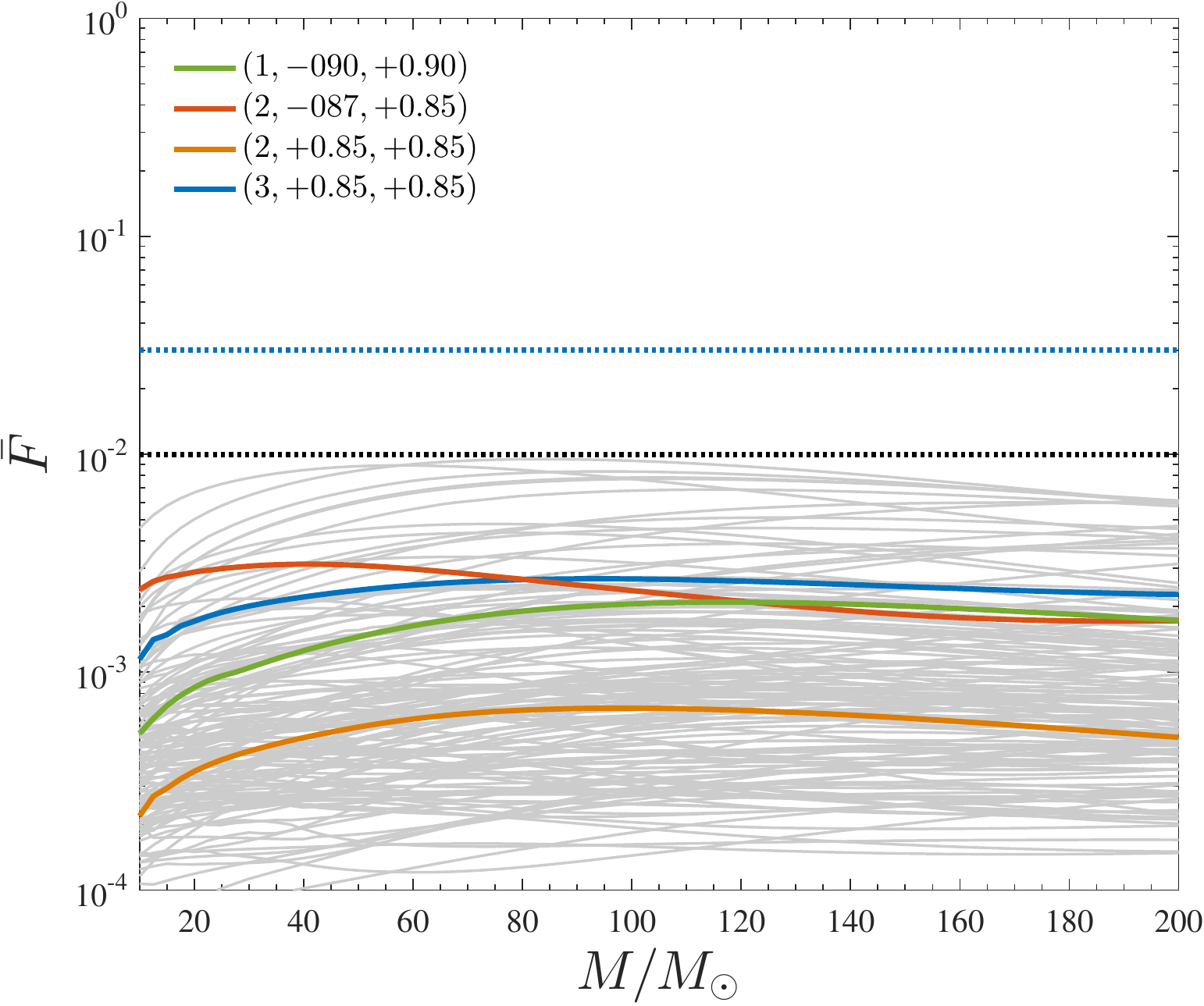}
\caption{\label{fig:barF_new_global} EOB/NR unfaithfulness with (i) the new
  fit for $c_3$ and (ii) a new interpolating fit for the NR NQC-extraction point
  that is used for $q>1$ and (iii) the postmmerger-ringdown description with QNMs for
  the $(3,+0.73,-0.85)$ and $(2,+0.87,-0.85)$ configurations.}
\end{figure}
To try to quantify the differences between
Figs.~\ref{fig:barF_new}-\ref{fig:outliers_bF}
and~\ref{fig:barF_new_global}, (though one can see that the
colored lines have slightly changed),
it is convenient to look at histograms of the maximal
faithfulness, $\max(F)\equiv \max(\bar{F}))-1$, that we display in
Fig.~\ref{fig:hist}.
\begin{figure}[t]
\center
\includegraphics[width=0.42\textwidth]{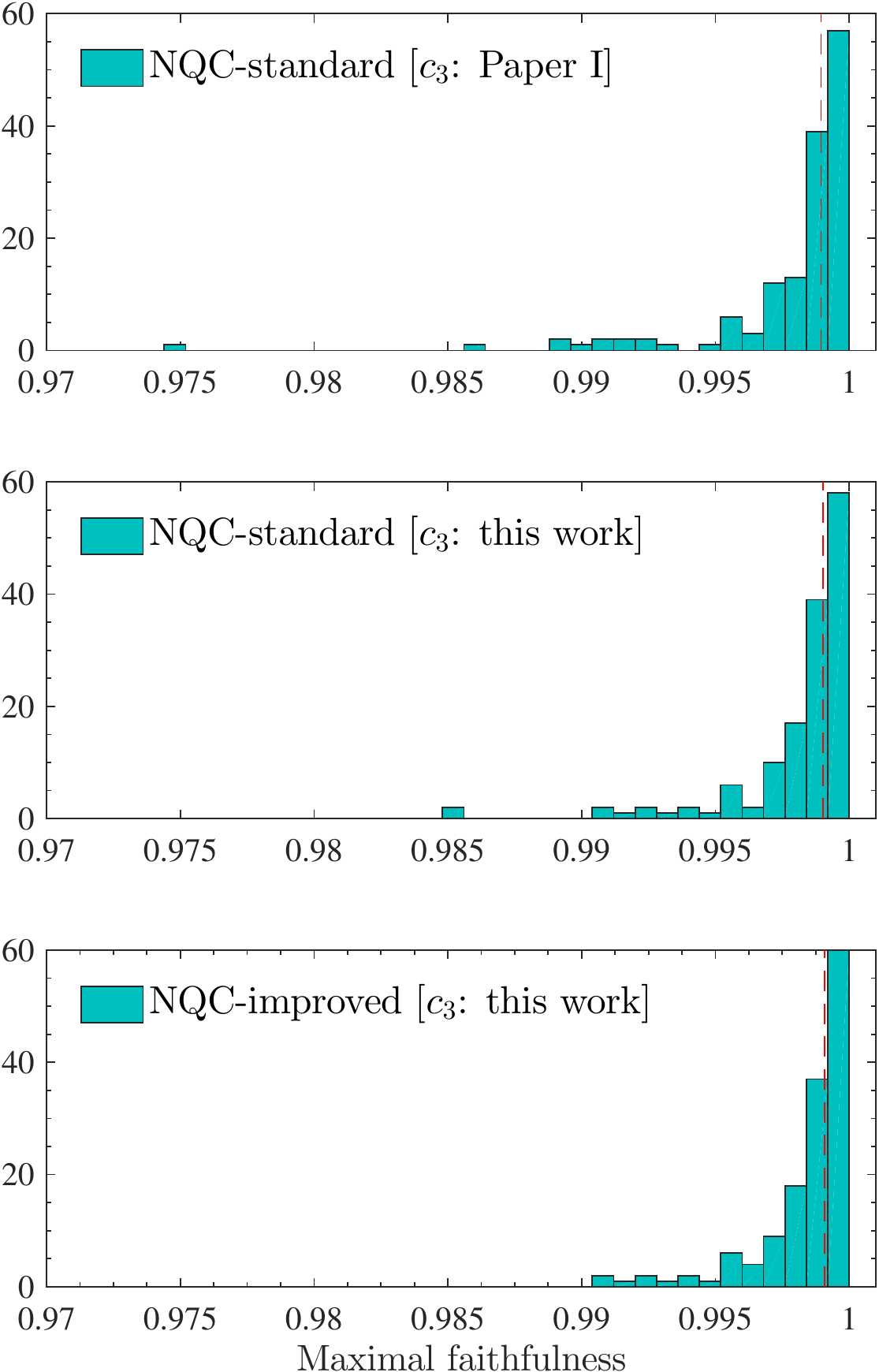}
\caption{\label{fig:hist} Improvement of the EOB/NR {\it faithfulness} by progressively
  including more NR information. Top panel: the model of Paper~I. Medium panel: effect of
  the new determination of $c_3$. Bottom panel: new determination of $c_3$, improved fit
  for the NQC point and superposition of QNMs to model the ringdown
  of SXS:BBH:0292 and SXS:BBH:0258.}
\end{figure}
The top panel corresponds to Fig.~\ref{fig:barF_0}, the medium to
Fig.~\ref{fig:barF_new} (with the two otuliers, SXS:BBH:0292 and SXS:BBH:025,
kept with the inaccurate postmerger-ringdown)
and the bottom to Fig.~\ref{fig:barF_new_global}. The values of the
medians are respectively, from top to bottom panels, $0.99895$, $0.99902$
and $0.99908$, and are indicated by the vertical dashed lines in the
figures. It is interesting to note that the use of a more accurate, and robust,
fit of the NQC point that incorporates more NR information,
though {\it absolutely needed} for theoretical reasons in order
to improve the EOB/NR consistency at merger, has essentially no
impact on this diagnostics.

\section{SXS datasets with $q=7$}
\label{sec:q7}
Let us now come back to the $q=7$ line in Fig.~\ref{fig:NQC_point_all}
to investigate the probable origin of this behavior. Before doing so,
let us however focus on the EOB/NR comparison for $(7,+0.4,0)$, SXS:BBH:0204,
whose merger behavior was showing the largest deviation with respect to the
global trend of the other datasets. The SXS:BBH:0204 dataset is the longest
\begin{figure}[t]
\center
\includegraphics[width=0.45\textwidth]{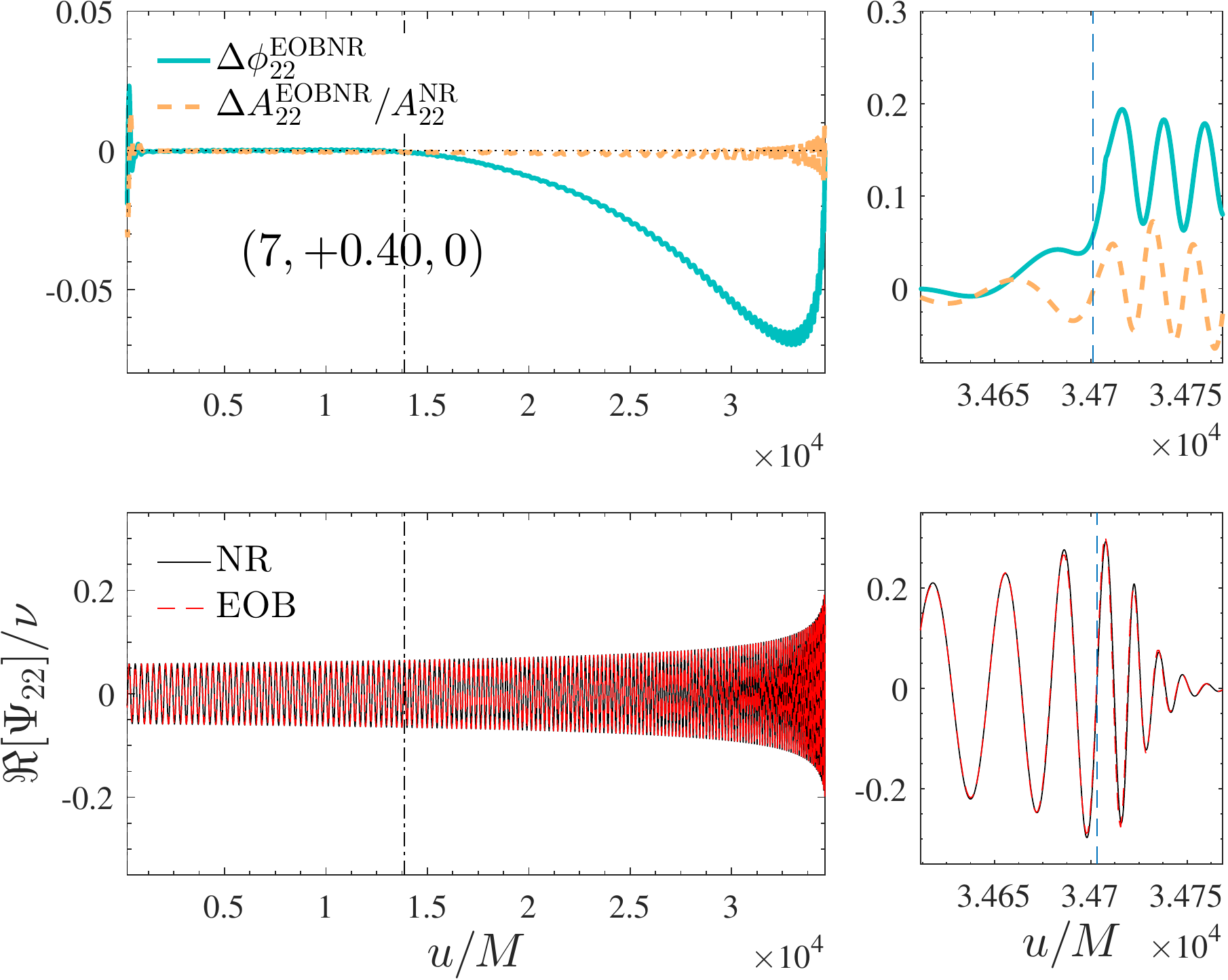}
\caption{\label{fig:phasing_q7_p04}EOB/NR phasing comparison for the 88.4 orbits
  long dataset SXS:BBH:0204. It is remarkable how the good phasing agreement is
  maintained up to merger, notably without using this waveform for calibration.}
\end{figure}
waveform currently present in the SXS catalog, with approximately 88.4 orbits from
the beginning fo the inspiral to merger and ringdown.
The standard phasing comparison is illustrated in Fig.~\ref{fig:phasing_q7_p04}.
The agreement all over is extremely good, with the phase difference barely
reaching the 0.1~rad value just a few orbits before merger. However, the top-right
panel of Fig.~\ref{fig:phasing_q7_p04} also illustrates rather dramatic phase
differences during postmerger-ringdown, with large-amplitude oscillations.
\begin{figure}[t]
\center
\includegraphics[width=0.42\textwidth]{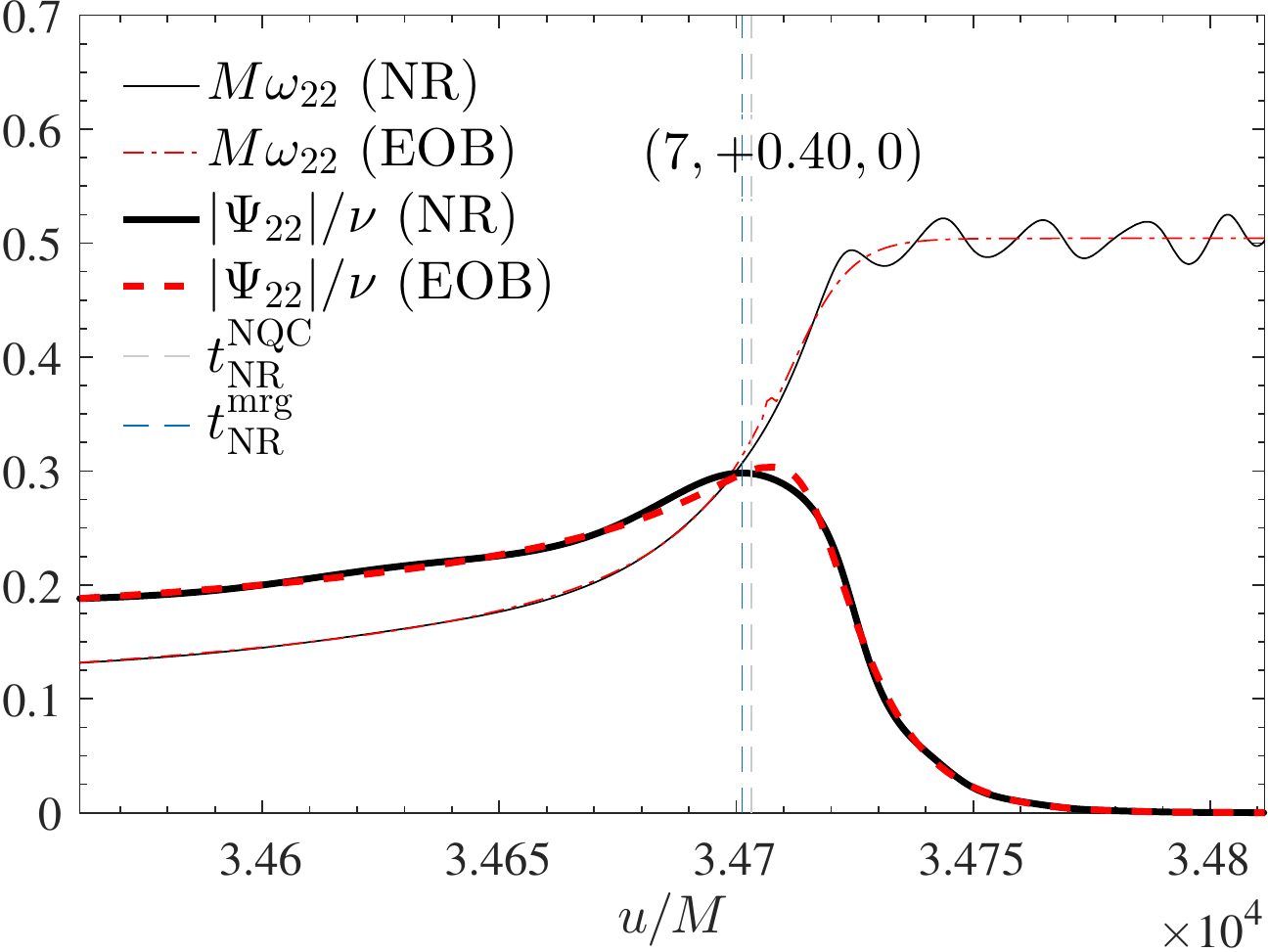} \\
\includegraphics[width=0.42\textwidth]{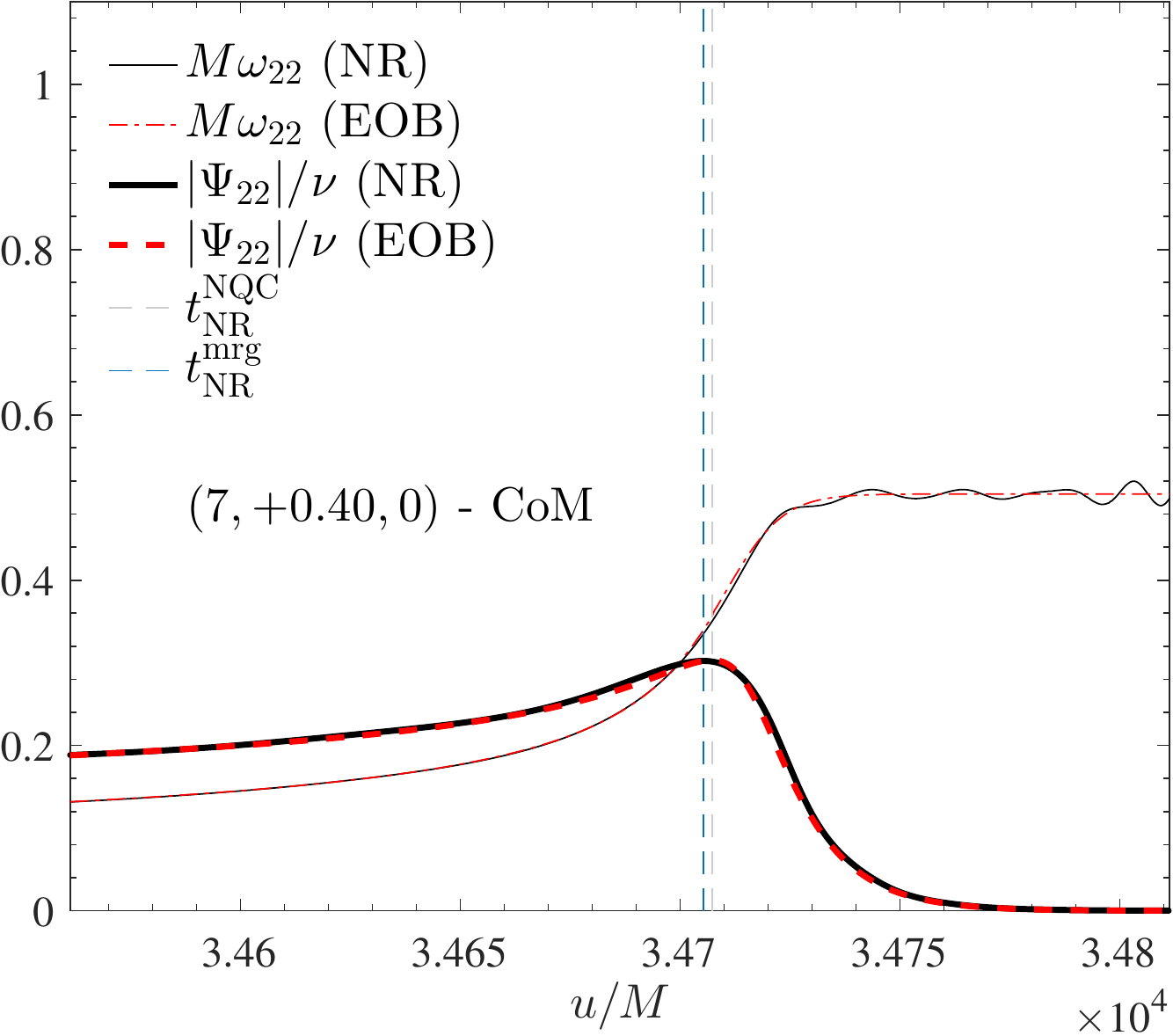}
\caption{\label{fig:Amp_omg_q7_p04}Effect of the motion of the center of
  mass on the behavior of the amplitude and frequency of SXS:BBH:0204
  of Fig.~\ref{fig:phasing_q7_p04}. Top panel: the EOB/NR
  disagreement between both modulus (around merger) and frequency (oscillations
  during the ringdown phase) is evident, and mostly due to the unphysical
  drift of the center of mass of the system. Once this effect is approximately
  corrected (as illustrated in Fig.~\ref{fig:CoMDriftq7} below),
  the NR waveform is visibly more consistent with the EOB one.}
\end{figure}
The analysis of the waveform amplitude and frequency, top panel
of Fig.~\ref{fig:Amp_omg_q7_p04} gives a better understanding
of what is going on. First, the modulus looks to have a quite
peculiar shape around merger (the peak is wide, with visible
oscillations) and also quite relevant oscillations occurr in
the frequency during the (expected) ringdown plateau. These oscillations
resamble the well-known beating between prograde and retrograde
modes, though such an amplitude looks weird here because this
type of behavior is typical only when the spins are anti-aligned with
the angular momentum, and large. So, it seems that there is
something unexpected in these waveform, that was also apparent
in the NQC plot shown before. These features illustrate clearly
that, as with all numerical relativity codes, the use of waveforms
requires some care and attention to detail due to the existence
and presence of unexpected features. Gravitational-wave data
generated from such codes are gauge dependent and may contain
artefacts from the numerical schemes used in the construction
of the initial data and subsequent evolution of the binary black holes. 
In particular, the SXS publicly available waveforms are known to contain
some features that must be accounted for when comparing to or calibrating
semi-analytical waveform models. For example, it was recently highlighted
that that SXS waveforms contain unmodeled features arising from the
displacement and drift of the center of mass, as seen in Fig.~\ref{fig:CoMDriftq7} ~\cite{Szilagyi:2015rwa}~\cite{Boyle:2015nqa}.  

The presence of such centre-of-mass (CoM) drifts has been associated to the
incomplete control of the ADM linear momentum ${\bf{P}}_{\rm{ADM}}$ in the
construction of initial data~\cite{Ossokine:2015yla}. The existence of
residual linear momenta leads to an overall drift of the CoM of the binary
leading to a number of unmodeled effects. Most notably, as the binary drifts
from the origin, the gravitational wave extraction spheres will move
off-center with respect to the center-of-mass inducing mode-mixing
between the spherical harmonic modes~\cite{Ossokine:2015yla}.
This coupling between the CoM drift and the outer boundary was
also identified in the long-duration, nonspinning, $q=7$ SXS
simulation~\cite{Szilagyi:2015rwa, Ossokine:2015yla}.
\textrm{SpEC} employs constraint preserving outer boundary conditions 
that are designed to work optimally for lower order spherical harmonics.
As the binary drifts or develops non-trivial recoils, the higher order
spherical harmonics can become more important and may lead to a runaway
acceleration of the CoM~\cite{Szilagyi:2015rwa, Ossokine:2015yla}. 
This effect was discussed for the $q=7$, nonspinning, run in which
the center-of-mass was observed to increase \textit{exponentially}
with time~\cite{Szilagyi:2015rwa}. The growth rate of the drift was
found to depend on the outer boundary radii behaving like $\sigma \propto R^{1.45}$
with timescales on the order of $\sim 2.6 \times 10^4 M$ for that run.
Notably, Ref.~\cite{Szilagyi:2015rwa} pointed out that the com drift
eventually yielded effects on the $\ell=m=2$ such to make the merger
part rather inaccurate.

\begin{figure}[t]
\center
\includegraphics[width=0.45\textwidth]{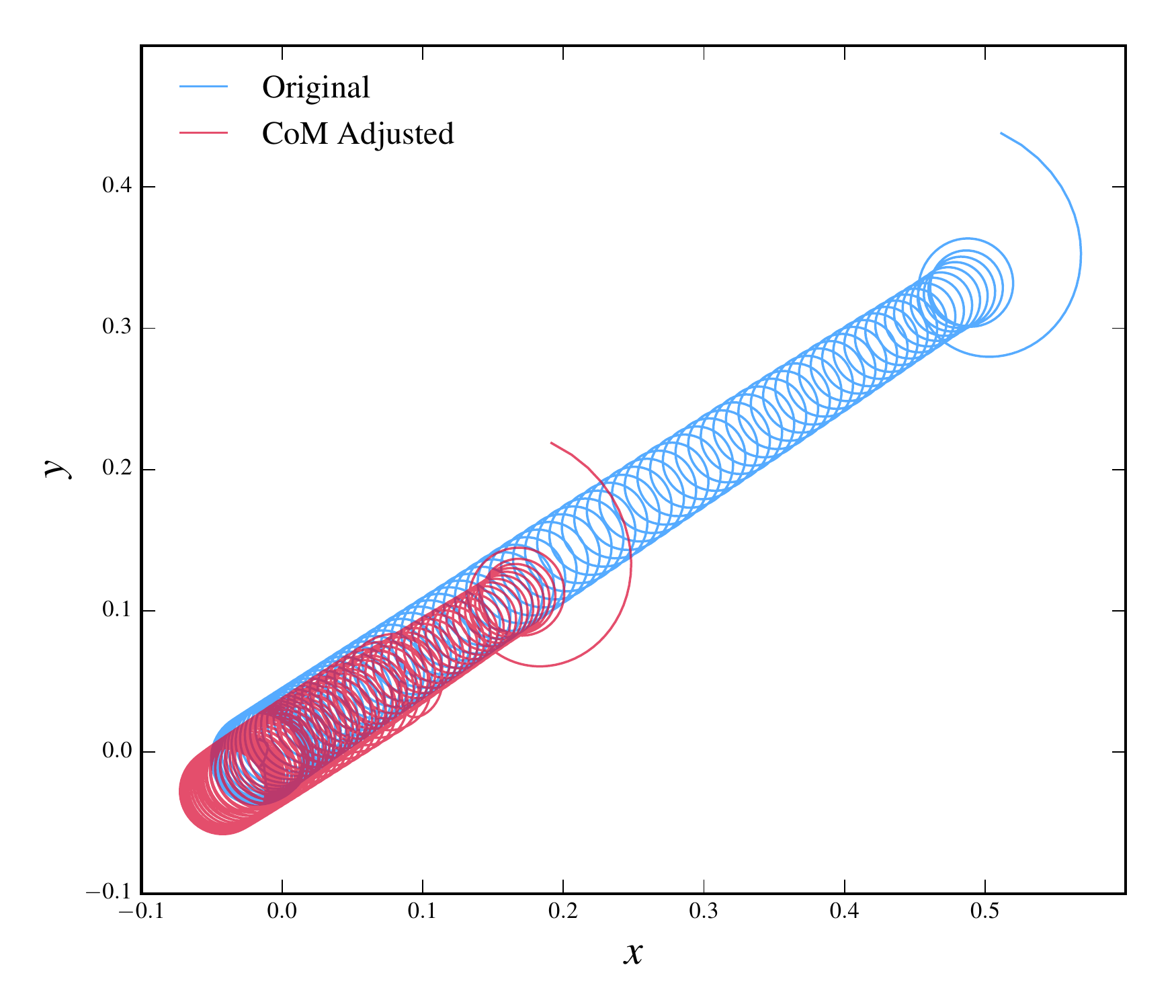}
\caption{\label{fig:CoMDriftq7}The impact of the CoM drifts can be mitigated by estimating the optimal boosts 
and translations from the horizon data, allowing us to correct for such drifts. Here we show the results of the procedure applied 
to the $(7,+0.4,0)$ SXS:BBH:0204 waveform.}
\end{figure}

A preliminary study of the impact of such drifts on all SXS waveforms
was presented in~\cite{Boyle:2015nqa}. Here it was seen that the
$(2,1)$, $(3,3)$ and $(3,1)$ modes exhibited distinct oscillations
that were not visible in the other waveform modes. These modes can
be seen to have the largest coupling to the dominant $(2,2)$ mode under
translations in the $x-y$ plane, suggesting that the unphysical oscillations
are a consequence of the mode-coupling induced by the drift of the
center-of-mass~\cite{Boyle:2015nqa}. At merger, it can be seen that power
is transferred between modes, leading to corrections above the \%-level
to the mode-amplitudes. For the \textrm{SXS:BBH:0004} simulation considered
in Ref.~\cite{Boyle:2015nqa}, the amplitude of the $(2,1)$ and $(3,1)$
modes as a fraction of the $(2,2)$ mode were seen to change by over
$1.3$\% and the amplitude of $(3,1)$ by $\sim 0.35$\%. However,
Ref.~\cite{Boyle:2015nqa} also introduced a
framework~\footnote{An implementation of these BMS transformations can
be found in the open-source python code
\textit{scri}: \url{https://github.com/moble/scri/}.}  
to account for such drifts by applying a BMS transformation 
on the Newman-Penrose Weyl scalar $\psi_4$ or, alternatively,
the transverse-traceless metric perturbation $h$. 
Schematically, the framework presented in~\cite{Boyle:2015nqa}
uses the coordinate positions and Christodoulou masses of
the black holes from  the horizon data in order to minimize
the average distance between the centre-of-mass and the origin
\begin{align}
  \label{eq:com_D}
\Xi ( \delta \bfx , \bfv ) = \int^{t_f}_{t_i} \, | \bfx_{\rm{CoM}} - (\delta \bfx + \bfv t) |^2 \, dt .
\end{align}
This allows one to estimate the optimal translation and
boost $(\delta \bfx , \bfv)$ that can be used to remove drifts 
from the numerical waveforms. Although the initial offset
of the SXS waveforms can be small, the length of the simulations means 
that even small boosts can grow into translations
that dominate over the initial offsets.
As noted in~\cite{Boyle:2015nqa}, the procedure detailed above will
be susceptible to a number of gauge effects but does
remove or mitigate unphysical effects that are not
anticipated based on our understanding of analytical waveforms.

An updated implementation of the quasi-equilibrium conformal
thin-sandwhich initial data formalism was introduced
in \cite{Ossokine:2015yla} partly with the aim of controlling the spurious
residual linear momenta present in the initial data. However,
for the current waveform catalog and subsequent simulations that
retain significant residual momenta  or develop non-trivial recoils,
the waveforms can be post-processed to help eliminate or
mitigate such initial offsets and overall drifts.

Coming back to the long 88.4 orbits SXS:BBH:0204 waveform,
it might be possible that the strange feature in the modulus
as well as the oscillations in the frequency during ringdown
are related to such unphysical drifts of the CoM. Before entering
this discussion let us recall that: (i) there is no mention
of the existence of these effects in papers of the SXS
collaboration where this waveform dataset was presented for the
first time; (ii) a simulation with the same parameters but
smaller initial separation, corresponding to 58 orbits
up to merger, SXS:BBH:0203, yields a waveform where the frequency
is perfectly flat (no oscillations) and the modulus has a
similarly smoother shaper much closer to the EOB one.
The SXS:BBH:204 waveform without the com drift yields the bottom
panel of Fig.~\ref{fig:Amp_omg_q7_p04}: the modulus is fully consistent,
at merger, with the EOB one; the small-amplitude oscillations in
the late inspiral part of the modulus have disappeared as well
as those in the post-merger part of the frequency.

We have verified that the same picture is true for the other
$q=7$ datasets although the importance of the com drift depend
on the simulation under consideration. For example, it is visible
for the $(7,\pm 0.4,0)$ data, more evidently for SXS:BBH:0204
and SXS:BBH:0206, and less, but still not negligibly, for
the shorter SXS:BBH:0203 and SXS:BBH:0205. By contrast, the effect
is practically absent for the $(7,\pm 0.6,0)$ datasets,
SXS:BBH:0202 and SXS:BBH:0207. This analysis justifies
our choice of not using $q=7$ data for the NQC fits.
\begin{figure}[t]
\center
\includegraphics[width=0.45\textwidth]{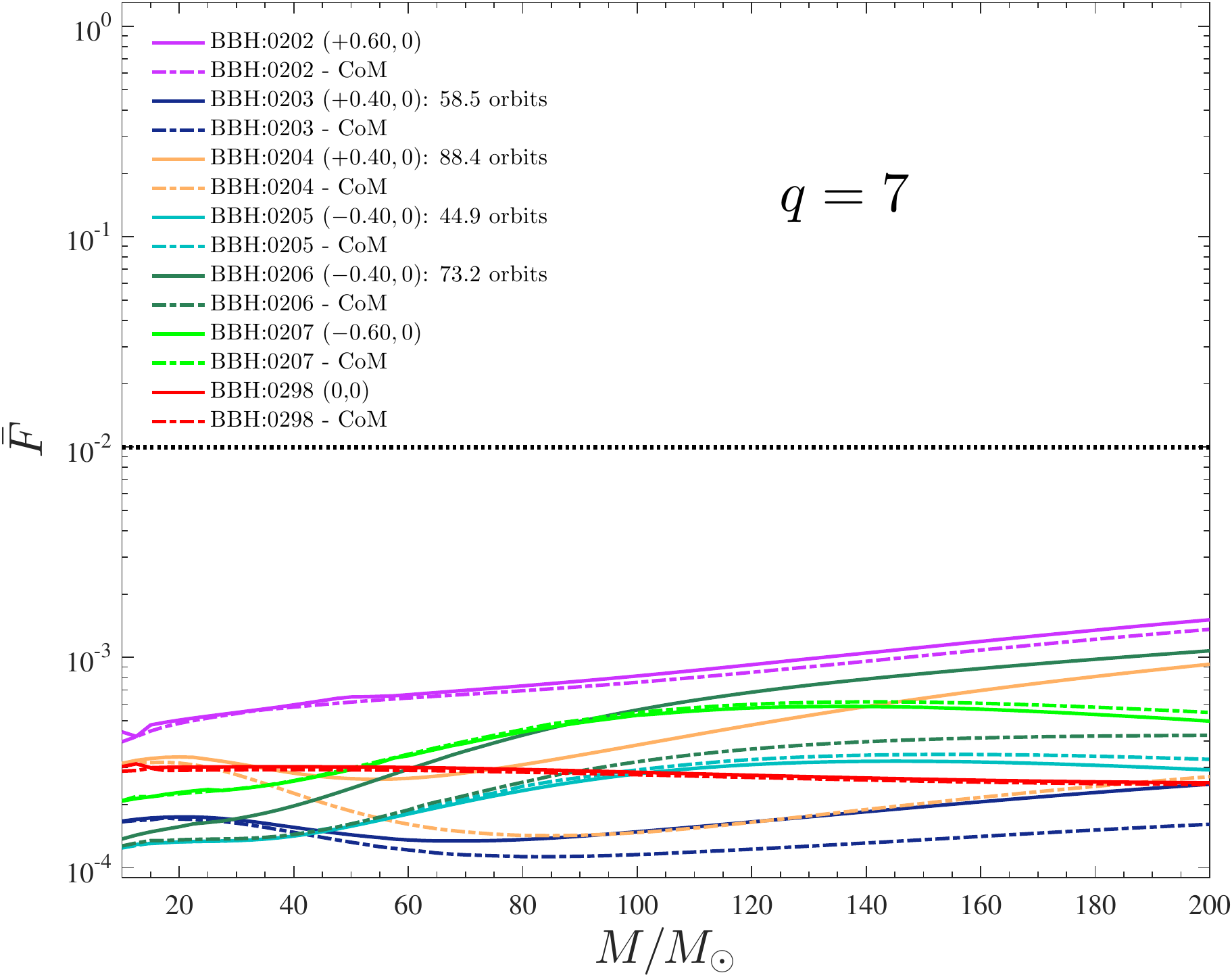}
\caption{\label{fig:barFq7}EOB/NR unfaithfulness for $q=7$ binaries.
  The removal of the CoM drift using Eq.~\eqref{eq:com_D} visibly reduces
  $\bar{F}$ for $(\pm 0.4,0)$ datasets.}
\end{figure}
Finally, it is interesting to evaluate the impact of the
com drift in the usual EOB/NR unfaithfulness comparison,
Fig.~\ref{fig:barFq7}, where one sees clearly the effect
of the removal of the com drift on the longest waveforms.
By contrast, as mentioned above, the correction is (essentially)
ineffective for the shortest one, except for SXS:BBH:0203,
where some effect is seen. The fact that for SXS:BBH:0202
the unfaithfulness $\bar{F}$ is found to grow monothonically
is mirroring slight inaccuracies in the postmerger interpolating fit,
analogously to what discussed in Fig.~\ref{fig:barF_new},
as well as the absence of prograde ($m<0$) modes (that are
somehow present for the negative spin datasets, see Ref.~\cite{Nagar:2016iwa})
in the modelization of the ringdown. Although the values
of $\bar{F}$ in Fig.~\ref{fig:barFq7} are well below the
standard target of $1\%$, it is remarkable that they
can be improved {\it not} by changing the analytical model,
but rather correcting systematics in the NR waveforms.

\section{BAM simulation with $q=8$}
\label{sec:q8-bam}
To anticipate future work, let us finally explore how our EOB model
performs in a region of the parameter space marginally outside the one
covered by the SXS waveform catalog. To do so, we compare the EOB prediction
wth one waveform with physical parameters $(8,+0.85,+0.85)$ that was previously
produced using the \rm{BAM} code~\cite{Bruegmann:2006at,Husa:2007hp,Gonzalez:2006md}
and employed to calibrate the IMRPhenomD model~\cite{Husa:2015iqa,Khan:2015jqa}.
This waveform is the same mentioned in Ref.~\cite{Bohe:2016gbl}
(at one resolution with eccentricity of $1.2\times 10^{-2}$)
and was checked to be in excellent agreement with an improved one
(smaller eccentricity, higher resolution) obtained simulated using the
{\rm Einstein Toolkit}. Following this statement of Ref.~\cite{Bohe:2016gbl}
we can thus consider this {\rm BAM} waveform accurate enough for testing
the EOB[NR] model we present here.
To be clear, we deal now with the new determination of $c_3$, but we rely on
the interpolating fit for the NQC-extraction point of Section~\ref{sec:spin0}
and not the one of Section~\ref{sec:NQC_improved}. We have verified that
the benefits of using the improved fits are negligible.

\begin{figure}[t]
\center
\includegraphics[width=0.42\textwidth]{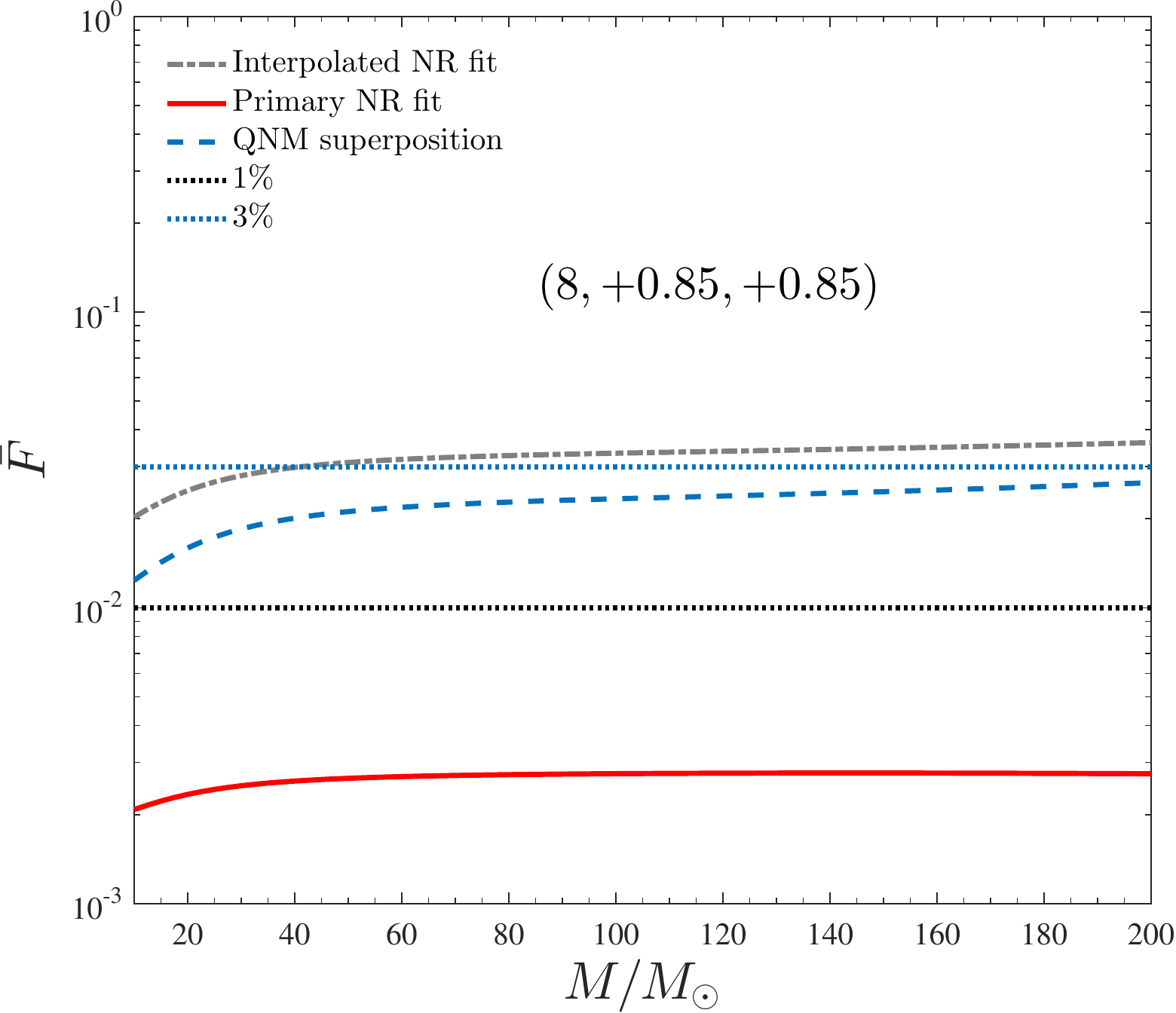}
\caption{\label{fig:barFq8}Unfaithfulness between EOB and the NR dataset $(8,+0.85,+0.85)$
  obtained using the BAM code. The growth of $\bar{F}$ is mirroring the inaccurate
  representation of the ringdown yielded by the interpolating postmerger fits
  of Ref.~\cite{Nagar:2016iwa}. If the primary NR fit is used, the model yields
  a more than acceptable result.}
\end{figure}

\begin{figure}[t]
\center
\includegraphics[width=0.42\textwidth]{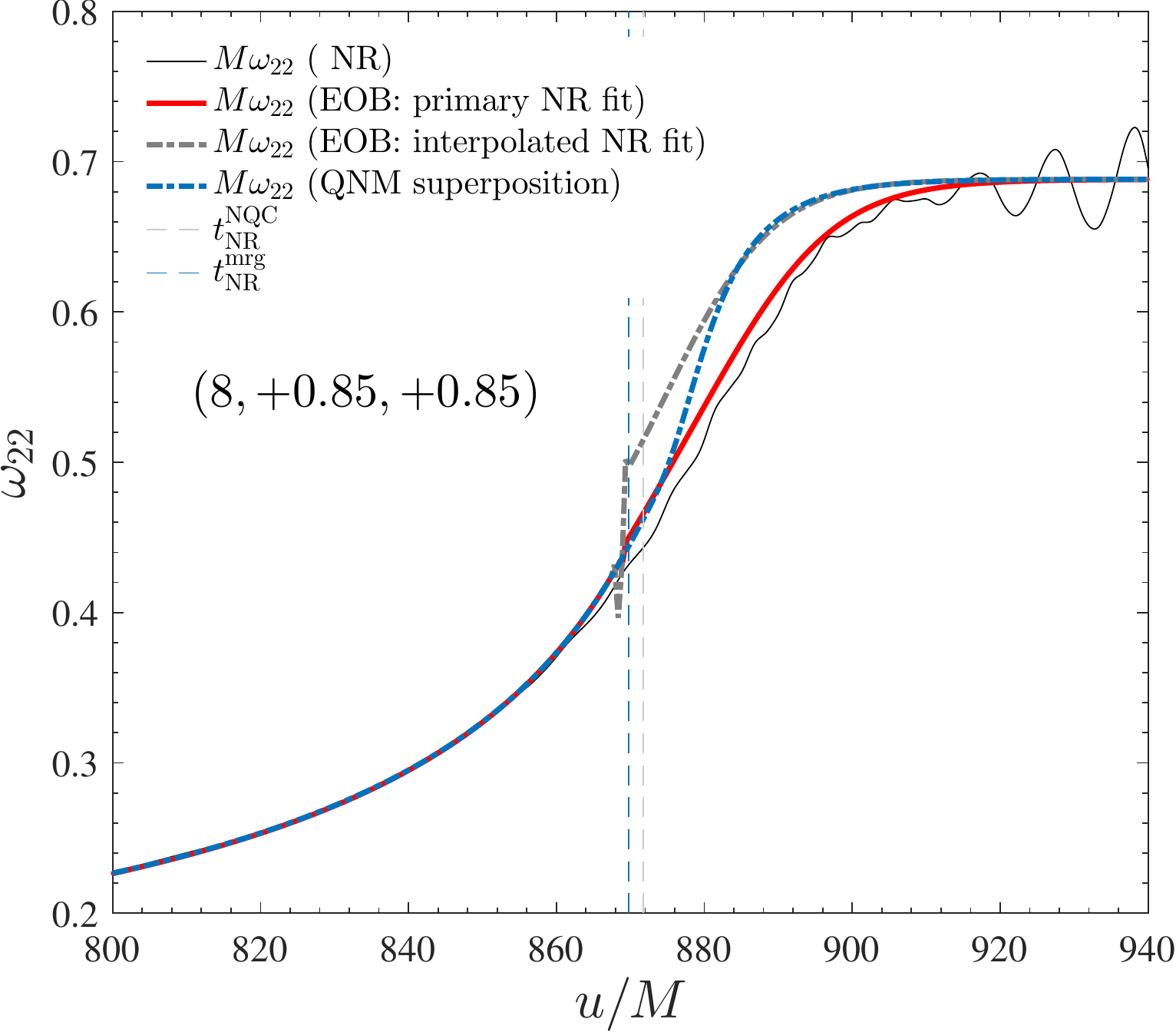}
\caption{\label{fig:q8p085p085_freq}Understanding the inaccuracy of the EOB waveform
  highlighted in Fig.~\ref{fig:barFq8}: EOB/NR comparison between the gravitational
  wave frequency obtained with different representations of the postmerger phase.
  If the superposition of 8~QNMs is unable to correctly reproduce the correct frequency
  behavior, there is no problem of doing it via the primary fit following Refs.~\cite{Damour:2014yha}.}
\end{figure}

Figure~\ref{fig:barFq8} illustrates the EOB/NR unfaithfulness obtained
with this model (dashed, dark-grey, line) that is starting above $1\%$
and eventually crosses the $3\%$ level. Note also that this waveform is
very short, so that the $\bar{F}$ plot largely emphasizes inaccuracies
of the late-inspiral, merger and ringdown. The result of Fig.~\ref{fig:barFq8}
is better understood by comparing the EOB and NR instantaneous GW frequencies,
$\omega_{22}$, Fig.~\ref{fig:q8p085p085_freq}, after aligning the waveforms on
the frequency interval $[\omega_L,\omega_R]=[0.3,0.35]$, i.e. during the plunge phase.
The interpolating postmerger fit provides a frequency at matching point
that is {\it too large} with respect to the actual value. This eventually
means that the EOB and NR waveforms dephase very fast after the matching
point, and explains the growth of $\bar{F}$ when $M$ is increased and the
merger part of the waveforms progressively moves towards the most sensitive
window of the detector. However, Fig.~\ref{fig:q8p085p085_freq} also illustrates
that the issue comes completely from the postmerger-ringdown part, while the behavior
of the EOB frequency provided by the combination of $c_3$ and of the NQC fit
(that, we remind, were determined independently of this dataset) is fully
consistent with the corresponding NR one.
To highlight this better, in Fig.~\ref{fig:q8p085p085_freq} we show, as a red line,
the description of the postmerger waveform obtained using the {\it primary fit}
instead of the interpolated one. Not surprisingly, in this case $\bar{F}$
also drops down to the $10^{-3}$ level (see Fig.~\ref{fig:barFq8}).
This fact indicates that the most urgent need to improve our model is to
build improved interpolating fits of the postmerger waveform,
following~\cite{Nagar:2016iwa}, that crucially incorporate NR waveform
data with large values of $q$ and large spins.
Finally, it is interesting to explore what can be achieved in this case by
using the old-fashioned description of the ringdown as a superposition of
QNMs with constant coefficients. As above, we use $N = 8$ QNMs and $\Delta t^{\rm QNM} = 0.5$.
Figures~\ref{fig:barFq8}-\ref{fig:q8p085p085_freq} show that an improvement with
respect to the interpolating fit case, though the frequency disagreement close to
the onset of the ringing of the fundamental model is still large enough that the
unfaithfulness is between $1\%$ and $3\%$. Although this straightforward
approach does not yield a fully satisfying solution to the problem, it still
represents a reasonable, and no costly, procedure that can be applied when the
interpolating fits looks clearly inaccurate. Also, this is the typical situation
that could be solved by the implementation of a ``pseudo-QNM'' frequency in order
to bridge the gap between the merger frequency $M\omega^{\rm mrg}\approx 0.45$
and $\omega_8\simeq 0.6$, that is the frequency of the eight overtone, i.e. the
lowest frequency included in the QNMs template.

\begin{figure}[t]
\center
\includegraphics[width=0.42\textwidth]{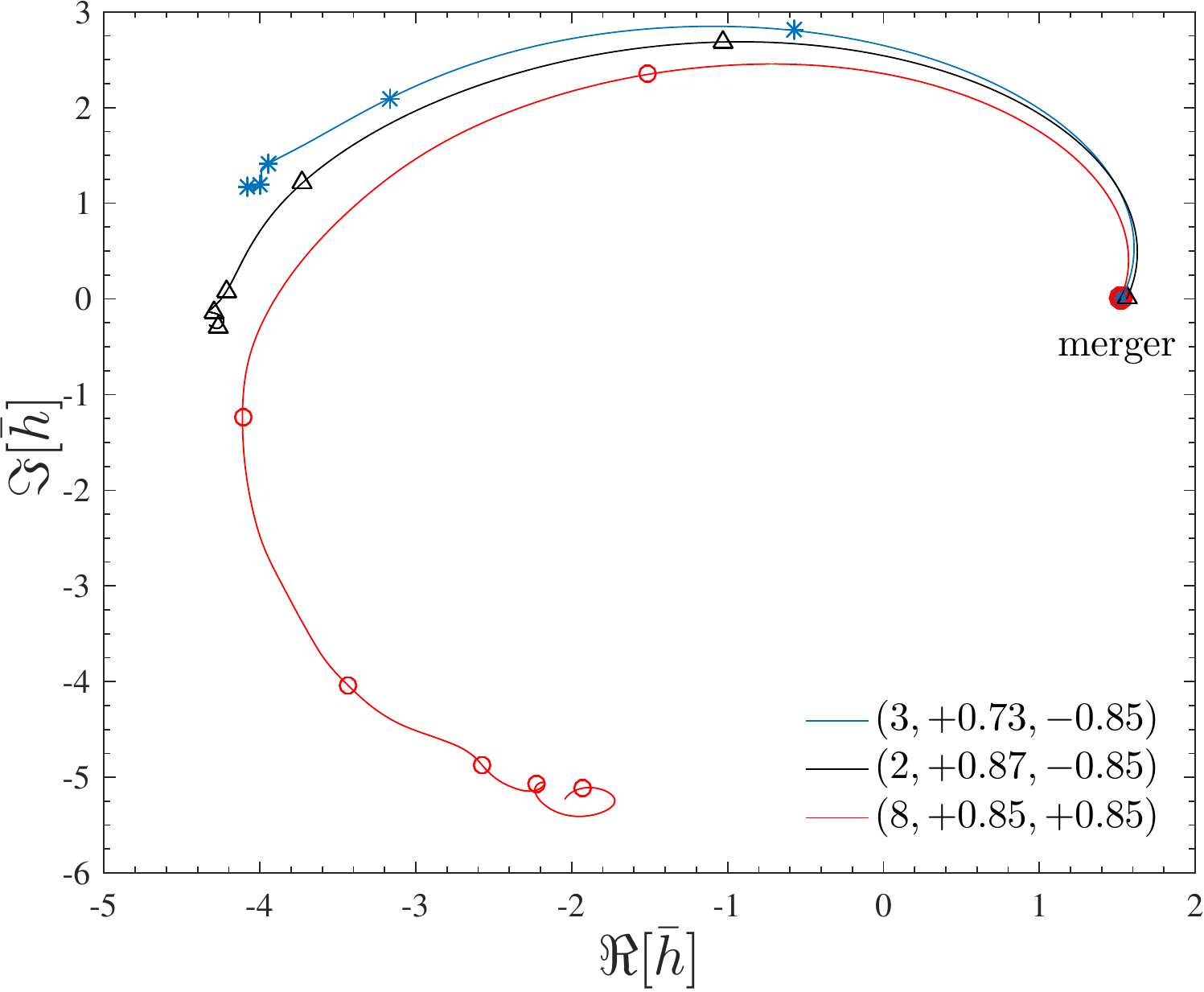}
\caption{\label{fig:barh}Behavior of the QNM-rescaled postmerger
  waveform $\bar{h}$, in the complex plane, for the three datasets
  considered discussed in the text. Each marker indicates intervals
  of $10M_{f}$ after the merger point of the respective waveform.
  The fact that the $(8,+0.85,+0.85)$ curve rotates by almost $2\pi$
  prevents it from being fully represented by a superposition of QNMs.
  The nonfactored waveform is dominated by the fundamental QNM only
  when the markers are approximately superposed. Note in this respect
  that the BAM waveform is more noisy than the SXS ones.}
\end{figure}
Let us finally comment on the reason why the ringdown description as a
linear combination of QNMs with constant coefficients was rather accurate
for $(2,+0.87,-0.87)$ and $(3,+0.73,-0.85)$ while it is not for
$(8,+0.85,+0.85)$. This is done, following Ref.~\cite{Damour:2014yha},
notably Sec.~III and Fig.~2 there, by inspecting the diagnostics given by
the postmerger waveform where the fundamental mode is factored out,
$\bar{h}\equiv e^{\sigma_1 \tau +{\rm i}\phi_{22}^{\rm mrg}}$. Here, $\sigma_1\equiv \omega_1 + \ii \alpha_1$ is the complex frequency of the fundamental mode, $h\equiv h_{22}$ is
the waveform, $\tau\equiv (t^{\rm NR}-t^{\rm NR}_{\rm mrg})/M_{f}$ is the
dimensionless time parameter that counts time in units of the final
BH mass $M_f$, and $\phi_{22}^{\rm mrg}$ is the value of the phase at merger (so
that $\bar{h}(0)$ is the real amplitude of the waveform $h$ at merger.
The rescaled waveform $\bar{h}$ is considered as a complex number. It is
shown in Fig.~\ref{fig:barh} for the three NR waveforms considered here.
It was noted in Ref.~\cite{Damour:2014yha}
that one can represent the ringdown as a linear superposition of QNMs with
constant coefficients only if $\bar{h}$ describes an {\it approximately straight line}
in the complex plane. This is so because the complex frequency differences
$\sigma_{21}\equiv \sigma_2-\sigma_1$, $\sigma_{31}\equiv \sigma_3-\sigma_1$,
$\dots$ are approximately real and positive, especially for high spins.
By contrast, when the spin of the final
BH gets high, the early postmerger part $\bar{h}$ is seen to strongly
rotate, which means that it cannot be represented as a standard superposition
of QNMs. This was discussed at length in Ref.~\cite{Damour:2014yha}, that
was focusing in particular on the case $(1,+0.97,+0.97)$ as an example of waveform where
one cannot use a linear combination of QNMs with constant coefficients
starting from the peak of the waveform (see also Ref.~\cite{Kamaretsos:2011um,Kamaretsos:2012bs} for similar conclusions obtained though using different diagnostics.)
Inspecting $\bar{h}$ for the three datasets considered here, one finds that
$\bar{h}$ is relatively ``straight'' for $(2,+0.87,-0.87)$ and $(3,+0.73,-0.85)$,
that correspond to $a_f\approx 0.84$ and $a_f\approx 0.836$ respectively,
while for $(8,+0.85,+0.85)$, that yield to $a_f\approx 0.89$, one has
a situation much closer to the $(1,+0.97,+0.97)$ ($a_f\approx 0.945$)
case discussed in Ref.~\cite{Damour:2014yha}.
The precise and quantitative understanding, using $\bar{h}$, of which
region of the parameter space can be easily represented by a superposition
of QNMs and when one has to resort to alternative description needs
precise investigations. As a rule of thumb, when the final BH spin is
not too extreme, and thus the postmerger-ringdown waveform relatively short,
resorting to a superposition of QNMs usually give reliable results.
Typically, this is the case when the black holes are nonspinning or
the spins are anti-aligned with the orbital angular momentum.
One the contrary, when the transition from inspiral to plunge and
merger is more adiabatic, e.g., high spins
{\it aligned} with the orbital angular momentum, the postmerger part
brings many cycles and this is usually inaccurately captured as a
superposition of QNMs because of the ``rotation'' of $\bar{h}$ described
above. However, there might be many intermediate situations (e.g., unequal
masses, one spin up and one spin down) where the straightforward
QNMs superposition can be accurate enough for analytical modeling
purposes. We plan to analyze these situations quantitatively in
future work.

\section{Conclusions}
\label{sec:conclusions}
We have tested the {\tt SEOBNR\_ihes} EOBNR model introduced in Ref.~\cite{Nagar:2015xqa}
(Paper~I) on the very large sample of 149 NR, spin-aligned, datasets publicly available through
the SXS catalog~\cite{SXS:catalog}. Of these 149 datasets, 96 became available
only recently, after October~31st~2016. Our most relevant findings can be summarized as follows:
\begin{itemize}
\item[1.] We found that, despite the rater small number (40) of datasets that were used
  to tune the model of Paper~I, without any additional change or tuning the model is
  found to perform rather well all over the full NR sample of 149 configurations.
  Looking as usual at the EOB/NR unfaithfulness $\bar{F}$, with total mass $M$ varying between
  $10M_\odot$ and $200M_\odot$, we find that $\max(\bar{F})< 1\%$ for all datasets except
  for four outliers that are in the range $1\%\lesssim\max(\bar{F})<3\%$.
  This result by itself probes the {\it robustness} of the theoretical construction
  of Ref.~\cite{Damour:2014sva} as well as of the NR information procedure of Paper~I.
  In addition, this is also illustrates that only a rather small amount of NR waveform
  data is effectively necessary to construct a model that would in principle yield an
  acceptable loss of events ($< 10\%$) if used for detection purposes.
\item[2.] We then proceeded by applying minimal changes to the model in order to improve
  it even further, aiming at having $\max{\bar{F}}<1\%$ all over the SXS catalog.
  To do so, we added three more simulations to those used to inform the model in
  Ref.~\cite{Nagar:2015xqa}, so to provide an improved determination of the
  next-to-next-to-next-to-leading spin-orbit effective spin-orbit parameter $c_3$
  that slightly modified the EOBNR spinning waveform of Paper~I generated away from
  the equal-mass, equal-spin regime. In practice, this amounts in
  {\it changing only four numerical coefficients} in the global fit of $c_3$ with respect
  to Paper~I, that are listed in Eqs.~\eqref{eq:p1_prime}-\eqref{eq:p4_prime}. The new
  evaluation of the unfaithfulness is again below $1\%$ except now for two (more)
  outliers that reacj the $1.1\%$ level at $200M_\odot$. We show that this is only
  due to the somehow inaccurate representation of the postmerger-ringdown waveform
  by means of the phenomenological fit of Ref.~\cite{Nagar:2016iwa} that was extrapolating
  from a small sample of sparse NR waveforms to the full parameter space. When a more accurate
  representation of the postmerger phase is used, e.g. using a standard superposition
  of $N=8$ QNMs, values of $\max(\bar{F})$ lower than $1\%$ are easily recovered
  (see Fig.~\ref{fig:outliers_bF}). 
\item[3.]We analyzed in detail the $q=7$ waveform data included in the SXS catalog and
  concluded that the longest waveforms presents features, around merger, that are due
  to the spurious motion of the center of mass and, as such, should probably not be
  used to inform the model. Still, once the unphysical features are (approximately)
  removed, the consistency of the NR waveform with the EOB prediction is fully
  recovered, with values of $\bar{F}$ well below the $0.1\%$ level.
\item[4.]We also compared the EOBNR prediction against a BAM waveform with parameters
  $(8,+0.85,+0.85)$ that was previously used in Ref.~\cite{Bohe:2016gbl}. We find, again,
  that the Achille's heel of the model is represented by the postmerger-ringdown interpolating
  fit, while the EOB/NR compatibility up to merger is more than satisfactory
  (see Fig.~\ref{fig:q8p085p085_freq}). In this particular case, due to the large value
  of the spin of the final BH, the standard superposition of QNMs is unable to reliably
  represent the postmerger phase, though it is able to lower $\bar{F}$ below the $3\%$ level.
  Our analysis suggests that the most urgent needs to improve our model is to produce
  more accurate representations of the postmerger waveform in order to improve the
  postmerger-ringdown models of Refs.~\cite{Nagar:2016iwa,Bohe:2016gbl}, in particular
  for large mass ratios and large spins. The old-style ringdown representation as a
  superposition of QNMs might still be effective and useful in some corners of the parameter
  space, but it will not work in more extreme regimes were the NR information is crucial.
  It would then be useful to have targeted, possibly short and inexpensive, NR simulations
  with the precise aim of accurately representing the postmerger-ringdown waveform
  all over the parameter space.
\item[5.]The model we presented here was informed and tested on NR waveforms up to mass
  ratio $q=10$ and {\it only up} to this mass ratio the model should be (conservatively)
  considered reliable. We have not included, on purpose, any information coming from
  test-particle waveform calculations~\cite{Taracchini:2014zpa,Harms:2014dqa} so to
  extrapolate the model more consistently to larger mass ratios.
  One of the reasons why we did so is that the EOB resummed waveform and radiation reaction
  used here~\cite{Damour:2014sva} is inconsistent with the one of Ref.~\cite{Harms:2014dqa},
  that was using higher-PN expressions. Still, the analytical expressions~\cite{Pan:2010hz}
  used in~\cite{Harms:2014dqa} proved inaccurate for high values of the spin of the central
  black hole, and more aggressive resummations of the waveform amplitude were proved to
  be needed~\cite{Nagar:2016ayt}. In future work we shall explore thoroughly the impact of
  such improved waveform resummation, that might possibly also simplify the determination of
  the NQC parameters and allow us to obtain simple fits for the NQC parameters instead of
  relying on fits fo the NR NQC point.In doing so, we will also analyze the impact of
  NR waveforms with $q>10$~\cite{Husa:2015iqa} in order to carefully check the blending
  of the present EOB waveforms with the test-particle limit ones.
\end{itemize}

\acknowledgments
We are grateful to M.~Hannam and S.~Husa for lending us their $(8,+0.85,+0.85)$ BAM
waveform discussed in Sec.~\ref{sec:q8-bam}.

\appendix
\section{Fits of the NR points used to determine NQC parameters}
\label{app:nqc}
The starting points for the iterative determination of the NQC parameters
are obtained as follows. Firstly, the waveform amplitude, frequency and
their time derivatives,
$(A_{22},\dot{A}_{22},\omega_{22},\dot{\omega}_{22})^{\rm NR}_{t_{\rm NQC}^{\rm NR}}$,
are extracted from the NR waveform at the time on the NR axis
$t_{\rm NQC}^{\rm NR}/M\equiv (t_{\rm mrg}^{\rm NR}+2)/M$. From these values, one performs fits
against the effective spin
\begin{align}
\hat{a}_0 = X_1\chi_1+X_2\chi_2.
\end{align}
Then, the coefficients of this fit are fitted as {\it linear} functions of $\nu$.
The dependence on $\hat{a}_0$ chosen depends on the quantity as well as
on the number of datapoints used. For example, for $q>1$ we fit the points with a
rational function of $\hat{a}_0$ that yields
\begin{align}
  \label{eq:omg22NQC0}
  \omega_{22}^{\rm NQC}=\frac{0.45584139\,\nu+0.27315247}{1+(0.75276414\,\nu-0.40081625)\hat{a}_0}  
\end{align}  
when fitting the points in Fig.~\ref{fig:NQC_0}, or
\begin{align}
    \label{eq:omg22NQCprime}
    \omega_{22}^{\rm NQC'}=\frac{0.46908067\,\nu+0.27022141}{1+(0.64131115\,\nu-0.37878384)\hat{a}_0}  
\end{align}
for fitting the points in Fig.~\ref{fig:NQC_point_all}. For $q=1$ we just adopt fourth-order polynomials,
and the outcome of the fit is given in Table~\ref{NQC_hybrid_q1}. The fits for $q>1$ corresponding
to the datapoints used in Fig.~\ref{fig:NQC_0} are given in Table~\ref{NQC_hybrid_ql1};
the alternative NQC-point fits that use more NR points, as illustrated in Fig.~\ref{fig:NQC_point_all}
and related text, are given in Table~\ref{NQC_global}.
\begin{table*}[t]
  \caption{\label{NQC_hybrid_q1} Fits of the NR quantities extracted at the NQC-extraction point,
    $t_{\rm NR}^{\rm NQC}/M=(t_{\rm mrg}^{\rm nR}+2)/M$ that are used for the iterative determination of
    the NQC parameters $(a_1,a_2,b_1,b_2)$ for $q=1$.}
\begin{ruledtabular}
\begin{tabular}{rl}
$A_{22}^{\rm NQC}         \quad  =\quad 0.00178195\, \hat{a}_0^4+0.00435589\,\hat{a}_0^3+0.00344489\,\hat{a}_0^2-0.00076165 \,\hat{a}_0+0.31973334$ \\
$\dot{A}_{22}^{\rm NQC}   \quad  = \quad 0.00000927\,\hat{a}_0^4-0.00024550\,\hat{a}_0^3+0.00012469\,\hat{a}_0^2 +0.00123845 \,\hat{a}_0-0.00195014 $ \\
$\omega_{22}^{\rm NQC}     \quad = \quad 0.00603482\,\hat{a}_0^4 + 0.01604555\,\hat{a}_0^3 +0.02290799\,\hat{a}_0^2  +0.07084587\,\hat{a}_0 +0.38321834$ \\
$\dot{\omega}_{22}^{\rm NQC}\quad = \quad 0.00024066\,\hat{a}_0^4+0.00038123\,\hat{a}_0^3 -0.00049714\,\hat{a}_0^2 + 0.00041219\,\hat{a}_0+0.01190548$ \\
\end{tabular}
\end{ruledtabular}
\end{table*}
\begin{table*}[t]
  \caption{\label{NQC_hybrid_ql1} Same as Table~\ref{NQC_hybrid_q1} but for $q>1$ and using the NR dataset discussed
    in Fig.~\ref{fig:NQC_0}. In this case, the corresponding $\omega_{22}^{\rm NQC}$ is given by Eq.~\eqref{eq:omg22NQC0} above.}
\begin{ruledtabular}
\begin{tabular}{r r}
$A_{22}^{\rm NQC}=$ & $(0.05385059\, \nu-0.00890942 )\, \hat{a}_0^2 +(-0.07942102\, \nu+ 0.02152423)\, \hat{a}_0 + (0.14805262\, \nu+ 0.28210487)$ \\
$\dot{A}_{22}^{\rm NQC}=$ &  $(0.00248472\, \nu-0.00033422 )\, \hat{a}_0^2 + (0.00105298\, \nu+ 0.00085160)\, \hat{a}_0 +(-0.00339257\, \nu-0.00110932)$ \\
$\dot{\omega}_{22}^{\rm NQC}=$ & $(-0.00177362\,\nu +0.00123900) \, \hat{a}_0 + (0.02424739\,\nu +0.00566504)$ \\
\end{tabular}
\end{ruledtabular}
\end{table*}
\begin{table*}
  \caption{\label{NQC_global} Same as Table~\ref{NQC_hybrid_q1} but for $q>1$ and using the NR dataset discussed
    in Fig.~\ref{fig:NQC_point_all}. In this case, the corresponding $\omega_{22}^{\rm NQC'}$ is given by Eq.~\eqref{eq:omg22NQCprime} above.}
\begin{ruledtabular}
 \begin{tabular}{cl}
    $A^{\rm NQC'}_{22}= $  & $(0.04680896\,\nu-0.00632114 )\, \hat{a}_0^3 +(0.06586192\, \nu-0.01180039)\, \hat{a}_0^2 +(-0.11617413\, \nu+ 0.02704959)\, \hat{a}_0  $ \\
                        & $+(0.15597465\, \nu+ 0.28034978)$\\
$\dot{A}_{22}^{\rm NQC'}$= & $(-0.00130824\, \nu+0.00006202 )\, \hat{a}_0^3  +(0.00199855\, \nu-0.00027474)\, \hat{a}_0^2+(0.00218838\, \nu+ 0.00071540)\, \hat{a}_0$\\
                        & $+(-0.00362779\, \nu-0.00105397)$\\
$\dot{\omega}_{22}^{\rm NQC'}$= &  $(0.00061175\,\nu + 0.00074001) \, \hat{a}_0  +(0.02504442\,\nu +0.00548217)$
\end{tabular}
\end{ruledtabular}
\end{table*}

\begin{table*}[t]
  \caption{\label{tab:nospin}Nonspinning NR datasets considered in this work. Datesets form 1 to 9 were used in the
    determination of $a_5^c$ in Paper~I and the related fits of the NQC parameters. The others are
    used here just for validation.}
\begin{center}
\begin{ruledtabular}
\begin{tabular}{ c|c cccccc }
  \# & Name & N orbits  &$\nu$ & $q$ & $\delta\phi^{\rm NR}_{\rm mrg}$ [rad] & $\Delta\phi^{\rm EOBNR}_{\rm mrg}$ [rad] & $\max(\bar{F})$ $[\%]$ \\
  \hline
1 & SXS:BBH:0066 & 28.1 & 0.25 & 1.00 & ... & $+0.0599$ & 0.10\\
2 & SXS:BBH:0002 & 32.4 & 0.25 & 1.00 & $-0.0633$ & $+0.1630$ & 0.47\\
3 & SXS:BBH:0007 & 29.1 & 0.24 & 1.50 & $-0.0183$ & $+0.0560$ & 0.07\\
4 & SXS:BBH:0169 & 15.7 & 0.22 & 2.00 & $-0.0267$ & $-0.1537$ & 0.05\\
5 & SXS:BBH:0030 & 18.2 & 0.19 & 3.00 & $-0.0857$ & $-0.0733$ & 0.04\\
6 & SXS:BBH:0167 & 15.6 & 0.16 & 4.00 & $-0.5076$ & $-0.0585$ & 0.04\\
7 & SXS:BBH:0056 & 28.8 & 0.14 & 5.00 & $+0.4405$ & $-0.0892$ & 0.03\\
8 & SXS:BBH:0166 & 21.6 & 0.12 & 6.00 & ... & $-0.2163$ & 0.03\\
9 & SXS:BBH:0063 & 25.8 & 0.10 & 8.00 & $+1.0118$ & $-0.4310$ & 0.04\\
10 & SXS:BBH:0185 & 24.9 & 0.08 & 9.99 & $+0.3499$ & $-0.5839$ & 0.07\\
\hline\hline
11 & SXS:BBH:0180 & 28.2 & 0.25 & 1.00 & $-0.4201$ & $+0.0709$ & 0.11\\
12 & SXS:BBH:0259 & 28.6 & 0.20 & 2.50 & $-0.0081$ & $+0.0403$ & 0.02\\
13 & SXS:BBH:0295 & 27.8 & 0.15 & 4.50 & $+0.2413$ & $-0.0254$ & 0.03\\
14 & SXS:BBH:0296 & 27.9 & 0.13 & 5.50 & $+0.4442$ & $-0.1595$ & 0.02\\
15 & SXS:BBH:0297 & 19.7 & 0.12 & 6.50 & $-0.0471$ & $-0.2669$ & 0.03\\
16 & SXS:BBH:0298 & 19.7 & 0.11 & 7.00 & $-0.0690$ & $-0.2874$ & 0.03\\
17 & SXS:BBH:0299 & 20.1 & 0.10 & 7.50 & $-0.0410$ & $-0.3097$ & 0.04\\
18 & SXS:BBH:0300 & 18.7 & 0.09 & 8.50 & $-0.0671$ & $-0.4308$ & 0.05\\
19 & SXS:BBH:0301 & 18.9 & 0.09 & 9.00 & $-0.1509$ & $-0.3616$ & 0.03\\
20 & SXS:BBH:0302 & 19.1 & 0.09 & 9.50 & $+0.0318$ & $-0.3947$ & 0.03\\
21 & SXS:BBH:0303 & 19.3 & 0.08 & 10.00 & $+0.3037$ & $-0.4566$ & 0.03
\end{tabular}
\end{ruledtabular}
\end{center}
\end{table*}

\begin{table*}[t]
  \caption{\label{tab:spinold}Spinning binaries: original spin-aligned SXS datasets used in Paper~I. The columns report: the number of the dataset; the name
    of the configuration on the SXS catalog; the symmetric mass ratio $\nu$; $(q,\chi_1,\chi_2)$, where $q=m_1/m_2$ is the mass ratio
    and $\chi_{1,2}$ are the dimensionless spins; the NR phase uncertainty at merger (when available), in radians, measured taking the
    difference between the two highest resolution levels (see text for details); the EOB/NR phase difference, in radians, at merger
    computed using EOB models informed with increasing amount of NR information (see text for details) and imposing an EOB/NR alignment
    during the early inspiral; the corresponding maximal unfaithfulnesses $\max{(\bar{F})}$, see Eq.~\eqref{eq:barF}.}
\begin{center}
\begin{ruledtabular}
\begin{tabular}{ l|lc c l c | c c c | c c l }
\# & Name & N orbits & $\nu$ & $(q,\chi_1,\chi_2)$ & $\delta\phi^{\rm NR}_{\rm mrg}$ & & $\Delta\phi^{\rm EOBNR}_{\rm mrg}$& & & $\max(\bar{F})$ $[\%]$ & \\ \hline
& & & & & & $c_3^{\rm old_{NQC}}$ & $c_3^{\rm new_{NQC}}$ & $c_3^{\rm new_{NQC'}}$ & $c_3^{\rm old_{NQC}}$ & $c_3^{\rm new_{NQC}}$ & $c_3^{\rm new_{NQC'}}$\\
\hline
\hline
22 & SXS:BBH:0004 & 30.2 & 0.25 & (1,$-0.50$,$0$) & $+0.0680$ & $+0.3731 $ & $+0.7742 $ & $+0.7742 $ & 0.07 & 0.09 & 0.09\\
23 & SXS:BBH:0005 & 30.2 & 0.25 & (1,$+0.50$,$0$) & $+0.2791$ & $+0.2930 $ & $+0.8445 $ & $+0.8445 $ & 0.04 & 0.15 & 0.15\\
24 & SXS:BBH:0156 & 12.4 & 0.25 & (1,$-0.95$,$-0.95$) & $-2.0991$ & $+0.3389 $ & $+0.3389 $ & $+0.3389 $ & 0.08 & 0.08 & 0.08\\
25 & SXS:BBH:0159 & 12.7 & 0.25 & (1,$-0.90$,$-0.90$) & $+0.3778$ & $+0.0746 $ & $+0.0746 $ & $+0.0746 $ & 0.08 & 0.08 & 0.08\\
26 & SXS:BBH:0154 & 13.2 & 0.25 & (1,$-0.80$,$-0.80$) & $-0.0058$ & $+0.1516 $ & $+0.1516 $ & $+0.1516 $ & 0.08 & 0.08 & 0.08\\
27 & SXS:BBH:0151 & 14.5 & 0.25 & (1,$-0.60$,$-0.60$) & $+0.1347$ & $-0.0082 $ & $-0.0082 $ & $-0.0082 $ & 0.05 & 0.05 & 0.05\\
28 & SXS:BBH:0148 & 15.5 & 0.25 & (1,$-0.44$,$-0.44$) & $+0.6981$ & $+0.8424 $ & $+0.8424 $ & $+0.8424 $ & 0.10 & 0.10 & 0.10\\
29 & SXS:BBH:0149 & 17.1 & 0.25 & (1,$-0.20$,$-0.20$) & $-0.9015$ & $+0.5001 $ & $+0.5001 $ & $+0.5001 $ & 0.14 & 0.14 & 0.14\\
30 & SXS:BBH:0150 & 19.8 & 0.25 & (1,$+0.20$,$+0.20$) & $-0.9593$ & $+1.0733 $ & $+1.0733 $ & $+1.0733 $ & 0.25 & 0.25 & 0.25\\
31 & SXS:BBH:0152 & 22.6 & 0.25 & (1,$+0.60$,$+0.60$) & $+0.3577$ & $-0.0211 $ & $-0.0211 $ & $-0.0211 $ & 0.05 & 0.05 & 0.05\\
32 & SXS:BBH:0155 & 24.1 & 0.25 & (1,$+0.80$,$+0.80$) & $+0.2657$ & $-0.3884 $ & $-0.3884 $ & $-0.3884 $ & 0.24 & 0.24 & 0.24\\
33 & SXS:BBH:0153 & 24.5 & 0.25 & (1,$+0.85$,$+0.85$) & $   \dots$ & $+0.1635 $ & $+0.1635 $ & $+0.1635 $ & 0.24 & 0.24 & 0.24\\
34 & SXS:BBH:0160 & 24.8 & 0.25 & (1,$+0.90$,$+0.90$) & $+0.7893$ & $+0.2250 $ & $+0.2250 $ & $+0.2250 $ & 0.46 & 0.46 & 0.46\\
35 & SXS:BBH:0157 & 25.2 & 0.25 & (1,$+0.95$,$+0.95$) & $+1.1689$ & $+0.2341 $ & $+0.2341 $ & $+0.2341 $ & 0.77 & 0.77 & 0.77\\
36 & SXS:BBH:0158 & 25.3 & 0.25 & (1,$+0.97$,$+0.97$) & $+1.2450$ & $+0.8158 $ & $+0.8158 $ & $+0.8158 $ & 0.69 & 0.69 & 0.69\\
37 & SXS:BBH:0172 & 25.4 & 0.25 & (1,$+0.98$,$+0.98$) & $+1.9682$ & $+0.4869 $ & $+0.4869 $ & $+0.4869 $ & 0.78 & 0.78 & 0.78\\
38 & SXS:BBH:0177 & 25.4 & 0.25 & (1,$+0.99$,$+0.99$) & $+0.3929$ & $+0.3418 $ & $+0.3418 $ & $+0.3418 $ & 0.83 & 0.83 & 0.83\\
39 & SXS:BBH:0178 & 25.4 & 0.25 & (1,$+0.994$,$+0.994$) & $-0.5200$ & $+0.3370 $ & $+0.3370 $ & $+0.3370 $ & 0.95 & 0.95 & 0.95\\
40 & SXS:BBH:0013 & 23.8 & 0.24 & (1.50,$+0.50$,$0$) & $   \dots$ & $+0.6178 $ & $+0.8734 $ & $+0.8450 $ & 0.14 & 0.20 & 0.19\\
41 & SXS:BBH:0014 & 22.6 & 0.24 & (1.50,$-0.50$,$0$) & $+0.1450$ & $+0.1839 $ & $+0.3484 $ & $+0.3413 $ & 0.04 & 0.04 & 0.04\\
42 & SXS:BBH:0162 & 18.6 & 0.22 & (2.00,$+0.60$,$0$) & $-0.6708$ & $-0.3531 $ & $-0.1951 $ & $-0.2199 $ & 0.05 & 0.12 & 0.11\\
43 & SXS:BBH:0036 & 31.7 & 0.19 & (3.00,$-0.50$,$0$) & $-0.0642$ & $+0.8544 $ & $+0.9466 $ & $+0.9408 $ & 0.03 & 0.03 & 0.03\\
44 & SXS:BBH:0031 & 21.9 & 0.19 & (3.00,$+0.50$,$0$) & $+0.0346$ & $+0.1774 $ & $+0.3294 $ & $+0.3288 $ & 0.05 & 0.06 & 0.06\\
45 & SXS:BBH:0047 & 22.7 & 0.19 & (3.00,$+0.50$,$+0.50$) & $   \dots$ & $+0.0811 $ & $-0.1858 $ & $-0.2240 $ & 0.06 & 0.15 & 0.09\\
46 & SXS:BBH:0046 & 14.4 & 0.19 & (3.00,$-0.50$,$-0.50$) & $   \dots$ & $+0.4618 $ & $+0.3490 $ & $+0.3434 $ & 0.08 & 0.05 & 0.07\\
47 & SXS:BBH:0110 & 24.2 & 0.14 & (5.00,$+0.50$,$0$) & $   \dots$ & $+0.0832 $ & $+0.2869 $ & $+0.2594 $ & 0.03 & 0.04 & 0.05\\
48 & SXS:BBH:0060 & 23.2 & 0.14 & (5.00,$-0.50$,$0$) & $   \dots$ & $+0.5855 $ & $+0.7012 $ & $+0.6994 $ & 0.04 & 0.04 & 0.04\\
49 & SXS:BBH:0064 & 19.2 & 0.10 & (8.00,$-0.50$,$0$) & $+0.8034$ & $+0.0499 $ & $+0.0286 $ & $+0.0001 $ & 0.06 & 0.06 & 0.06\\
50 & SXS:BBH:0065 & 34.0 & 0.10 & (8.00,$+0.50$,$0$) & $-2.9703$ & $+0.9736 $ & $+0.8511 $ & $+0.7970 $ & 0.02 & 0.03 & 0.04
\end{tabular}
\end{ruledtabular}
\end{center}
\label{tab:fitting}
\end{table*}

\begin{table*}[t]
  \caption{\label{tab:q1}Same structure of Table~\ref{tab:spinold}, but referring to new spin-aligned
    configurations, with $q<2$, not used in Paper~I. The datasets in bold, that give the largest maximal
    unfaithfulness when the $c_3^{\rm old_{\rm NQC}}$ determination is used, are the only ones used to
    provide the improved determination of $c_3$ that yield the  $c_3^{\rm new_{\rm NQC}}$ and $c_3^{\rm new_{\rm NQC'}}$
    values for $\Delta\phi^{\rm EOBNR}_{\rm mrg}$ and $\max(\bar{F})$.}
\begin{center}
\begin{ruledtabular}
\begin{tabular}{l|l c c l c | c c c | c c l }
$\#$ & Name & N orbits & $\nu$ & $(q,\chi_1$,$\chi_2)$ & $\delta\phi_{\rm NR}^{\rm mrg}$ & & $\Delta\phi^{\rm EOBNR}_{\rm mrg}$ & & & $\max(\bar{F})$ $[\%]$ & \\ \hline
  & & & & & & $c_3^{\rm old_{NQC}}$ & $c_3^{\rm new_{NQC}}$ & $c_3^{\rm new_{NQC'}}$ & $c_3^{\rm old_{NQC}}$ & $c_3^{\rm new_{NQC}}$ & $c_3^{\rm new_{NQC'}}$\\
  \hline
  \hline
51 & SXS:BBH:0212 & 28.6 & 0.25 & (1,$-0.80$,$-0.80$) & $-3.1754$ & $+1.6613 $ & $+1.6613 $ & $+1.6613 $ & 0.08 & 0.08 & 0.08\\
52 & SXS:BBH:0215 & 25.8 & 0.25 & (1,$-0.60$,$-0.60$) & $+2.1069$ & $+0.9068 $ & $+0.9068 $ & $+0.9068 $ & 0.04 & 0.04 & 0.04\\
53 & SXS:BBH:0170 & 15.5 & 0.25 & (1,$+0.44$,$+0.44$) & $+0.0278$ & $-0.4039 $ & $-0.4039 $ & $-0.4039 $ & 0.07 & 0.08 & 0.08\\
54 & SXS:BBH:0228 & 23.5 & 0.25 & (1,$+0.60$,$+0.60$) & $+2.4480$ & $+0.0888 $ & $+0.0888 $ & $+0.0888 $ & 0.05 & 0.05 & 0.05\\
55 & SXS:BBH:0230 & 24.2 & 0.25 & (1,$+0.80$,$+0.80$) & $-1.3471$ & $-0.2300 $ & $-0.2300 $ & $-0.2300 $ & 0.25 & 0.25 & 0.25\\
56 & SXS:BBH:0223 & 23.3 & 0.25 & (1,$+0.30$,$+0.00$) & $+0.1170$ & $+0.9346 $ & $+1.1444 $ & $+1.1444 $ & 0.12 & 0.21 & 0.21\\
57 & SXS:BBH:0222 & 23.6 & 0.25 & (1,$-0.30$,$+0.00$) & $+0.1361$ & $+0.6240 $ & $+0.7713 $ & $+0.7713 $ & 0.05 & 0.06 & 0.06\\
58 & SXS:BBH:0225 & 23.5 & 0.25 & (1,$+0.40$,$+0.80$) & $+2.5723$ & $+2.2891 $ & $+0.9983 $ & $+0.9983 $ & 0.47 & 0.09 & 0.09\\
59 & SXS:BBH:0224 & 22.9 & 0.25 & (1,$+0.40$,$-0.80$) & $+1.8737$ & $+2.2534 $ & $+1.3953 $ & $+1.3953 $ & 0.29 & 0.17 & 0.17\\
60 & SXS:BBH:0221 & 22.7 & 0.25 & (1,$-0.40$,$+0.80$) & $+1.7979$ & $+3.0846 $ & $+1.9880 $ & $+1.9880 $ & 0.80 & 0.44 & 0.44\\
61 & SXS:BBH:0220 & 25.7 & 0.25 & (1,$-0.40$,$-0.80$) & $+1.1940$ & $+1.9703 $ & $+1.3057 $ & $+1.3057 $ & 0.13 & 0.09 & 0.09\\
62 & SXS:BBH:0226 & 22.9 & 0.25 & (1,$+0.50$,$-0.90$) & $+2.4759$ & $+2.5333 $ & $+1.4938 $ & $+1.4938 $ & 0.34 & 0.17 & 0.17\\
63 & SXS:BBH:0219 & 22.4 & 0.25 & (1,$-0.50$,$+0.90$) & $+1.0612$ & $+3.4057 $ & $+2.1497 $ & $+2.1497 $ & 0.93 & 0.48 & 0.48\\
64 & SXS:BBH:0218 & 29.1 & 0.25 & (1,$-0.50$,$+0.50$) & $+0.4240$ & $+1.5324 $ & $+1.5324 $ & $+1.5324 $ & 0.17 & 0.17 & 0.17\\
65 & SXS:BBH:0227 & 23.1 & 0.25 & (1,$+0.60$,$+0.00$) & $+2.3492$ & $-0.1544 $ & $+0.6109 $ & $+0.6109 $ & 0.04 & 0.12 & 0.12\\
66 & SXS:BBH:0217 & 22.7 & 0.25 & (1,$-0.60$,$+0.60$) & $+1.6009$ & $+1.1389 $ & $+1.1389 $ & $+1.1389 $ & 0.19 & 0.19 & 0.19\\
67 & SXS:BBH:0216 & 23.6 & 0.25 & (1,$-0.60$,$+0.00$) & $+2.2546$ & $-0.0606 $ & $+0.3930 $ & $+0.3930 $ & 0.07 & 0.05 & 0.05\\
68 & SXS:BBH:0214 & 24.4 & 0.25 & (1,$-0.62$,$-0.25$) & $+1.8662$ & $-0.9374 $ & $-0.6141 $ & $-0.6141 $ & 0.11 & 0.08 & 0.08\\
69 & SXS:BBH:0229 & 23.1 & 0.25 & (1,$+0.65$,$+0.25$) & $+2.3005$ & $-0.6130 $ & $+0.1326 $ & $+0.1326 $ & 0.10 & 0.04 & 0.04\\
70 & SXS:BBH:0213 & 22.3 & 0.25 & (1,$-0.80$,$+0.80$) & $+1.2991$ & $+1.0759 $ & $+1.0760 $ & $+1.0760 $ & 0.20 & 0.20 & 0.20\\
71 & {\bf SXS:BBH:0232} & 23.9 & 0.25 & (1,$+0.90$,$+0.50$) & $+1.7528$ & $-2.4962 $ & $-1.3867 $ & $-1.3867 $ & 2.49 & 0.89 & 0.89\\
72 & {\bf SXS:BBH:0231} & 23.1 & 0.25 & (1,$+0.90$,$+0.00$) & $+2.8942$ & $-2.1568 $ & $-0.7340 $ & $-0.7340 $ & 1.39 & 0.20 & 0.20\\
73 & SXS:BBH:0211 & 22.3 & 0.25 & (1,$-0.90$,$+0.90$) & $+2.2000$ & $+1.0183 $ & $+1.0301 $ & $+1.0301 $ & 0.14 & 0.21 & 0.21\\
74 & SXS:BBH:0210 & 24.3 & 0.25 & (1,$-0.90$,$+0.00$) & $+1.1308$ & $-0.8632 $ & $-0.0140 $ & $-0.0140 $ & 0.22 & 0.07 & 0.07\\
75 & SXS:BBH:0209 & 27.0 & 0.25 & (1,$-0.90$,$-0.50$) & $-2.2779$ & $+0.0777 $ & $+0.6918 $ & $+0.6918 $ & 0.11 & 0.07 & 0.07\\
76 & SXS:BBH:0306 & 12.6 & 0.25 & (1.31,$+0.96$,$-0.90$) & $+0.7815$ & $-1.2673 $ & $-0.3053 $ & $-0.2937 $ & 0.29 & 0.40 & 0.39\\
77 & SXS:BBH:0025 & 22.4 & 0.24 & (1.50,$+0.50$,$-0.50$) & $+0.1348$ & $+0.7431 $ & $+1.1521 $ & $+1.1795 $ & 0.20 & 0.28 & 0.39\\
78 & SXS:BBH:0019 & 20.4 & 0.24 & (1.50,$-0.50$,$+0.50$) & $-0.3626$ & $-0.1380 $ & $+0.1427 $ & $+0.1416 $ & 0.05 & 0.04 & 0.04\\
79 & SXS:BBH:0016 & 30.7 & 0.24 & (1.50,$-0.50$,$+0.00$) & $-0.0311$ & $+0.6344 $ & $+0.8033 $ & $+0.7960 $ & 0.03 & 0.03 & 0.03
\end{tabular}
\end{ruledtabular}
\end{center}
\label{tab:fitting}
\end{table*}

\begin{table*}[t]
  \caption{\label{tab:q2}Same as Table~\ref{tab:q1}, but referring to $q=2$, spin-aligned, BBH configurations.
    The dataset in bold, SXS:BBH:0257, is the only one used of this set to provide the new determination of
    $c_3$ that yields the  $c_3^{\rm new_{\rm NQC}}$ values for $\Delta\phi^{\rm EOBNR}_{\rm mrg}$ and $\max(\bar{F})$.}
\begin{center}
\begin{ruledtabular}
\begin{tabular}{l|l c c l c | c c c | c c r }
$\#$ & Name & N orbits & $\nu$ & $(q,\chi_1$,$\chi_2)$ & $\delta\phi_{\rm NR}^{\rm mrg}$ & & $\Delta\phi^{\rm EOBNR}_{\rm mrg}$ & & & $\max(\bar{F})$ $[\%]$ & \\ \hline
  & & & & & & $c_3^{\rm old_{NQC}}$ & $c_3^{\rm new_{NQC}}$ & $c_3^{\rm new_{NQC'}}$ & $c_3^{\rm old_{NQC}}$ & $c_3^{\rm new_{NQC}}$ & $c_3^{\rm new_{NQC'}}$\\
  \hline
  \hline
80 & SXS:BBH:0234 & 27.8 & 0.22 & (2,$-0.85$,$-0.85$) & $-0.5179$ & $+3.3569 $ & $+1.9117 $ & $+1.9039 $ & 0.21 & 0.13 & 0.13\\
81 & SXS:BBH:0235 & 25.1 & 0.22 & (2,$-0.60$,$-0.60$) & $+0.7081$ & $+1.8290 $ & $+1.1071 $ & $+1.0990 $ & 0.11 & 0.08 & 0.08\\
82 & SXS:BBH:0238 & 32.0 & 0.22 & (2,$-0.50$,$-0.50$) & $+0.2792$ & $+2.2538 $ & $+1.6376 $ & $+1.6295 $ & 0.09 & 0.06 & 0.06\\
83 & SXS:BBH:0240 & 23.5 & 0.22 & (2,$-0.30$,$-0.30$) & $+0.0472$ & $+1.0541 $ & $+0.8448 $ & $+0.8386 $ & 0.06 & 0.05 & 0.05\\
84 & SXS:BBH:0243 & 23.3 & 0.22 & (2,$-0.13$,$-0.85$) & $+2.0139$ & $+2.7099 $ & $+1.8501 $ & $+1.8432 $ & 0.32 & 0.21 & 0.22\\
85 & SXS:BBH:0248 & 23.2 & 0.22 & (2,$+0.13$,$+0.85$) & $+0.9247$ & $+2.4351 $ & $+1.3149 $ & $+1.3294 $ & 0.36 & 0.23 & 0.23\\
86 & SXS:BBH:0251 & 23.5 & 0.22 & (2,$+0.30$,$+0.30$) & $+0.0170$ & $+1.3405 $ & $+1.0557 $ & $+1.0214 $ & 0.14 & 0.09 & 0.08\\
87 & SXS:BBH:0253 & 28.8 & 0.22 & (2,$+0.50$,$+0.50$) & $+0.1010$ & $+1.4502 $ & $+0.4482 $ & $+0.4217 $ & 0.15 & 0.06 & 0.06\\
88 & SXS:BBH:0256 & 23.9 & 0.22 & (2,$+0.60$,$+0.60$) & $+0.4916$ & $+1.6769 $ & $+0.1673 $ & $+0.1489 $ & 0.16 & 0.10 & 0.11\\
89 & {\bf SXS:BBH:0257} & 24.8 & 0.22 & (2,$+0.85$,$+0.85$) & $+0.3056$ & $+3.7223 $ & $+0.4311 $ & $+0.5929 $ & 1.04 & 0.08 & 0.07\\
90 & SXS:BBH:0247 & 22.6 & 0.22 & (2,$+0.00$,$+0.60$) & $+1.4523$ & $+1.3997 $ & $+1.0516 $ & $+1.0256 $ & 0.12 & 0.11 & 0.16\\
91 & SXS:BBH:0246 & 22.9 & 0.22 & (2,$+0.00$,$+0.30$) & $-4.2316$ & $+4.2002 $ & $+4.0612 $ & $+4.0646 $ & 0.07 & 0.06 & 0.06\\
92 & SXS:BBH:0245 & 23.0 & 0.22 & (2,$+0.00$,$-0.30$) & $+0.0053$ & $+1.2993 $ & $+1.2268 $ & $+1.2112 $ & 0.12 & 0.11 & 0.07\\
93 & SXS:BBH:0244 & 23.2 & 0.22 & (2,$+0.00$,$-0.60$) & $+1.7648$ & $+1.9083 $ & $+1.5937 $ & $+1.5893 $ & 0.23 & 0.19 & 0.19\\
94 & SXS:BBH:0250 & 23.2 & 0.22 & (2,$+0.30$,$+0.00$) & $+0.0114$ & $+1.0907 $ & $+1.1286 $ & $+1.1359 $ & 0.13 & 0.13 & 0.13\\
95 & SXS:BBH:0249 & 23.2 & 0.22 & (2,$+0.30$,$-0.30$) & $+0.0207$ & $+1.0894 $ & $+1.3016 $ & $+1.2862 $ & 0.13 & 0.23 & 0.16\\
96 & SXS:BBH:0242 & 23.1 & 0.22 & (2,$-0.30$,$+0.30$) & $+0.0381$ & $+0.2622 $ & $+0.4012 $ & $+0.3991 $ & 0.04 & 0.04 & 0.04\\
97 & SXS:BBH:0241 & 23.1 & 0.22 & (2,$-0.30$,$+0.00$) & $+0.0372$ & $+0.5318 $ & $+0.5254 $ & $+0.5607 $ & 0.03 & 0.03 & 0.03\\
98 & SXS:BBH:0252 & 22.5 & 0.22 & (2,$+0.37$,$-0.85$) & $-2.5921$ & $+1.6158 $ & $+1.6861 $ & $+1.6855 $ & 0.48 & 0.51 & 0.51\\
99 & SXS:BBH:0239 & 22.2 & 0.22 & (2,$-0.37$,$+0.85$) & $-2.8781$ & $-0.1778 $ & $-0.1285 $ & $-0.1272 $ & 0.11 & 0.11 & 0.11\\
100 & SXS:BBH:0255 & 23.3 & 0.22 & (2,$+0.60$,$+0.00$) & $-0.2908$ & $+0.0670 $ & $+0.1870 $ & $+0.2071 $ & 0.05 & 0.12 & 0.11\\
101 & SXS:BBH:0254 & 22.9 & 0.22 & (2,$+0.60$,$-0.60$) & $-0.0309$ & $+0.0725 $ & $+0.8745 $ & $+0.8826 $ & 0.44 & 0.61 & 0.60\\
102 & SXS:BBH:0237 & 22.6 & 0.22 & (2,$-0.60$,$+0.60$) & $-1.3705$ & $-0.8173 $ & $-0.4980 $ & $-0.4676 $ & 0.17 & 0.09 & 0.09\\
103 & SXS:BBH:0236 & 23.4 & 0.22 & (2,$-0.60$,$+0.00$) & $-0.2104$ & $-0.2146 $ & $-0.1289 $ & $-0.1348 $ & 0.06 & 0.05 & 0.06\\
104 & SXS:BBH:0258 & 22.8 & 0.22 & (2,$+0.87$,$-0.85$) & $+0.7293$ & $-1.8064 $ & $-0.5120 $ & $-0.4985 $ & 0.98 & 1.49 & 0.43\\
105 & SXS:BBH:0233 & 22.0 & 0.22 & (2,$-0.87$,$+0.85$) & $-1.4586$ & $-1.8336 $ & $-1.3124 $ & $-1.3166 $ & 0.54 & 0.29 & 0.31
\end{tabular}
\end{ruledtabular}
\end{center}
\end{table*}

\begin{table*}[t]
  \caption{\label{tab:q3}Same as Table~\ref{tab:q2}, but referring to spin-algined BBHs with $3\leq q\leq 7$. We stress that {\it none}
    of the datasets listed here was used in the determination of {\it any} value of $c_3$, but rather they only served to validate
    the model.}
\begin{center}
\begin{ruledtabular}
\begin{tabular}{l| l c l l c | c c c | c c r }
$\#$& Name &N orbits &$\nu$ & $(q,\chi_1$,$\chi_2)$ & $\delta\phi_{\rm NR}^{\rm mrg}$ & & $\Delta\phi^{\rm EOBNR}_{\rm mrg}$ & & & $\max(\bar{F})$ $[\%]$ & \\ \hline
  & & & & & & $c_3^{\rm old_{NQC}}$ & $c_3^{\rm new_{NQC}}$ & $c_3^{\rm new_{NQC'}}$ & $c_3^{\rm old_{NQC}}$ & $c_3^{\rm new_{NQC}}$ & $c_3^{\rm new_{NQC'}}$\\
  \hline
  \hline
106 & SXS:BBH:0260 & 25.8 & 0.19 & (3,$-0.85$,$-0.85$) & $-2.8231$ & $+1.3837 $ & $+1.0296 $ & $+0.9924 $ & 0.16 & 0.14 & 0.14\\
107 & SXS:BBH:0264 & 23.4 & 0.19 & (3,$-0.60$,$-0.60$) & $-1.9449$ & $+0.8182 $ & $+0.6367 $ & $+0.6341 $ & 0.09 & 0.08 & 0.08\\
108 & SXS:BBH:0267 & 23.4 & 0.19 & (3,$-0.50$,$-0.50$) & $-0.1123$ & $+1.4304 $ & $+1.2883 $ & $+1.2977 $ & 0.07 & 0.04 & 0.07\\
109 & SXS:BBH:0270 & 22.8 & 0.19 & (3,$-0.30$,$-0.30$) & $+0.0119$ & $+0.7593 $ & $+0.6971 $ & $+0.6930 $ & 0.05 & 0.04 & 0.04\\
110 & SXS:BBH:0283 & 23.5 & 0.19 & (3,$+0.30$,$+0.30$) & $-0.0063$ & $+0.9727 $ & $+0.8730 $ & $+0.8708 $ & 0.04 & 0.03 & 0.03\\
111 & SXS:BBH:0286 & 24.1 & 0.19 & (3,$+0.50$,$+0.50$) & $-0.3122$ & $+0.5992 $ & $+0.2557 $ & $+0.2143 $ & 0.06 & 0.15 & 0.09\\
112 & SXS:BBH:0291 & 24.5 & 0.19 & (3,$+0.60$,$+0.60$) & $+1.3163$ & $-0.1624 $ & $-0.5959 $ & $-0.6302 $ & 0.14 & 0.32 & 0.34\\
113 & SXS:BBH:0293 & 25.6 & 0.19 & (3,$+0.85$,$+0.85$) & $+1.1508$ & $+0.0013 $ & $-0.2404 $ & $-0.1945 $ & 0.20 & 0.29 & 0.27\\
114 & SXS:BBH:0278 & 22.8 & 0.19 & (3,$+0.00$,$+0.60$) & $+1.8831$ & $+1.0782 $ & $+0.9450 $ & $+0.9411 $ & 0.10 & 0.10 & 0.10\\
115 & SXS:BBH:0277 & 22.9 & 0.19 & (3,$+0.00$,$+0.30$) & $+0.0142$ & $+0.8898 $ & $+0.8964 $ & $+0.8525 $ & 0.05 & 0.05 & 0.04\\
116 & SXS:BBH:0276 & 23.0 & 0.19 & (3,$+0.00$,$-0.30$) & $-0.0680$ & $+1.2736 $ & $+1.2354 $ & $+1.2310 $ & 0.10 & 0.10 & 0.10\\
117 & SXS:BBH:0275 & 22.6 & 0.19 & (3,$+0.00$,$-0.60$) & $-0.8362$ & $+1.7759 $ & $+1.6612 $ & $+1.6568 $ & 0.18 & 0.16 & 0.16\\
118 & SXS:BBH:0279 & 22.6 & 0.19 & (3,$+0.23$,$-0.85$) & $-0.2303$ & $+1.8845 $ & $+1.9554 $ & $+1.9144 $ & 0.26 & 0.37 & 0.27\\
119 & SXS:BBH:0274 & 22.4 & 0.19 & (3,$-0.23$,$+0.85$) & $-1.0771$ & $+0.2083 $ & $+0.2309 $ & $+0.2277 $ & 0.13 & 0.13 & 0.13\\
120 & SXS:BBH:0280 & 23.6 & 0.19 & (3,$+0.27$,$+0.85$) & $+0.6050$ & $+1.4186 $ & $+0.8745 $ & $+0.8733 $ & 0.28 & 0.32 & 0.31\\
121 & SXS:BBH:0273 & 22.9 & 0.19 & (3,$-0.27$,$-0.85$) & $-1.3195$ & $+1.9565 $ & $+1.6140 $ & $+1.5855 $ & 0.12 & 0.14 & 0.09\\
122 & SXS:BBH:0282 & 23.3 & 0.19 & (3,$+0.30$,$+0.00$) & $-0.0463$ & $+0.9797 $ & $+1.0543 $ & $+1.0064 $ & 0.06 & 0.09 & 0.09\\
123 & SXS:BBH:0281 & 23.2 & 0.19 & (3,$+0.30$,$-0.30$) & $+0.0210$ & $+1.1088 $ & $+1.2655 $ & $+1.2626 $ & 0.10 & 0.18 & 0.18\\
124 & SXS:BBH:0272 & 22.7 & 0.19 & (3,$-0.30$,$+0.30$) & $-0.0059$ & $+0.1127 $ & $+0.1981 $ & $+0.1941 $ & 0.04 & 0.03 & 0.03\\
125 & SXS:BBH:0271 & 22.5 & 0.19 & (3,$-0.30$,$+0.00$) & $+0.0003$ & $+0.4184 $ & $+0.4446 $ & $+0.4403 $ & 0.03 & 0.03 & 0.03\\
126 & SXS:BBH:0285 & 23.8 & 0.19 & (3,$+0.40$,$+0.60$) & $+1.2523$ & $+0.9367 $ & $+0.5573 $ & $+0.5566 $ & 0.12 & 0.14 & 0.14\\
127 & SXS:BBH:0284 & 22.8 & 0.19 & (3,$+0.40$,$-0.60$) & $-0.4849$ & $+1.1027 $ & $+1.4184 $ & $+1.4149 $ & 0.29 & 0.26 & 0.26\\
128 & SXS:BBH:0269 & 22.3 & 0.19 & (3,$-0.40$,$+0.60$) & $-0.9915$ & $-0.3676 $ & $-0.2150 $ & $-0.2181 $ & 0.07 & 0.05 & 0.05\\
129 & SXS:BBH:0268 & 22.9 & 0.19 & (3,$-0.40$,$-0.60$) & $-0.8497$ & $+1.2626 $ & $+1.0233 $ & $+1.0199 $ & 0.09 & 0.08 & 0.08\\
130 & SXS:BBH:0174 & 35.5 & 0.19 & (3,$+0.50$,$+0.00$) & $-0.6660$ & $+1.2089 $ & $+1.4297 $ & $+1.4301 $ & 0.09 & 0.14 & 0.13\\
131 & SXS:BBH:0045 & 21.0 & 0.19 & (3,$+0.50$,$-0.50$) & $   \dots$ & $+0.0866 $ & $+0.5459 $ & $+0.4961 $ & 0.29 & 0.37 & 0.37\\
132 & SXS:BBH:0290 & 24.2 & 0.19 & (3,$+0.60$,$+0.40$) & $+0.1787$ & $-0.3282 $ & $-0.5082 $ & $-0.5447 $ & 0.11 & 0.17 & 0.18\\
133 & SXS:BBH:0289 & 23.8 & 0.19 & (3,$+0.60$,$+0.00$) & $+1.8791$ & $-0.5607 $ & $-0.3307 $ & $-0.3699 $ & 0.09 & 0.08 & 0.06\\
134 & SXS:BBH:0288 & 23.5 & 0.19 & (3,$+0.60$,$-0.40$) & $+1.9707$ & $-0.4652 $ & $+0.0664 $ & $+0.0241 $ & 0.30 & 0.40 & 0.40\\
135 & SXS:BBH:0287 & 23.5 & 0.19 & (3,$+0.60$,$-0.60$) & $-0.2407$ & $-0.2999 $ & $+0.3673 $ & $+0.3176 $ & 0.45 & 0.61 & 0.61\\
136 & SXS:BBH:0263 & 22.0 & 0.19 & (3,$-0.60$,$+0.60$) & $-0.9226$ & $-1.1034 $ & $-0.8608 $ & $-0.8634 $ & 0.18 & 0.12 & 0.12\\
137 & SXS:BBH:0266 & 22.0 & 0.19 & (3,$-0.60$,$+0.40$) & $+1.3841$ & $-0.9093 $ & $-0.7034 $ & $-0.7060 $ & 0.13 & 0.09 & 0.09\\
138 & SXS:BBH:0262 & 22.5 & 0.19 & (3,$-0.60$,$+0.00$) & $+1.2380$ & $-0.3905 $ & $-0.2917 $ & $-0.2943 $ & 0.05 & 0.04 & 0.04\\
139 & SXS:BBH:0265 & 23.4 & 0.19 & (3,$-0.60$,$-0.40$) & $+2.1253$ & $+0.3689 $ & $+0.3143 $ & $+0.3114 $ & 0.06 & 0.05 & 0.05\\
140 & SXS:BBH:0292 & 23.9 & 0.19 & (3,$+0.73$,$-0.85$) & $-0.6610$ & $-1.4035 $ & $-0.4570 $ & $-0.4977 $ & 1.05 & 1.45 & 0.18\\
141 & SXS:BBH:0261 & 21.5 & 0.19 & (3,$-0.73$,$+0.85$) & $-0.6331$ & $-1.6990 $ & $-1.3883 $ & $-1.3902 $ & 0.33 & 0.22 & 0.22\\
142 & SXS:BBH:0061 & 34.5 & 0.14 & (5,$+0.50$,$+0.00$) & $   \dots$ & $+0.5422 $ & $+0.8040 $ & $+0.7322 $ & 0.03 & 0.04 & 0.05\\
143 & SXS:BBH:0208 & 49.9 & 0.14 & (5,$-0.90$,$+0.00$) & $+8.8938$ & $+3.2712 $ & $+3.6577 $ & $+3.6624 $ & 0.08 & 0.08 & 0.08\\
144 & SXS:BBH:0202 & 62.1 & 0.11 & (7,$+0.60$,$+0.00$) & $+1.2013$ & $-0.1358 $ & $-0.1405 $ & $-0.2271 $ & 0.15 & 0.15 & 0.18\\
145 & SXS:BBH:0203 & 58.5 & 0.11 & (7,$+0.40$,$+0.00$) & $+8.0859$ & $+1.7976 $ & $+1.7958 $ & $+1.7585 $ & 0.02 & 0.02 & 0.03\\
146 & SXS:BBH:0204 & 88.4 & 0.11 & (7,$+0.40$,$+0.00$) & $+16.4975$ & $+0.0118 $ & $+0.0098 $ & $-0.0472 $ & 0.09 & 0.09 & 0.09\\
147 & SXS:BBH:0205 & 44.9 & 0.11 & (7,$-0.40$,$+0.00$) & $+4.4086$ & $+0.9674 $ & $+0.9671 $ & $+0.9641 $ & 0.03 & 0.03 & 0.03\\
148 & SXS:BBH:0206 & 73.2 & 0.11 & (7,$-0.40$,$+0.00$) & $+11.0775$ & $-0.4651 $ & $-0.4655 $ & $-0.4959 $ & 0.11 & 0.11 & 0.11\\
149 & SXS:BBH:0207 & 36.1 & 0.11 & (7,$-0.60$,$+0.00$) & $-1.3594$ & $+0.2969 $ & $+0.2974 $ & $+0.2680 $ & 0.03 & 0.06 & 0.06
\end{tabular}
\end{ruledtabular}
\end{center}
\label{tab:fitting}
\end{table*}

\bibliography{refs20170320}      

\end{document}